\documentclass{emulateapj}
\usepackage{natbib}
\usepackage{lscape}

\slugcomment{Accepted for publication in ApJ}

\shorttitle{Gamma-ray Spectral Variability of Blazars}
\shortauthors{Nandikotkur et al.}

\begin{document}

\title{Does the Blazar Gamma-ray Spectrum Harden with Increasing Flux? - Analysis of Nine Years of EGRET Data.}

\author{Giridhar~Nandikotkur\altaffilmark{1,2}, Keith~M.~Jahoda\altaffilmark{2}, R.~C.~Hartman\altaffilmark{2}, R.~Mukherjee\altaffilmark{3}, P.~Sreekumar\altaffilmark{4}, M.~B\"ottcher\altaffilmark{5}, R.~M.~Sambruna\altaffilmark{2}, and Jean~H.~Swank\altaffilmark{2}}

\altaffiltext{1}{Department of Physics, University of Maryland at College Park, College Park, MD 20742; giridhar@milkyway.gsfc.nasa.gov}
\altaffiltext{2}{Exploration of the Universe Division, NASA Goddard Space Flight Center, Greenbelt, MD, 20771}
\altaffiltext{3}{Department of Physics and Astronomy, Barnard College and Columbia University , New York, NY 10027}
\altaffiltext{4}{Space Astronomy and Instrumentation Division, ISRO Satellite Centre, Bangalore, Karnataka, 560017, India}
\altaffiltext{5}{Astrophysical Institute,Department of Physics and Astronomy, Ohio University,
Athens, OH 45701}

\begin{abstract}
The Energetic Gamma Ray Experiment Telescope (EGRET) on the 
Compton Gamma Ray Observatory (CGRO) discovered gamma-ray emission from more 
than 67 blazars during its nine-year lifetime. We conducted an exhaustive 
search of the EGRET archives and selected all the blazars 
that were observed multiple times and were bright 
enough to enable a spectral analysis using standard power-law models.
The sample consists of 18 flat-spectrum radio quasars (FSRQs),
6 low-frequency peaked BL~Lac objects (LBLs) and 2 high-frequency peaked 
BL~Lac objects (HBLs). We do not detect any clear pattern in the variation
of spectral index with flux. Some of the blazars do not 
show any statistical evidence for spectral variability. The spectrum
hardens with increasing flux in a few cases. There is also 
evidence for a flux-hardness anticorrelation at low fluxes in five 
blazars. The well observed blazars (3C~279, 3C~273, PKS~0528+134, 
PKS~1622-297, PKS~0208-512) do not show any overall trend in 
the long-term spectral dependence on flux, but the sample shows a  mixture 
of hard and soft states. We observed a previously unreported spectral hysteresis at weekly 
timescales in all three FSRQs for which data from flares 
lasting for $\sim$(3-4) weeks were available. All three sources show 
a counterclockwise rotation despite the widely different flux profiles. 
We analyze the observed spectral behavior 
in the context of various inverse Compton mechanisms believed 
to be responsible for emission in the EGRET energy range. Our analysis uses the 
EGRET skymaps that were regenerated to include the
changes in performance during the mission.

\end{abstract}


\section{Introduction}
\label{intro}
 Blazars are a class of active galactic nuclei (AGNs) characterized by highly luminous 
and rapidly variable continuum emission at all observed frequencies from radio to gamma-rays. 
Very long basline interferometry(VLBI) structures of these sources reveal compact cores with jet-like 
features which often show evidence of superluminal motion \citep{VC}.
The broadband spectral energy distribution (SED) of these sources shows two peaks. It has been 
widely accepted, in the scenario of leptonic models, that the lower-frequency peak is due to 
synchrotron emission from relativistic plasma  moving along the jet away from the core of the 
AGN while the second
 peak is attributed to inverse-Compton scattering of relativistic electrons by soft ambient photons,
produced either internal or external to the jet. 
These ``seed-photons'' for inverse-Compton emission could come from synchrotron 
emission itself as postulated by synchrotron-self Compton (SSC) models 
\citep{ghi85,mgc92,mg85,bm96}, or they could be entering the jet directly from the
 accretion disk as in the ECD (external Compton scattering of direct 
disk radiation) models \citep{der92,ds93}, or they could reach the jet 
after being re-scattered by surrounding broad-line-region (BLR) clouds as in
 the ECC (external Compton scattering from clouds) models \citep{sik94,bl95,dss97}.
 In addition, the BLR could also reflect the synchrotron photons back into the 
jet to undergo inverse-Compton scattering (external-reflection-Compton model; \citet{gm96}). 
Finally, the seed photons could be produced by the infrared (IR) dust that surrounds 
the blazar nucleus (external Compton from infrared dust-ERC(IR); \citet{sik02,blaz00}). 
The dust is more concentrated in a torus that lies in the equatorial plane of the blazar \citep{wag95}. 
Quite often, a combination of these models is required to fit the broadband spectrum of a blazar through
 the entire range of frequencies from radio to gamma-rays.

Observationally, the class of blazars includes flat-spectrum radio quasars
(FSRQs) and  BL~Lac objects. FSRQs have strong and broad  
optical emission lines while the lines are weak in BL~Lac objects. The 
 position of the peaks in a broadband SED allows a further division of BL~Lac objects into 
two categories: low-frequency-peaked BL~Lacs (LBLs)
and high-frequency peaked BL~Lacs (HBLs). The first peak is at infra-red/optical frequencies 
for {\it red blazars} which could be either the FSRQs or the Low-frequency-peaked blazars (LBLs) 
and at UV/X-rays for  {\it blue blazars} or the 
High-frequency-peaked blazars (HBLs). The second peak is in the gamma-ray range 
(MeV-GeV) for LBLs \& FSRQs and in the TeV range for HBLs.
HBLs are much lower in overall luminosity than FSRQs with 
LBLs somewhere between \citep{fos98}.

During its nine year lifetime, EGRET has detected GeV-range emission from more than 67 blazars and a number of them 
have been observed multiple times \citep{har99}. The EGRET energy range 
(30 MeV-10 GeV) lies near the maximum or on the falling portion of the inverse-Compton peak for FSRQs and
on the rising portion of the peak in the case of HBLs and it lies somewhere in between for LBLs.
A continuity in the observed spectral properties of BL~Lacs and FSRQs 
has been postulated by \citet{fos98} with the gamma-ray spectral index getting progressively harder
from FSRQs to HBLs. While this trend is expected of the average spectral properties of these
sources, previous studies have suggested a hardening of the gamma-ray spectral index
in FSRQs with an increase in flux. This was reported for individual blazars in \citet{muk95,sre96,bloom97,sta03}
and was also observed in the combined data from 18 brightest blazars \citep{sre01}. This feature, coupled with the
fact that the average spectral index of $2.15\pm0.04$ measured for blazars \citep{muk97} 
is quite close to the spectral index of $2.10\pm 0.03$ \citep{sre98} for 
diffuse gamma-ray background, is used  to attribute the extragalactic 
gamma-ray background to emission from unresolved blazars \citep{stec96}.

With the EGRET's calibration finalized and its archive now complete, the behavior of 
gamma-ray spectral index can be studied in detail across different epochs
and over a broad range of flux. This paper presents the results of such an effort and is 
organized as follows. We reanalyzed the entire blazar data from the EGRET mission for this 
project. Section \ref{sso} describes the data and \S \ref{anal} discusses the 
analysis procedure. We examine the spectral properties of different source classes,
the long term and the short term spectral variability in \S \ref{res}, 
discuss the implications of the results in \S \ref{disc}, and summarize in \S \ref{conc}.

\section{Source Selection and Observations}
\label{sso}
CGRO was launched on April 5 1991 and it re-entered the earth's atmosphere on June 4, 2000.
One of the four instruments on board was EGRET that was sensitive in the energy range 30 MeV-10 GeV.
The 3EG contains the basic 
results (flux and spectral indices) from analysis of all observations 
until the end of Cycle 4 (1995 October 3). \citet{muk97} 
presented summary results for all blazars detected through the end of Cycle 4 
and included the spectral indices for blazars that were detected at a 
significance greater than $6\sigma$. Although there were
very few new detections after Cycle 4 (e.g. PKS~2255-282 and Mrk~501), eight blazars were observed 
multiple times in Cycles 5-9.  Spectral analysis results after Cycle 4 are available only for 
PKS~0528+134 \citep{muk99II}, which contains results through the end of Cycle 6.

EGRET viewing periods (VPs) ranged in duration from 3 to 20 days 
but they were usually a week long. Sometimes an object was observed 
during two or more contiguous viewing periods, either as a part of the observing 
schedule or because the object was in an extremely active state. EGRET was operated with a 
narrow field of view for most of the latter half of the mission (Cycle 5 onward) to conserve gas lifetime, 
thus limiting the number of accessible targets. The details (viewing period number, 
start and end dates) of the viewing periods 
(after Cycle 4) are listed in columns 1-3 of Table \ref{Tab-2}. Columns 4 \& 5, respectively, 
list the sources that were in the field of view (FOV) during that time, and their
off-axis viewing angle. Information for viewing periods prior to Cycle~5 is listed in 3EG. 

We have analyzed all nine years of data for all the blazars seen by EGRET, and these objects 
are listed in Table \ref{Tab-5}. The sample consists of 98 sources, 
67 of which are confirmed identifications. The 31 ``possible'' identifications are marked by a 
question mark in column 2, and the more common names of the sources are listed in column 3. 
The distribution consists of 66~flat-spectrum radio quasars~(FSRQ), 
17~LBLs, 4~HBLs, 
10~flat-spectrum radio sources~(FSRS) and 1~radio galaxy. The classifications (listed in column~10 of Table~\ref{Tab-5})
have been adopted from 3EG and \citet{ghi98}. The 66 flat-spectrum radio quasars have been
further classified into 19 high-polarization quasars (FSRQ(HP)), 15 low-polarization quasars (FSRQ(LP)).
Polarization information could not be obtained for the remaining. 
Twenty-six of the 97 sources were observed multiple times and were bright enough during 
those observations to yield a spectral index. These sources are marked by a ``Y''in column~9.

\section{Analysis}
\label{anal}
EGRET was a spark chamber telescope with an effective area of   
$1000$ $cm^2$ at 150 MeV, $1500$ $cm^2$ between 500 MeV-1 GeV, which decreased
gradually to about $700$ $cm^2$ at 10 GeV. \
The off-axis sensitivity decreased as an approximate Gaussian with a 
FWHM of $\sim$$20^o$. The sensitivity beyond $30^o$ was less 
than $15\%$ of the on-axis sensitivity. Details of the instrument and calibration can be found in \citet{kan88,hug80,tho93}; and \citet{epo99}. 
During its nine year lifetime, the spark chamber gas was refilled multiple times \citep{bert01}, 
and for most of the latter half of the mission (Cycle 5 onward), EGRET was operated with a 
narrow FOV($18^o$ useful radius) to conserve gas lifetime. The detection efficiency of 
EGRET varied throughout the mission due to aging of the spark-chamber gas
between refills and a hardware failure in 1997. An energy dependent effect was also 
observed in the degradation. The method used to calibrate the efficiency up to Cycle 4 \citep{epo99} 
did not deal with this energy-dependence adequately. The results in this paper are derived from 
EGRET maps produced using the calibration described in \citet{bert01}.

One of the standard EGRET data products for any viewing period is a pair of maps 
showing gamma-ray arrival directions from the observed sky-region in the energy intervals
30-100 MeV and $>$100 MeV. For the work presented here, these maps were 
used in conjunction with a list of all 
the known EGRET-sources, to determine simultaneously the counts from all sources in the 
field of view and their significance of detection in the two energy-intervals 
through a method of maximum likelihood. Details of the 
maximum likelihood method and the process of determination of the significance of detection 
can be found in \citet{mat96} and \citet{epo99}. All sources that were detected at 
a significance $<~2\sigma$ in the energy interval $>$100 MeV were eliminated from the list 
and the process was repeated again to determine the counts and fluxes (along with the associated errors)
for the remaining sources.

If the source of interest was detected at a significance $>4 \sigma$ 
in the energy interval $>$100 MeV, then a four point spectrum was determined using counts 
recorded in the energy intervals (in MeV) 30-100, 100-300, 300-1000 and 1000-10000. 
The points were fitted with a single power law of the form 
$F(E) = k(E/E_o)^{-\alpha}$  photons cm$^{-2}$ s$^{-1}$ MeV$^{-1}$ where $F(E)$ is the flux, $\alpha$
the photon spectral index, $E$ the photon energy, $E_{o}$ the energy normalization factor and $k$ a
coefficient of normalization.

\begin{figure}
\plottwo{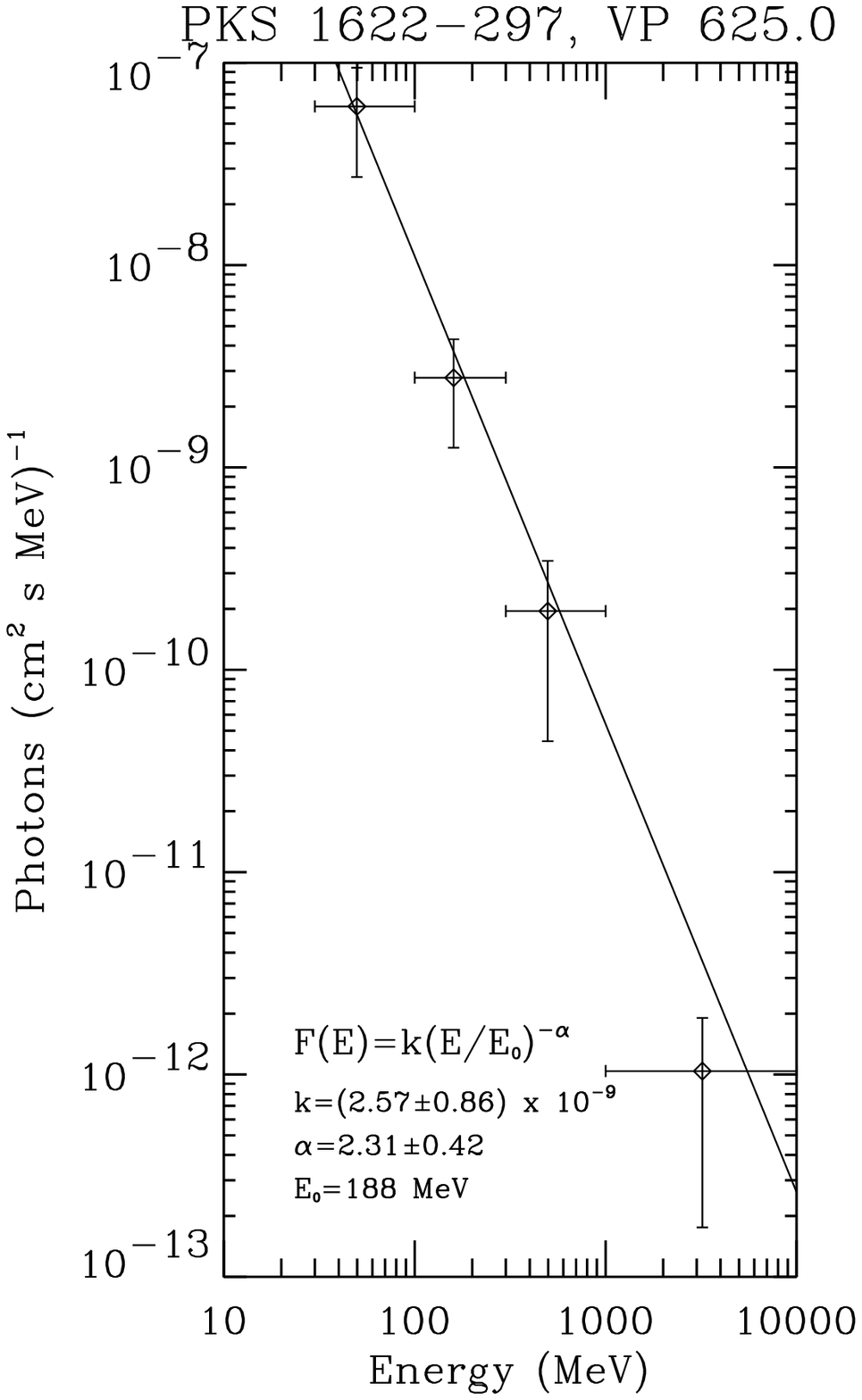}{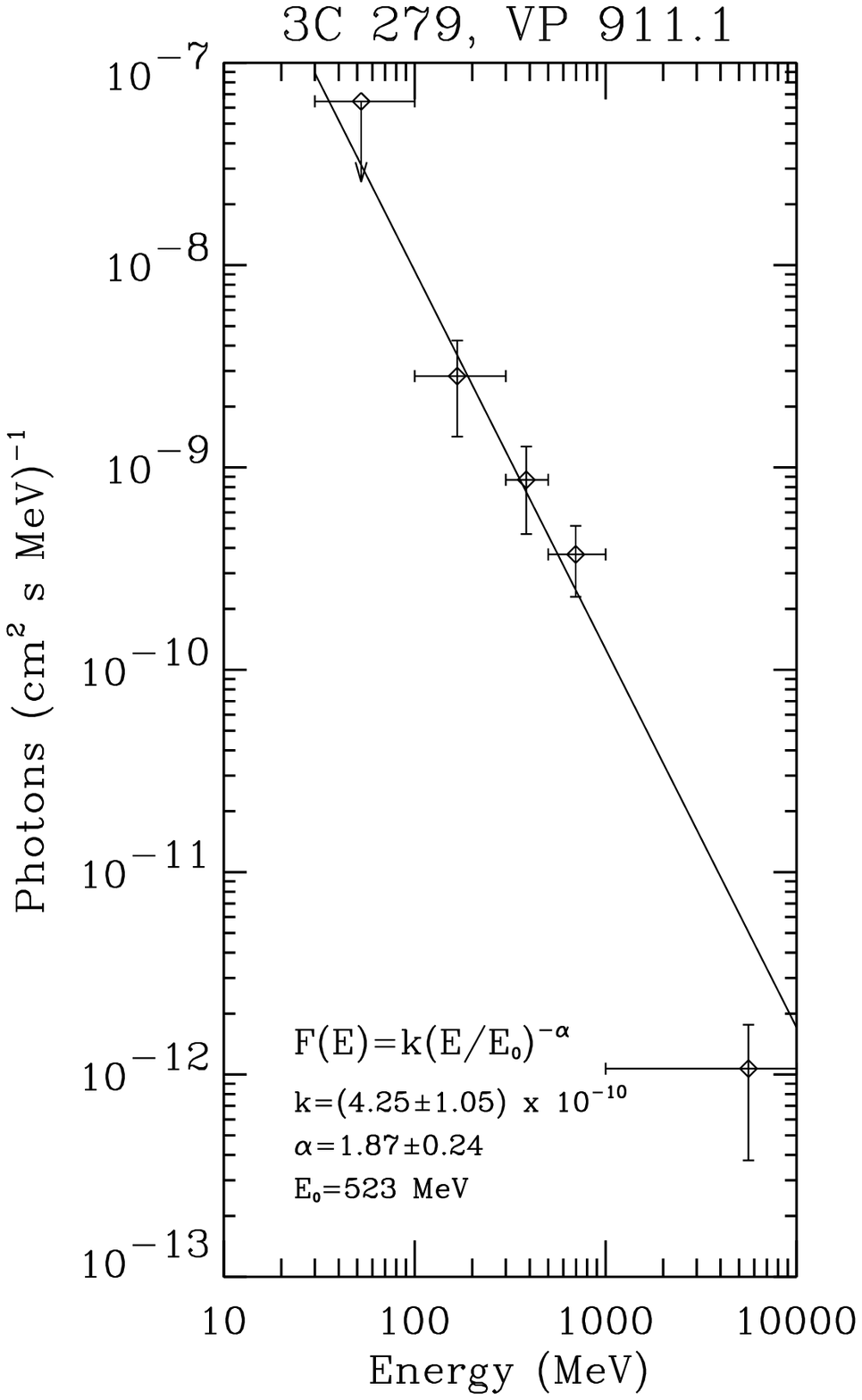}
\caption{Sample EGRET spectra of blazars. Left: PKS~1622-297 in August 1997. Right: 3C~279 during the flare in February 2000. The straight line is a power law fit to the data. The fit parameters (discussed in section \ref{anal}) are included in the figure.
\label{egsamp1}}
\end{figure}

If the overall significance of detection of the source was greater than $6\sigma$, the
energy intervals with a strong detection were further split up into smaller intervals 
(for which standard EGRET maps exist) to determine the spectral index. For the strongest sources, 
the standard 10 intervals 30-50, 50-70, 70-100, 100-150, 150-300, 
300-500, 500-1000, 1000-2000, 2000-4000, 4000-10000 (all in MeV) were utilized. 
Most of the spectra after Cycle 5 had to be determined using 4-5 energy intervals (when a source 
was not undergoing a flare). Figure \ref{egsamp1} shows sample four-point spectra from 
PKS~1622-297 and a 5-point spectrum from 3C~279. Analysis using the new maps has constrained the spectral 
indices better (lower errors) for a majority of the sources. Previous EGRET spectral analyses required a 
minimum of 6-sigma significance of source detection and used 10 energy bins to calculate the spectrum. 
We have used a slightly different approach, lowering the cutoff to 4~sigma. This does not affect 
the quality of the spectral analysis since we are using only 4-5 energy intervals  
for computing the spectral indices for faint sources, giving us better statistics 
in each bin, and lowering the errors. In addition, we found that the 
spectral index was within the error bars of the index calculated using 10 energy bins.

Some of the blazars considered here were part of extended campaigns. If 
the source was not very bright during such times,
adjacent viewing periods were combined. The analysis process was 
then repeated with the combined data, and an attempt was made to extract the 
spectrum. The longest period for which a source was in EGRET's field of view 
continuously was 49 days (7 viewing periods), for 3C~273 and 3C~279. 
Sometimes, all the observations during a cycle had to be combined to obtain 
a reliable detection and spectrum. 

We have done a complete spectral analysis for all the blazars detected by EGRET
using the recalibrated data products. Table \ref{Tab-5} (column 6) shows their 
average photon spectral indices. For the bright blazars that were observed multiple times,
we used the sample mean and standard deviation (of mean) as the spectral index. For 
the rest, we used the spectral index from all the data available unless a source was bright 
during one of the observations and was almost inactive 
during the rest of the viewing periods. Column~7 lists the mean flux ($>100$ MeV) recorded for these sources in units of
$10^{-8}$ photons cm$^{-2}$ sec$^{-1}$. Table \ref{Tab-3} 
lists the results of spectral analyses for sources which yielded more than 
one spectral index value. Columns 5, 6 \& 7 list the spectral index, flux and the detection significance, 
respectively. The viewing periods that were combined to get the spectra are listed in 
column~3 while their corresponding starting dates are listed in column 2 
in the same order. For identification purposes, each of these observations is 
labeled in the spectral index vs. flux plot shown in Figure \ref{egspec1}, with the labels listed in column 4 of Table \ref{Tab-3}.

\section{Results}
\label{res}
\subsection{Gamma-ray spectral distribution}
\label{specgdist}

Since a classification of blazars was based on the location 
of the synchrotron peak, we searched the literature for multiwavelength 
fits to data from all the blazars detected by EGRET, in order to determine the frequency of their 
synchrotron peaks and to examine its dependence on the gamma-ray spectral index.
Multiwavelength fits to the broadband spectrum 
($log(\nu F_\nu)$ vs $log(\nu)$) from 
simultaneous data are available for more than one epoch for:
3C~279 \citep{hart01,ballo02}, BL~Lac \citep{bott00}, 3C~273 \citep{kat02}, 
PKS~2155-304 \citep{chap99,kat00}, Mrk~421 \citep{tak00,kraw01}, 
Mrk~501 \citep{tav01,kat99,kraw02,pet00}, PKS~0528+134 \citep{muk99II}.
For the rest of the blazars, we used values from \citet{ghi98} \& \citet{von95} 
that are compilations of multiwavelength data (simultaneous and non-simultaneous) from literature 
and corresponding broadband model-fits. In cases where there is more than one fit available, 
or when a clear determination of the peak was not possible, the peak frequency was fixed 
at the average value and the error was calculated from one of the extremes. 
The logarithm of synchrotron peak frequency values have been listed in column 8 of 
Table ~\ref{Tab-5}.

\begin{figure}
\plotone{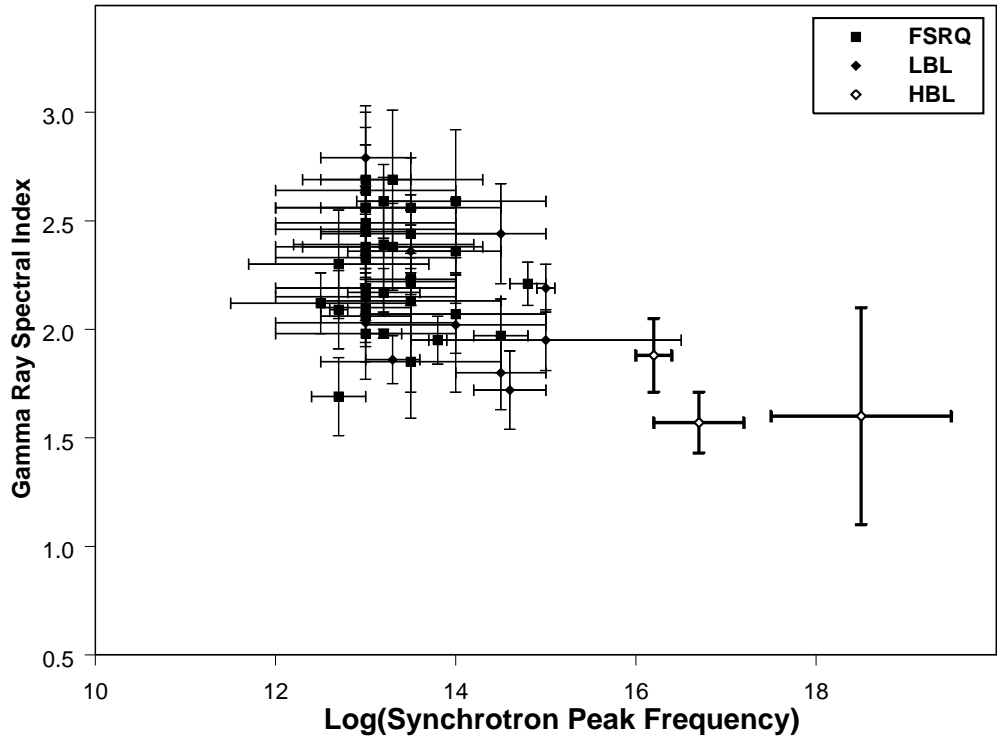}
\caption{Spectral index vs Log(Synchrotron Peak Frequency) of blazars. Classifications are based 
on \citet{ghi98} and \citet{hart97}. The sample of sources shown in the plot consists of 
37 FSRQs, 10 LBLs and 3 HBLs. The data for the plot comes from Table~\ref{Tab-5}}
\label{egpeak}
\end{figure}

The plot of gamma ray spectral index vs log synchrotron peak frequency for the blazars in our 
sample is shown in Figure \ref{egpeak}. The sample of sources shown in the plot consists of 
37 FSRQs, 10 LBLs and 3 HBLs. The blazars for which we could not find the 
synchrotron peak frequency in the literature have been excluded from the figure.
Since FSRQs have the lowest synchrotron-peak frequency and 
the EGRET energy range lies on the decreasing portion of their 
inverse Compton peak (in a plot of the broadband spectral 
energy distribution), they are expected to have soft spectral indices. 
HBLs have the highest synchrotron peak frequency and the 
EGRET-range lies on the rising portion of their 
inverse Compton peak. Consequently, they are expected to have hard spectral indices.
LBLs lie somewhere in between. Under this unified-blazar paradigm, a plot of 
gamma-ray spectral indices vs. synchrotron peak frequencies
should have a smooth variation from FSRQs to LBLs to HBLs \citep{fos98}.
A plot similar to that shown in Figure \ref{egpeak} was made in the past 
for 27 blazars using data through Cycle 4 \citep{lin99}.
A comparison of our data with this work shows that some of the spectral 
indices obtained by us are different, due to the
availability of more data and the recalibration of 
the raw data products, as described earlier (see section 3). 

We obtained a mean spectral index of 2.26$\pm$0.03 for the 66~FSRQs, 2.14$\pm$0.08 for
the 17~LBLs, 1.68$\pm$0.09 for the 3~HBLs and 2.48$\pm$0.1 for the 10 other flat spectrum radio sources (FSRS). 
The spectral index for FSRQs with high polarization (HP)
and low polarization (LP) was 2.19$\pm$0.06 and 2.32$\pm$0.06 respectively.
The spectral index increases across HBLs, LBLs, FSRQs(HP) and  FSRQs(LP). This is 
consistent with the prediction that the spectral properties of blazars form
a well defined sequence from HBLs to LBLs to FSRQs (HP,LP) \citep{ghi98,fos98}.

\subsection{Spectral variability with Flux}
\label{fluxvar}
\subsubsection{Long term spectral variability}
\label{lterm}

We searched for variability in the spectral index of all the blazars 
(for which two or more spectral indices could be calculated) 
using the $\chi^2$ test and the results are listed in Table~\ref{Tab-4}. Column 2 contains
the sample mean ($\Gamma_{\mu}$) and the standard deviation of the mean ($\sigma$) for each blazar. 
The $\chi^2_{red}$ value obtained from fitting the sample of spectral indices with a line of constant mean $\Gamma_{\mu}$
is listed in column 3. Column 4 lists the degrees of freedom (DOF) (which is one less than the sample size), 
while column 5 contains the confidence level for the presence of spectral variability.
We do not detect any statistical evidence for spectral variability in 16 of the 26 blazars.
The confidence levels for the presence of spectral variability are low ($<80\%$), mostly due to the large error bars on the spectral indices. 

We looked for spectral variability correlated with flux 
using the Pearson's correlation coefficient. The correlation coefficient is listed in column 6 of Table \ref{Tab-4}.
The coefficient, which could be calculated only in cases where there were three or more observations, is
negative when the spectral index (positive) hardens with increasing flux. 
The dependence of spectral index on flux is not uniform across all the blazars.
The index hardens with increasing flux in some cases,
softens in others, and in the rest does not vary with flux.
Using a cutoff of 0.8 for the correlation coefficient, we found the spectral index
to be  correlated with flux  in 10 of the 26 blazars (including those with 
two observations where there was visual evidence). The spectrum hardened
with increasing flux in 6 of them while the spectrum softened in the remaining 4.  
Only five sources satisfied both the spectral variability and the 
index-flux correlation criteria: PKS~0537-441, 1222+216 (4C~21.35), PKS~1633+382 (4C+38.41), 2200+420 (BL~Lac) and
2230+114. We discuss some individual sources below.

1253-055 (3C~279): This object shows spectral variability at a confidence level of 99.99$\%$ and
shows marginal evidence for hardening with increase in flux.
The spectral states at a flux $>$ $70 \times10^{-8}$  photons cm$^{-2}$ s$^{-1}$ span more 
than 85$\%$ of the range of fluxes observed. These states
do not show any overall trend in the 
spectral index  vs. flux space (correlation coefficient of
0.05) and do not show any significant evidence for 
spectral variability (confidence level of $58\%$). 
The quiescent states from Cycles 3 and 4 have a softer spectral index 
when compared with the average value of 1.96 while the quiescent state from
Cycle 6 has a harder spectral index.

PKS~0208-512: We do not see any overall trend for this source, in spite of a strong evidence for 
spectral variability (confidence of 98$\%$). However, the spectral index does show evidence of 
hardening with increasing flux (correlation coefficient of -0.95) at fluxes higher than $60 \times 10^{-8}$ 
photons cm$^{-2}$ s$^{-1}$. A similar trend was also observed in this source by \citet{sta03} who combined 
simultaneous data from the Compton Telescope (COMPTEL; 0.75-30 MeV) and EGRET. They obtained a correlation coefficient 
of -0.78 between the spectral index in the 0.75 MeV - 10 GeV range and the flux ($>$100 MeV) 
recorded in the EGRET energy range. 

We observe a softening in the spectral index (coefficient of +0.95) as the flux increases, 
at fluxes lower than $80 \times 10^{-8}$ units. There is an indication
of this effect at lower fluxes in \citet[see their Figure 4]{sta03}, but the large error bars do not justify a separate fit.
Moreover, PKS~0208-512 has been categorized as an ``MeV-blazar'' and the spectrum from these sources 
shows a break between 1-20 MeV \citep{sik02,ski97,blo95,coll97}. Hence, a single power law does not adequately describe
the entire energy range form 0.75 MeV-10 GeV. Flux anti-correlations between COMPTEL and EGRET could 
also be expected for MeV-blazars (observed in case of PKS~0528+134, also a possible MeV blazar; \citet{coll97}).
But a reanalysis of the 1993 COMPTEL data by \citet{sta03} lowered the significance of the only detection
of this source in the MeV energy range, with no detections in its many subsequent observations.
Consequently, the association of PKS~0208-512 with MeV-blazars is questionable. But the unique nature of 
the spectral dependence on flux (initial softening and subsequent hardening) in the EGRET energy range, 
makes this strong gamma-ray source an interesting candidate for future observations.

PKS~0528+134: Previously published results for this object \citep{muk96}
showed a correlation of -0.85 between spectral index and flux using data from viewing periods 
0.2-0.5 (combined), 1.0, and 213.0. The same combination of viewing periods using 
recalibrated data did not show any evidence of spectral hardening. 
Inclusion of data through viewing period 420.0 decreased the 
correlation to -0.5 \citep{muk99II}. We obtained a correlation coefficient
-0.5 for the complete data which included observations from Cycles 5 and 6.  
The large error bars yield a low confidence of spectral variability of
$67 \%$ despite a spread in the values.

PKS~0537-441, PKS~1633+382 (4C+38.41): Spectral indices for these objects 
harden with increasing flux (correlation coefficients of
-0.97 \& 0.99) and show spectral variability at a confidence of 
87$\%$ and 98$\%$ respectively. 

2200+420 (BL~Lac) \& 2230+114 (CTA 102):  The spectrum hardens with increasing flux in these sources. 
The correlation coefficient was not calculated in these cases as there were only two observations.

Some of the FSRQs and LBLs show spectra that appear to soften with increasing flux. 
This can be seen in PKS~1222+216 (4C~21.35), PKS~1219+285 (ON~231), and, also in PKS~0208-512 
and S5~0716+714 at low fluxes.\\

HBLS; 1101+384 (Mrk~421) \& PKS~2155-304: The spectral index for Mrk~421 hardens as flux increases  
with a coefficient of -0.997. PKS~2155-304, however, shows a softening in spectral index 
with increasing flux (correlation coefficient of +0.99). This seems to be in contradiction
to standard scenario wherein the gamma-ray emission from HBLs is from SSC mechanism \citep{bot02,li00}.
HBLs are not very bright at EGRET energies and some of the spectral states show 
deviations from a simple power-law fit. Consequently, the error bars on the spectral indices
are large and the confidence level of spectral variability from the $\chi^2$ test is low.

\subsubsection{Short term spectral variability: \it{Spectral hysteresis during a flare}}
\label{sterm}

Blazars vary on multiple timescales. The smallest timescale of resolution 
(based on the light curves) for a gamma-ray flares is 
about $\sim 3-8$ hours \citep{mat97,nan97}. A study of spectral hysteresis (in relation to flux) 
during flares provides crucial insights into factors related to
the comoving electron dynamics (electron acceleration and cooling) \citep{kus00,li00,bot02}. 
Since we have extracted spectral indices from data accumulated over a 
complete viewing period (usually 7 days long), we cannot track spectral changes 
during a flare. Some of the sources, however, were observed for 
more than 2 consecutive viewing periods when they were flaring, either
because the observations were planned a priori, or because the schedule was 
changed to track the flare. The flux history in such cases could show an event 
occurring at larger timescales. Figure \ref{egspec2} shows the light curve and 
spectral index vs. flux plots of the flare for the three sources for which we had data from 
four contiguous viewing periods. We discuss the individual sources below.

PKS~1622-297: This source was observed from June 6, 1995 (VP~421.0) 
until July 7, 1995 (VP~423.5) for 4 consecutive viewing periods when it underwent 
a large flare. \citet{mat97} split the observations 
into 10 unequal intervals with the size of the intervals ranging from 1 day to 10 days,
A plot between flux in the two intervals- 100~MeV-300~MeV (x) and $>$~300~MeV (y) showed a
clockwise progression in time. Previously published spectral index 
values \citep{muk97} for the four viewing periods during this flare 
show evidence for spectral variability at a confidence level 
of $97\%$ ($\chi^2$ of 9.81 with 3 degrees of freedom) 
but a large fraction ($68\%$) of $\chi^2$ was due to the 
hard spectral index of 1.72$\pm$0.15 during viewing period 422.0, which recorded the largest 
flux. We obtained a spectral index of 2.22$\pm$0.11 for this viewing period
with the recalibrated maps. A $\chi^2$ test yielded confidence level of $51\%$,
with most of the contribution from viewing period 421.0. The spectral indices 
trace out a counterclockwise loop (as time progresses) in the spectral index-flux space. 

PKS~1406-076: Four out of the five points for this object 
are from 4 consecutive viewing periods from December 22, 1992 (VP~204.0) 
till February 2, 1993 (VP~207.0). While the light curve shows two successive flares,
the large error bars on the spectral indices reduce the variability in spectral
index to a confidence of $62\%$.  The spectral indices show a counterclockwise progression
during the period.

PKS~0528+134: Observations of this object in April-May 1991
spanned 36 days. The source was observed from April 4, 1991 to 
May 4, 1991 and then for 2 weeks from May 16, 1991, and a week from June 6, 
1991. These points are marked with their viewing-periods number (0.2-0.5, 1.0 \& 2.1)
in Figure~\ref{egspec1}. The $\chi^2$ test shows spectral variability at 
a confidence level of 73$\%$, which is the largest among the three sources. The light curve in Figure~\ref{egspec2} shows
the contiguous viewing periods 0.2-0.5 to be a part of a single event and the spectral
index traces out an intertwined loop in counterclockwise direction as in the other cases.

Spectral hysteresis is a commonly observed phenomenon in HBLs at X-ray wavelengths. It has been reported
at multiple timescales ranging from hours \citep{tak96,sem93,gli06} to seconds \citep{cui04} indicating
the presence of scale invariance in the spectral evolution at X-ray wavelengths.  We have 
observed hysteresis at weekly timescales in all the three FSRQs for which data from at least 4 contiguous 
viewing periods was available. This effect has previously never been observed in FSRQs in gamma-rays at 
these timescales. The light curve in Figure~\ref{egspec2} shows completely different flare 
profiles for the 3 blazars. The flare in PKS~0528+134 is a combination of a slow rising phase accompanied
by a faster decaying phase while the profile in PKS~1622-297 is exactly the opposite with a fast rise and
a comparatively slower decay. The light curve in the case of PKS~1406-076 shows multiple events- a flare consisting 
of a uniform rise and decay, and a second flare that was partially captured. Although the
flux profiles are quite different, all the three sources show a
counterclockwise rotation in the spectral index vs. flux space as time progresses. 
	
\section{Discussion}
\label{disc}
Under a leptonic jet paradigm, the broadband spectral energy distribution of blazars is modeled 
by considering synchrotron and inverse Compton emission from a blob of $e^{+}e^{-}$ 
plasma moving relativistically along the jet axis. The seed photons for the 
inverse Compton process could come from the jet (SSC), the accretion disk (ECD),
the broad line region clouds (ECC) or the infra-red dust in the surrounding torus ERC(IR) (see \S \ref{intro}
for the references for various emission mechanisms).

\subsection{Spectral variability in FSRQs and LBLs}
External-Compton process appears to play an important role 
in FSRQs and LBLs \citep{ghi98}. In a broadband 
spectral energy distribution, the peak emission frequency for ERC(IR), ECD and ECC 
processes increases in that order. The latter two have their peak frequency in the EGRET energy 
range and affect spectral variations in a more direct way. However, not all FSRQs have a strong 
infra-red component since it depends on the size of the torus and the density of the circum-nuclear dust. 
The inverse Compton component due to SSC emission does not have a 
pronounced peak. The plateau of the SSC emission extends from hard X-rays 
(few hundred keV) to the GeV gamma-ray region.

When the source is in a low-intermediate state, emission from 
the SSC process is at least as important as that from 
the three ERC processes, and the photon spectrum could be hard or soft, 
based on relative fraction of flux from these processes. Model fits 
to the broadband spectrum from 3C~279 by \citet{hart01} \& \citet{ballo02} 
and PKS~0528+134 by \citet{muk99II} show that moderate-large flares require a greater 
contribution from the ERC processes and have a higher bulk Lorentz factor $\Gamma$
for the particles. 

As the source undergoes a flare, the 
energy density of the IR field and the broad line field (BEL), both of which are proportional to 
$\Gamma^2$ \citep{sik02} as measured 
in the comoving frame of the plasma, also increase. The flares might also be related to 
the injection of energetic particles near the base of the jet, where the 
external radiation fields have a higher density than further out. 
Consequently, the contribution of external Compton process increases. 
Although emission from all processes increases, the ECD and ECC processes 
affect the EGRET energy range more. However, the extent of contribution from the ERC(IR) 
process, whose peak lies below the EGRET energy range \citep{blaz00}, 
could be one of the factors that determines how soft the 
EGRET spectrum is during the low-intermediate states as it affects the 
lower energies of the EGRET range. Broadband fits for
3C~279 from various epochs in \citet[their fig. 2]{hart01} show the 
low-intermediate states (Cycle 3, 4 \& 6 in our Figure \ref{egspec1}) to have comparable 
fluxes from SSC, ECD, and ECC processes, with the EGRET data lying on the falling portion of 
the SSC emission plateau. The 0.3-30 MeV region, which is just below EGRET energy-range,
shows a significant excess (in the broadband fits for Cycles 3 and 4) that could be 
caused by the ERC(IR) process (not included in the model). Hence the resultant EGRET spectra are soft. 
The enhanced MeV emission could also explain the softening 
in the spectral index seen from low to intermediate fluxes in PKS~0208-512, which has 
been categorized as an ``MeV-blazar'' \citep{ski97,blo95} and more recently questioned by \citet{sta03}.
The existence of MeV-blazars as a separate class has been disputed due of the dearth of 
sources. However the Swift Burst Alert Telescope has recently detected 
bright emission, and rapid variability at short timescales (down to 1-2 ks) from the optically faint quasar 
J0746 in the 15-195 keV range leading to the inclusion of this blazar in the MeV-blazar class \citep{sam06}.

While the ERC(IR) process could play an important role during the low-intermediate states, 
there is a substantial increase in emission from the ECD and ECC process during large flares, 
as shown in the case of 3C~279 by \cite{hart01}. The EGRET spectrum during these flares is hard or 
soft based on the relative contribution from these two processes. 
Since the IR emitting region is very large and far away, the ERC(IR) process 
does not vary over timescales as short as those of the other two process
and does not play a significant role in the spectral variations seen at short time scales 
during flares that typically last a few days.

Spectral hysteresis: The three sources (PKS~1622-297, PKS~0528+134, PKS~1406-076)
that were observed for an extended time during flares, show evidence of 
counterclockwise spectral hysteresis. A possible explanation for this may be  
the emergence of an external Compton component near the onset of the flare, 
where an ERC component at MeV-GeV energies might be dominant. 
As the flare evolves and the emission region moves out, the 
intrinsically harder SSC radiation might take over 
(as it takes time $\sim$ R/c for the internal radiation field to build up), leading to
hardening of the spectrum as the flux is already decreasing. This would work in 
situations in which the acceleration time scale is shorter than the
cooling time scale, and the cooling time scale is of the order of the 
dynamical time scale, R/c.

\subsection{Spectral variability in HBLs- a possible external component?}
Inverse Compton emission in the case of HBLs is usually attributed to the SSC mechanism.
The EGRET energy range lies on the rising portion the inverse Compton bump for Mrk 421.
As the source undergoes a flare, the maximum flux at the peak increases, leading to increased 
contribution at higher frequencies on the low energy tail of the inverse Compton bump causing
the spectra to harden. In PKS~2155-304, however, we observed the opposite trend of spectrum softening
with increasing flux. This is in contradiction with the standard SSC interpretation \citep{kus00,li00,bot02}.
The soft spectral index (2.22$\pm$0.46) is from the flare in November
1997 (viewing period 701.0 in Figure \ref{egspec1}). In a broadband distribution 
(plot of $\nu F_{\nu}$ vs $\nu$), a photon spectral index of 2 is a horizontal line. The 
rising portion of inverse-Compton peak has a spectral index less than 2 while the decreasing 
portion of the synchrotron peak would have a photon spectral index of greater than 2.
If the frequency range of observations is close to the frequency where these two branches meet,
it is possible for the synchrotron branch to move in to the observed energy range during flares
leading to a spectral index that is higher than 2.0. But the EGRET energy range is 
quite far from where the two branches meet in case of PKS~2155-304 \citep[see their fig. 7]{kat00}. 
Extending the synchrotron emission up to EGRET energies would come close to the theoretical 
limit for synchrotron emission from leptons for Doppler factors usually seen in such sources. The observed trend 
in PKS~2155-304 could be an indication for a quasi-
external Compton component expected from a decelerating-jet model \citep{geo03}, 
where synchrotron emission from a previous, slower
component may provide an additional target photon field for Compton
scattering. Alternatively, this could be a signature of proton synchrotron
emission in a hybrid leptonic/hadronic model \citep{muc03} since one would expect that there might be a non-negligible proton fraction present in the jet.

While the varying dominance of ERC components (or an absence thereof) 
can explain some of the gamma-ray spectral variability observed so far, it is just one of the 
many possible scenarios.  Gamma-ray spectral variability could arise out of a combination of factors 
that are both internal and external to the plasma. The internal factors affect the 
two energy cutoffs of the particle injection spectrum, the injection spectral index, the injection energy,
the magnetic field, the bulk Lorentz factor of plasma and the particle density.
The external factors are: the energy density of the infra-red field due 
to the dusty torus, the energy density of the 
broad emission line region, and the level of accretion disk activity. 
The current gamma-ray data, however, do not allow us to effectively explore the large
parameter space as the error bars on the 
spectral indices are usually high. This is due to the limitations of EGRET's sensitivity.
The Gamma-ray Large Area Space telescope (GLAST)\footnote{http://glast.gsfc.nasa.gov} 
with a higher sensitivity and greater duty cycle of coverage for individual sources should be 
able to determine spectral indices more accurately. In addition to the large number of 
parameters that go into the jet models, the  long integration 
times of EGRET imply that we are averaging over substantial and completely arbitrary 
sections of individual outbursts, or adding contributions from multiple smaller outbursts.
A study of spectral hysteresis during individual flares could indicate if we are dealing with
global, structural changes (including, e.g., a change of the bulk Lorentz factor), or with factors 
related to the co-moving electron dynamics (electron acceleration/cooling) \citep{kus00,li00,bot02}.
The EGRET data do not allow us to study spectral evolution during a flare. 
This would be an area of study ideal for GLAST.

\subsection{Blazars as a source of extragalactic gamma-ray background}
\citet{stec96} postulated that the {\it entire} diffuse 
extragalactic gamma-ray background (EGRB) can be attributed to emission from 
unresolved blazars, based on two factors:~(1) The spectral index for diffuse 
gamma-ray background is $2.10\pm 0.03$ \citep{sre98}, which is quite close
to the previously published value of the average spectral index ($2.15\pm0.04$) for
all observed blazars \citep{muk97} and (2) The preliminary concave shape of the 
extragalactic gamma-ray background, determined prior to 1995, 
could be well fitted by the diffuse emission calculated from the blazar 
luminosity function \citep[their Figure 3]{stec96}. However, the EGRB spectrum 
published in \citet{sre98} shows deviations from a powerlaw 
index of 2.1$\pm$0.03 at low and high energies. The curvature is less prominent but
it cannot be ruled out. In addition, the flare-state spectra had to be harder 
than the quiescent state spectra for a good fit\citep[their Figure 3]{stec96} . 

We obtained a value of 2.25$\pm$0.03 (2.22$\pm$0.03 for FSRQ+LBL+HBL)
for the average spectral index of all the blazars observed by EGRET and a median value of 2.25. 
64 of the blazars have spectral indices $\ge$2.1 while 33 of them have spectral indices that were lower.

A more current background spectrum generated using the finalized EGRET data from Cycles 1-4, based 
on a model different than that used in \citet{sre98}, was published by \citet{stro04} and is shown 
in Figure \ref{egrb}. The broadband spectrum from 30 MeV -50 GeV shows a clear break at 2 GeV. The spectrum is steeper 
than the spectrum in \citet{sre98} at energies below 2 GeV. 
The spectrum has a concave curvature due to the break and the rise 
beyond 10 GeV. Although the whole range from 30 MeV -50 GeV
shows substantial deviations from a power law fit, we find that the points 
below 2 GeV can be fitted well with a power law of slope 2.24$\pm$0.01 
and a correlation coefficient of 0.99. The slope is very close
to our value for the average blazar spectral index of 2.25$\pm$0.03 (superposed in Figure \ref{egrb}). 
 
As we noted earlier, if the EGRB has to be entirely due to blazars, then 
the flare-state spectra must be harder than the quiescent state spectra 
to produce concave shape in the diffuse background spectrum \citep{stec96}. We did not 
observe any strong evidence for the flaring states 
to have a harder spectrum in case of the  well-observed blazars. The spectrum hardened with 
increasing flux in some of them while it softened in some others. For some blazars, both trends were
observed. Consequently, any discussion of blazars as sole contributors to 
diffuse extragalactic background depends on the similarity of the blazar spectrum to 
that of the EGRB. The proximity of two indices
below 2 GeV certainly makes blazars a prime candidate for contributing to the diffuse background.

The current blazar data shows some deviations from power law during flares in some cases, but at 
energies below 2 GeV. However the data do not have enough statistics to attempt a detailed analysis using
broken power law models. Although the blazar spectrum cannot be measured accurately above 2~GeV due to 
EGRET's limited sensitivity there is no a priori reason to expect a sharp break in the blazar spectrum at
 2 GeV along with an increasing contribution at energies higher. This suggests
the necessity of an increased contribution by other sources to the extragalactic gamma-ray background 
at higher energies. The current estimates for blazar contribution to the diffuse background emission range from 
nearly 100\% \citep{stec96} to 25\% \citep{chi98,muc00}. GLAST, with its higher sensitivity 
and larger energy range (up to 100 GeV) would be able to measure the diffuse 
background more accurately and help narrow down the class of sources
contributing to it.

\section{Conclusions}
\label{conc}
We analyzed all nine years of EGRET data for blazars and noted the following.
\begin{enumerate}
\item
The sample contained 98 sources: 66~FSRQs, 17~LBLs, 4~HBLs, 10~FSR sources and 1~radio galaxy.
We obtained a mean spectral index of 2.26$\pm$0.03 for FSRQs, 2.14$\pm$0.08 for
LBLs, 1.68$\pm$0.09 for HBLs (spectral index could not be calculated for one of them), and 2.48$\pm$ 0.1 for flat spectrum radio sources. 
The gamma-ray spectral index shows a transition 
from FRSQs to LBLs to HBLs with FSRQs having the softest spectral index and HBLs 
having the hardest.
\item
We did not observe any clear correlation between the gamma-ray spectral index 
and flux. A majority of blazars did not show any overall trend. The spectra
hardened with increasing flux in some, while it softened in some energy intervals
for few others.  For those blazars where the spectra varied and did not show an overall trend, 
the sample consisted of a mixture of hard and soft states. 
\item
We observed a previously unreported counterclockwise hysteresis at weekly timescales in the spectral index 
vs. flux space. The effect was consistently seen in the flare data from all the 3 FSRQs which were
observed for at least 4 contiguous viewing periods during the flare. The flux profiles of these sources 
were very different from each other.
\item
It is difficult to understand clearly and categorize the 
observed gamma-ray spectral variability (or a lack thereof) 
in blazars due to the large error bars on spectral indices and the long integration times
needed to get the spectral information.
\item
Gamma-ray spectral variability can arise out of a combination of
several physical parameters that are both internal and external to the jet.
The current data do not have the required energy and time resolution
to narrow down the parameter-space used in the models due to EGRET's limitations.
GLAST should be able to provide more accurate spectral information on shorter timescales.
\item   
 It is reasonable to expect that there might be 
deviations from power law behavior in the gamma-ray photon spectrum, given the 
many possible types of emission mechanisms in blazars.
There is some evidence of this in the EGRET data during large flares
but insufficient statistics prevent a detailed analysis using broken power law models.  
This would be an ideal area for GLAST to study.
\item
HBLs are faint at EGRET energies. This results in a 
low confidence of spectral variability even when there is a strong 
correlation between spectral index and flux. 
With the enhanced energy resolution and sensitivity
of GLAST,  as well as the new atmospheric Cherenkov 
telescopes (e.g. H.E.S.S., VERITAS, MAGIC and CANGAROO), it should be 
possible to detect more HBLs at gamma-ray energies and 
determine their spectral indices more accurately.
\item
We obtained a value of 2.25$\pm$0.03 for the average spectral index of all 
the blazars observed by EGRET. This is very close to the spectral index 
of 2.24$\pm$0.01 for the extragalactic gamma-ray background observed below 2 GeV
which make blazars as one of the significant contributors to the EGRB. But
the break in the broadband background spectrum at 2 GeV and a subsequent increase ($>$10 GeV)  
suggests the necessity of an increased contribution by other sources at higher energies.
\end{enumerate}
We would like to thank Olaf Reimer and Andrew Strong for their inputs on the extragalactic gamma ray background
and providing us with the data for Figure \ref{egrb}. G.N. would like to thank the 
Leon Herreid Foundation whose fellowship made a part of this research possible. 
G.N. would also like to thank Sangeeta Parashar for editing the manuscript.

\clearpage

\begin{figure}
\begin{minipage}[b]{0.4\textwidth}
\centering
\includegraphics[width=3.6in]{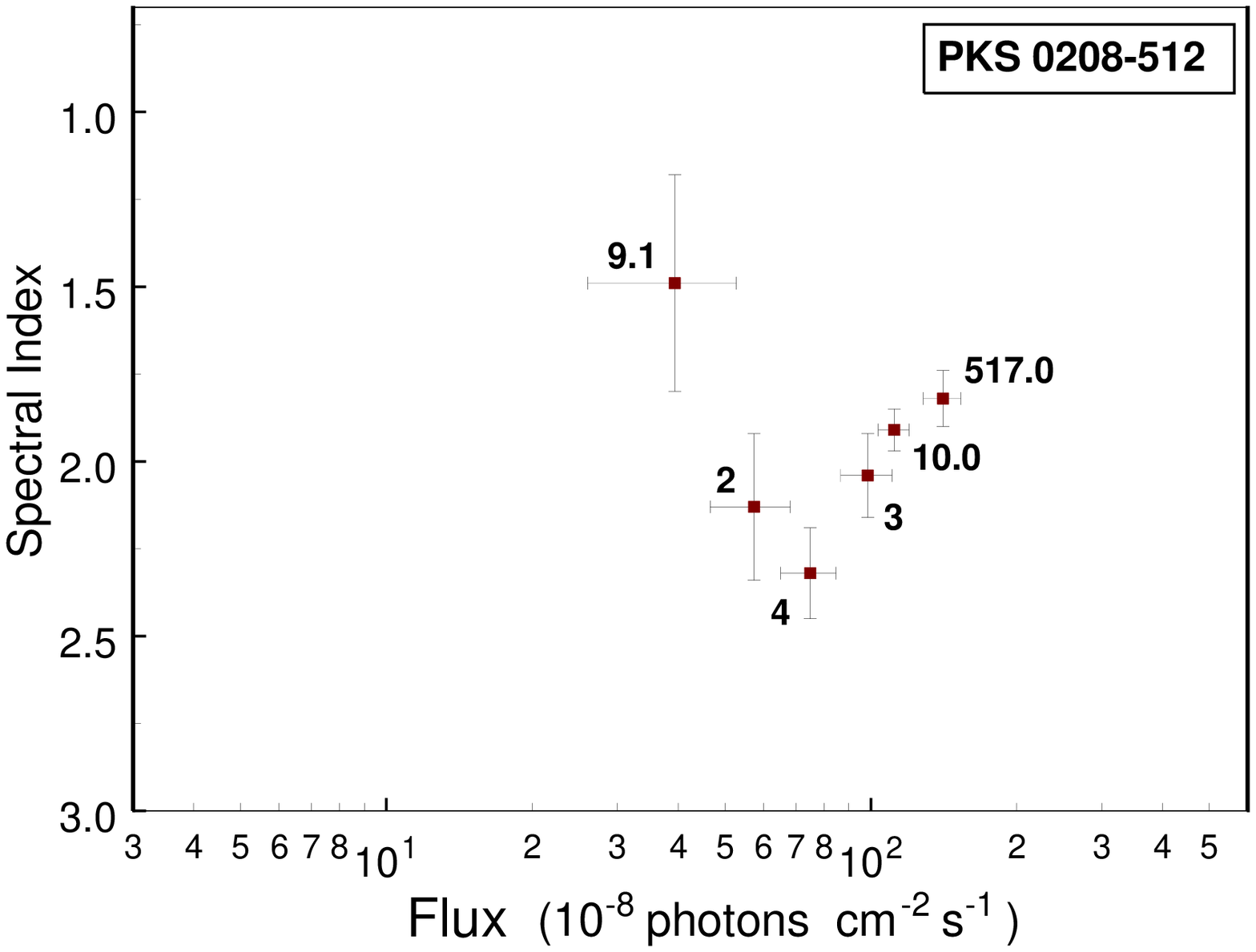}
\end{minipage}
\hspace{0.8in}
\begin{minipage}[b]{0.4\textwidth}
\centering
\includegraphics[width=3.6in]{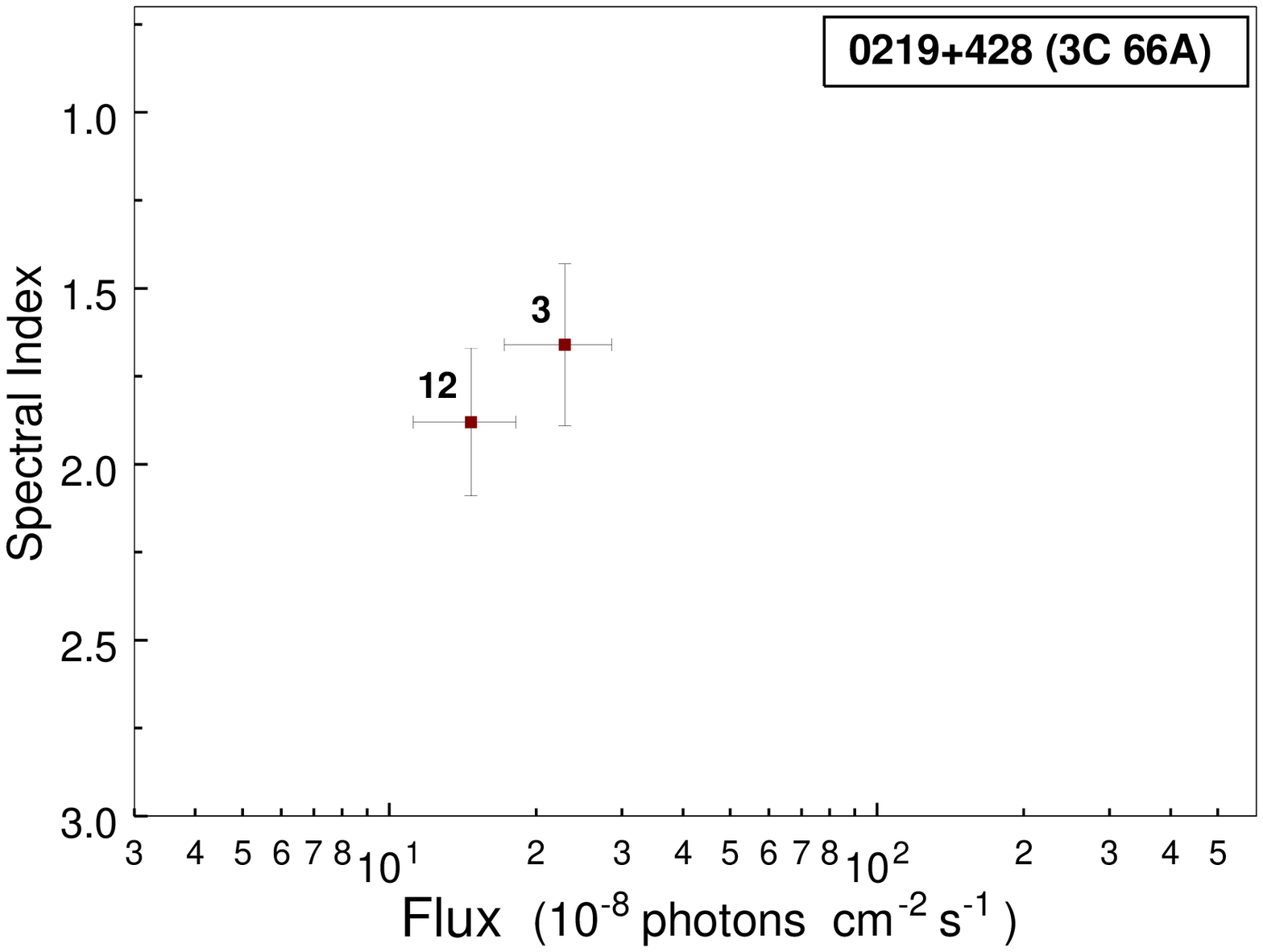}
\end{minipage}

\begin{minipage}[b]{0.4\textwidth}
\centering
\includegraphics[width=3.6in]{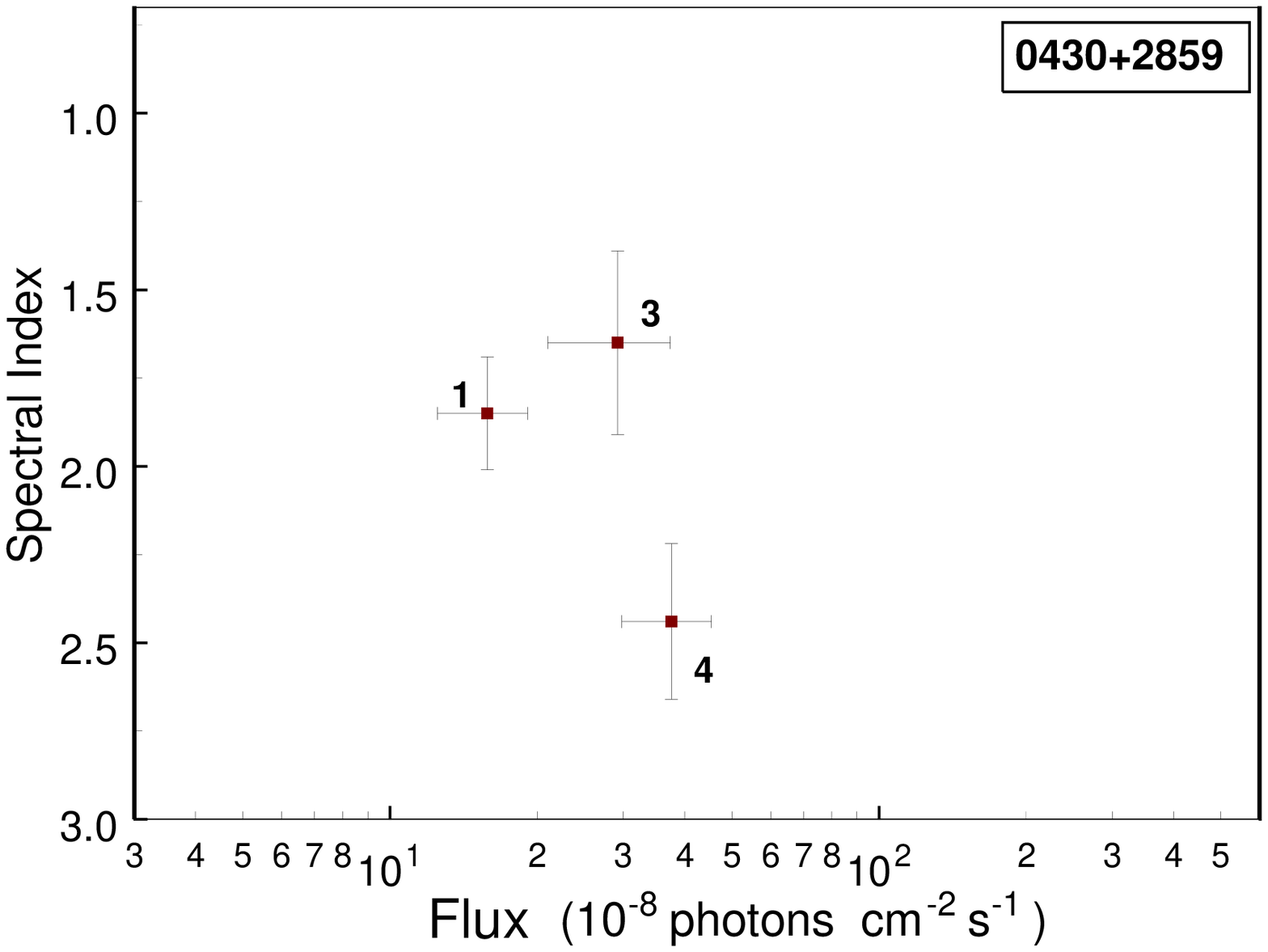}
\end{minipage}
\hspace{0.8in}
\begin{minipage}[b]{0.55\textwidth}
\includegraphics[width=3.6in]{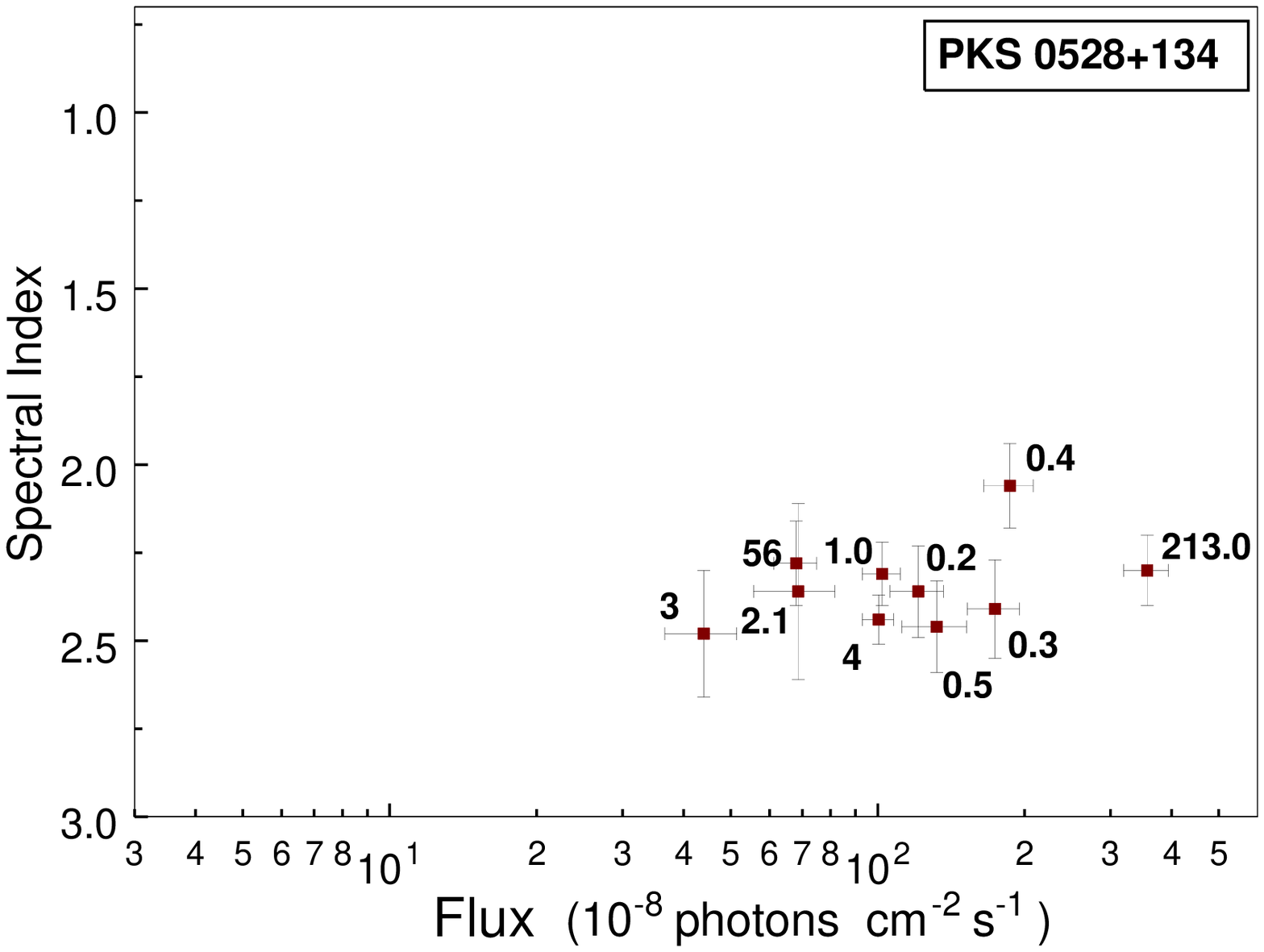}
\end{minipage}
\begin{minipage}[b]{0.4\textwidth}
\centering
\includegraphics[width=3.6in]{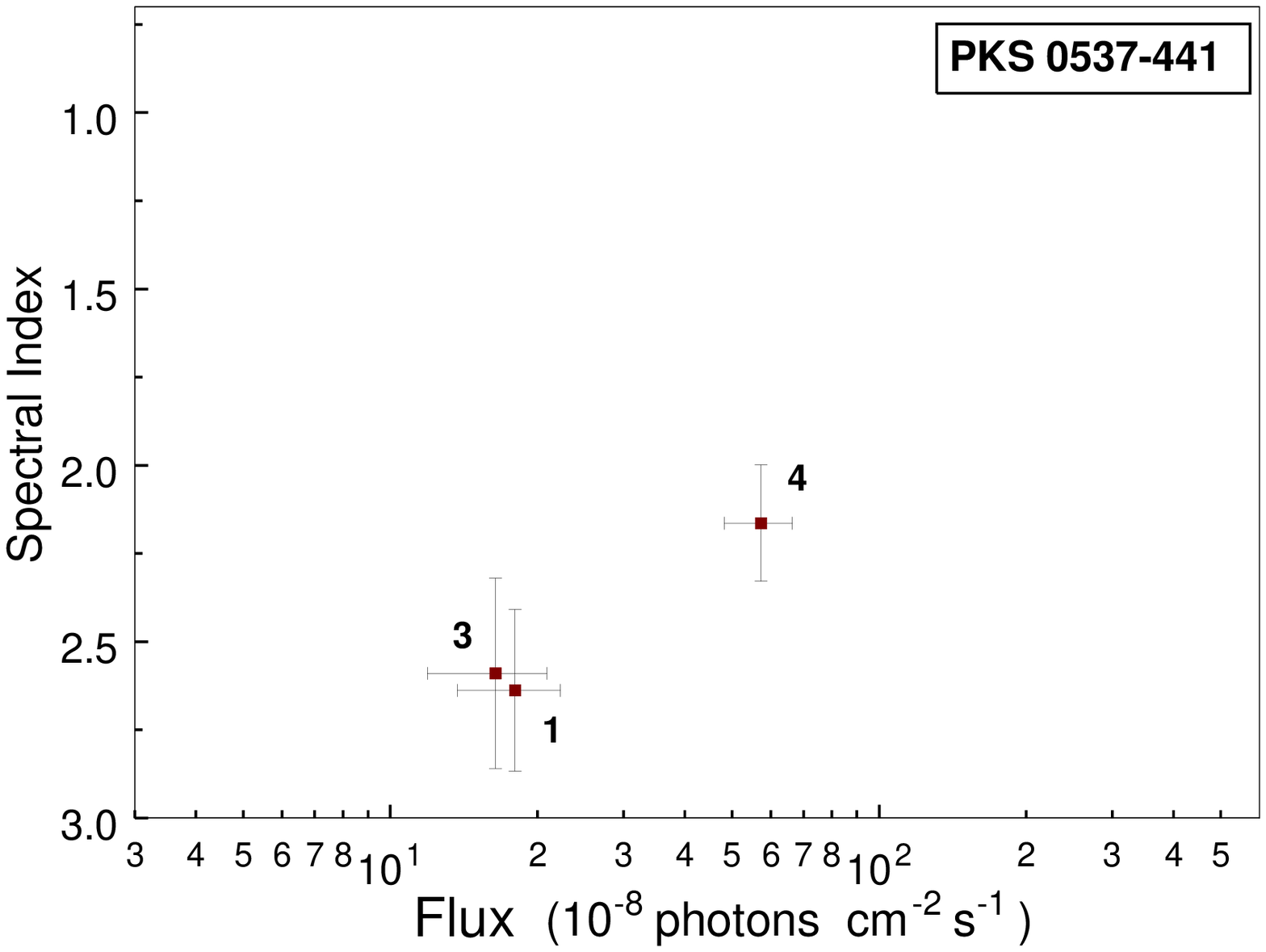}
\end{minipage}
\hspace{0.8in}
\begin{minipage}[b]{0.4\textwidth}
\centering
\includegraphics[width=3.6in]{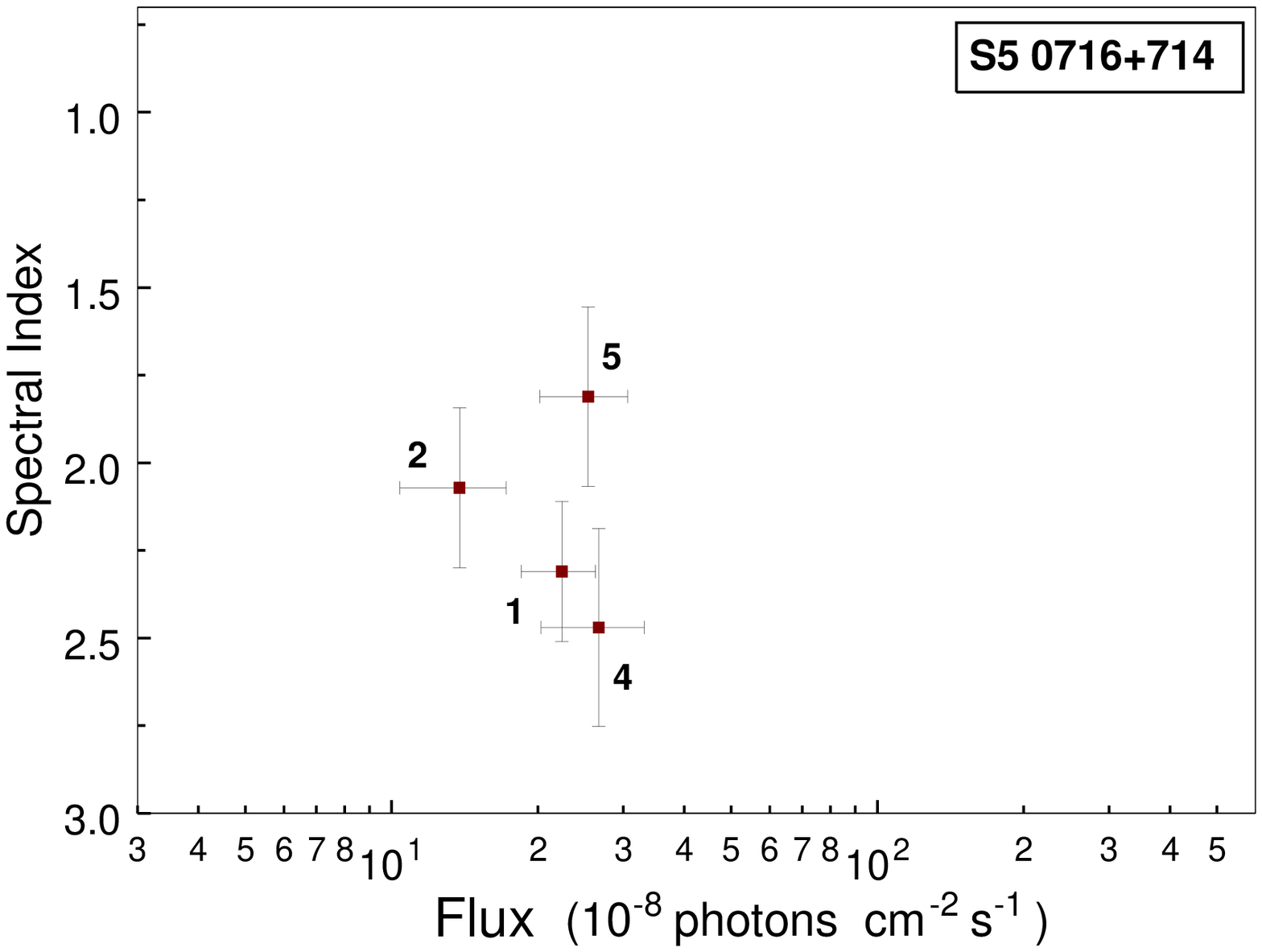}
\end{minipage}

\caption{Variation of photon spectral index in the 30~MeV-10~GeV energy range with gamma-ray flux ($>$100 MeV) in units of $10^{-8}$  photons cm$^{-2}$ s$^{-1}$. The spectral index is obtained from observations that are either one viewing period long (labeled by decimal numbers), or a combination of the viewing periods during one or more Cycle of observations (labeled by integers that show the Cycle(s) being combined). The details of the viewing periods/Cycles used in the analysis are given in Table \ref{Tab-3}.}

\label{egspec1}
\end{figure}

\begin{figure}

\begin{minipage}[b]{0.4\textwidth}
\centering
\includegraphics[width=3.6in]{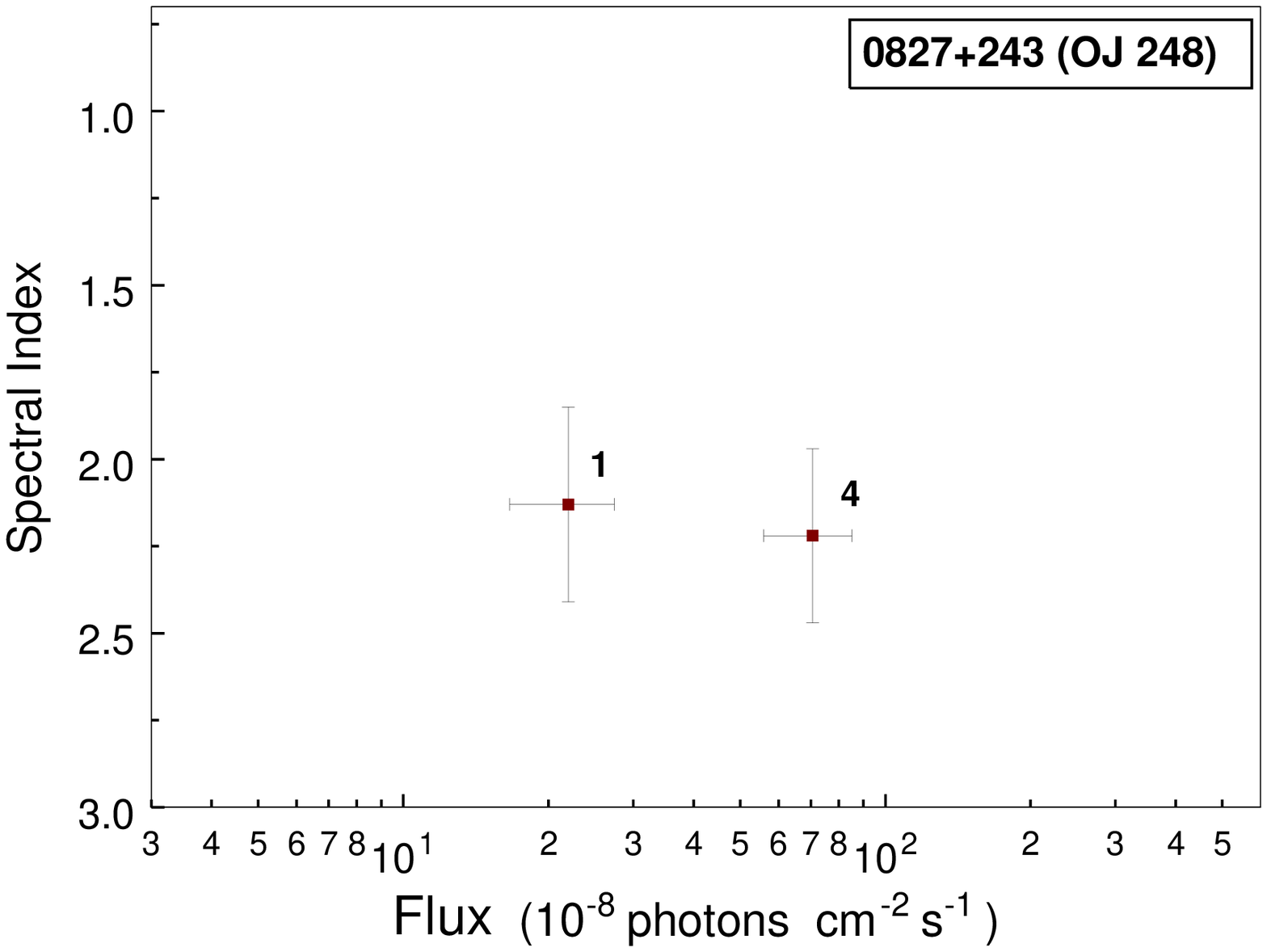}
\end{minipage}
\hspace{0.8in}
\begin{minipage}[b]{0.4\textwidth}
\centering
\includegraphics[width=3.6in]{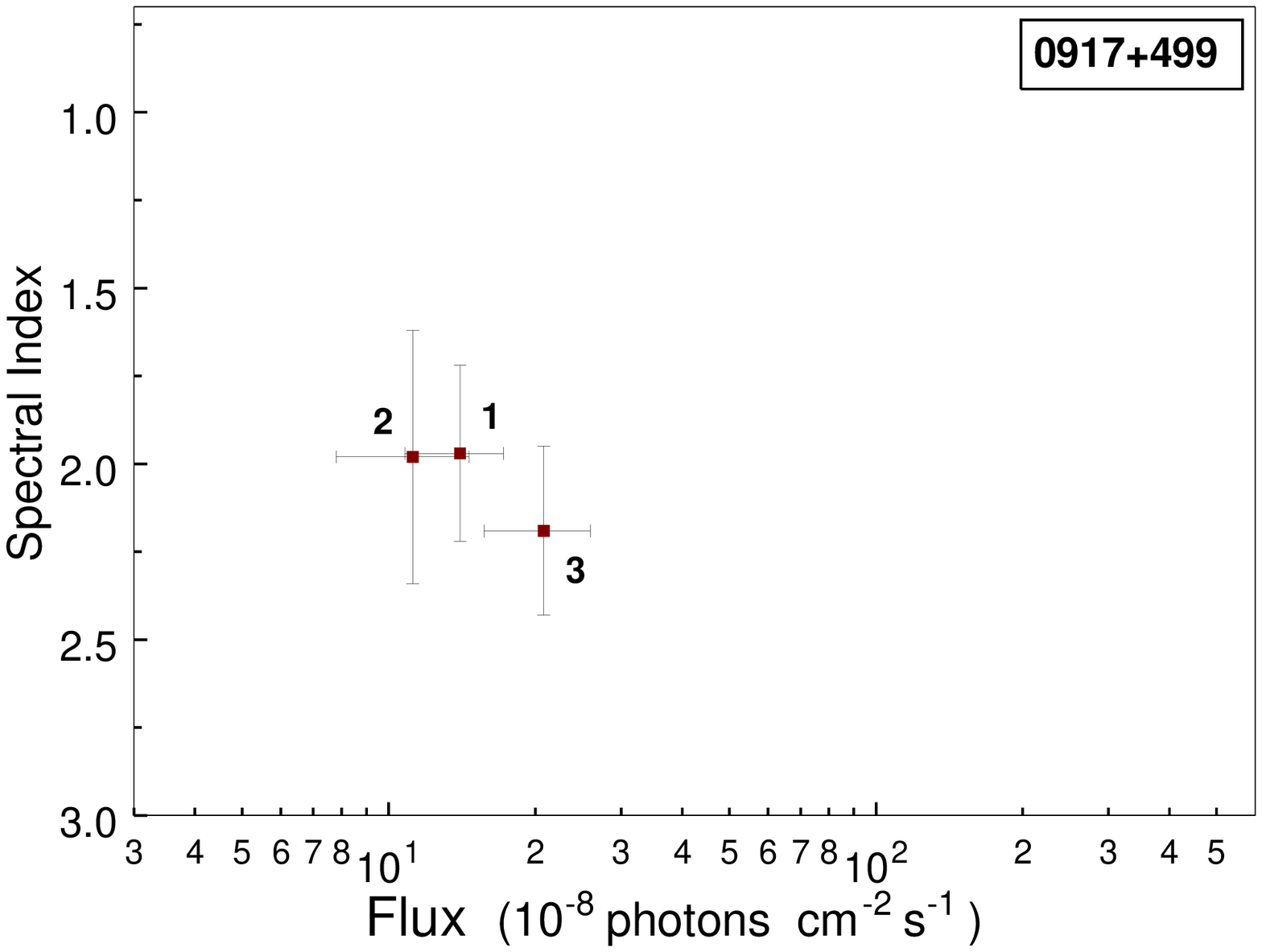}
\end{minipage}

\begin{minipage}[b]{0.4\textwidth}
\centering
\includegraphics[width=3.6in]{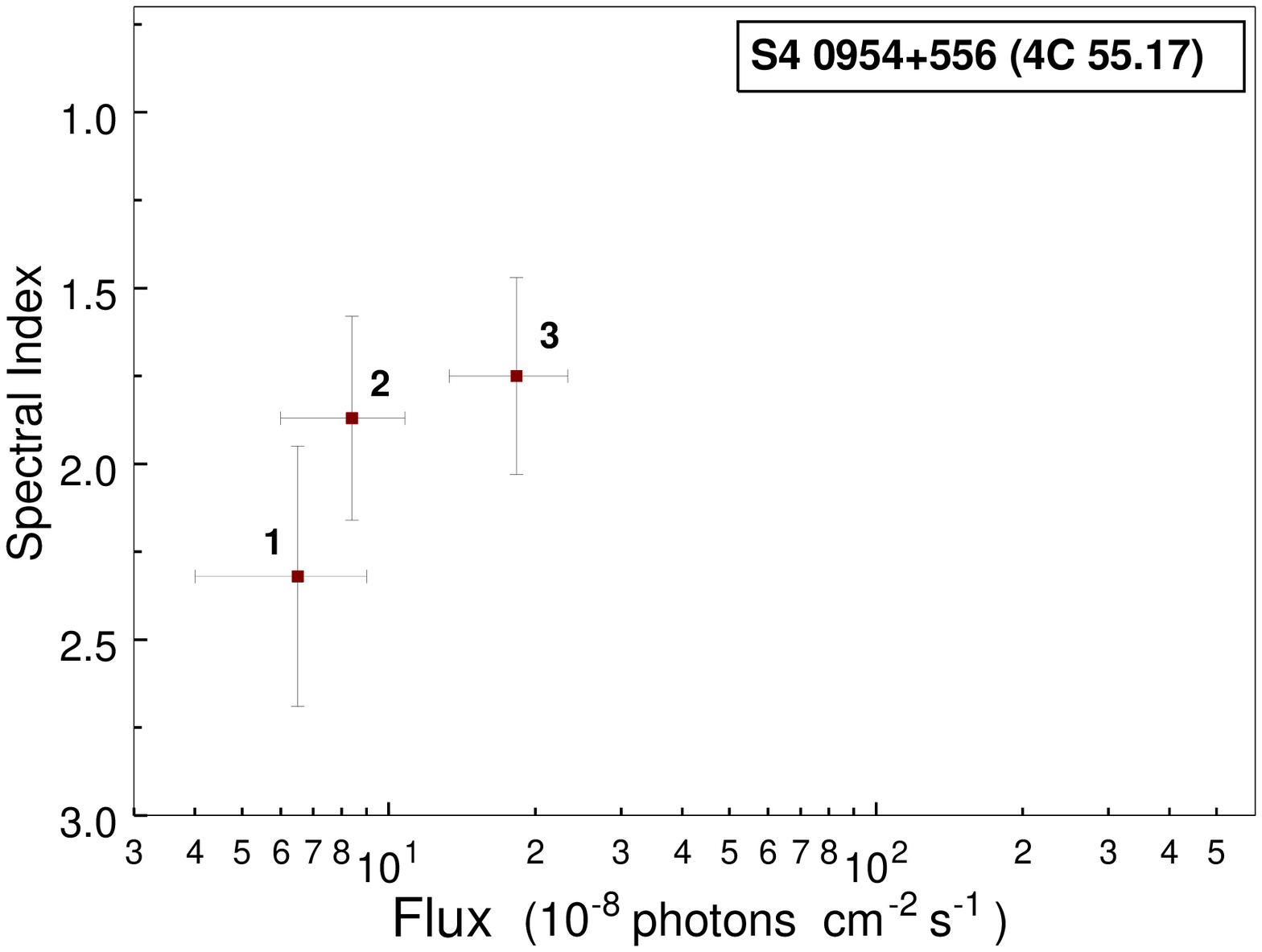}
\end{minipage}
\hspace{0.8in}
\begin{minipage}[b]{0.4\textwidth}
\centering
\includegraphics[width=3.6in]{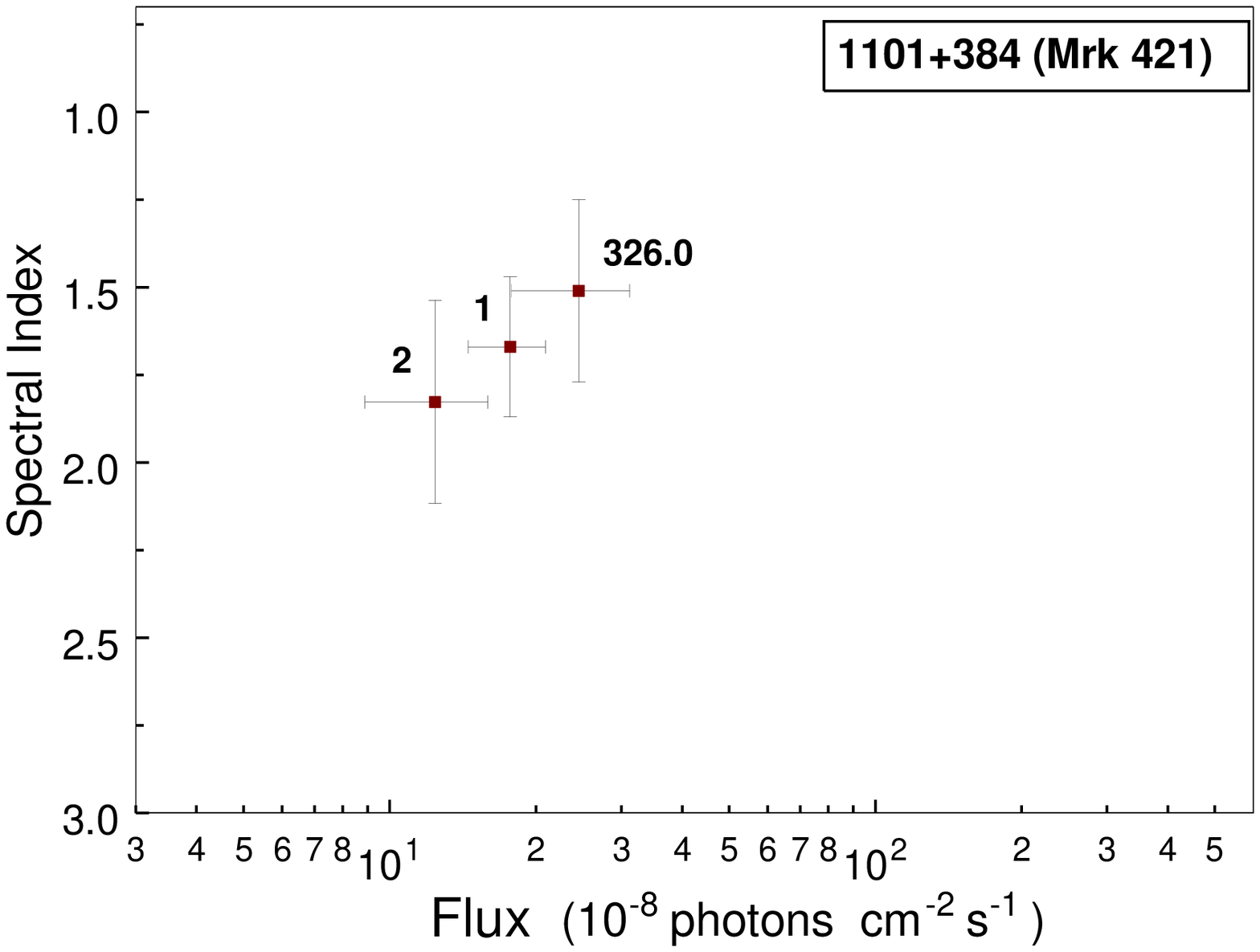}

\end{minipage}

\begin{minipage}[b]{0.4\textwidth}
\centering
\includegraphics[width=3.6in]{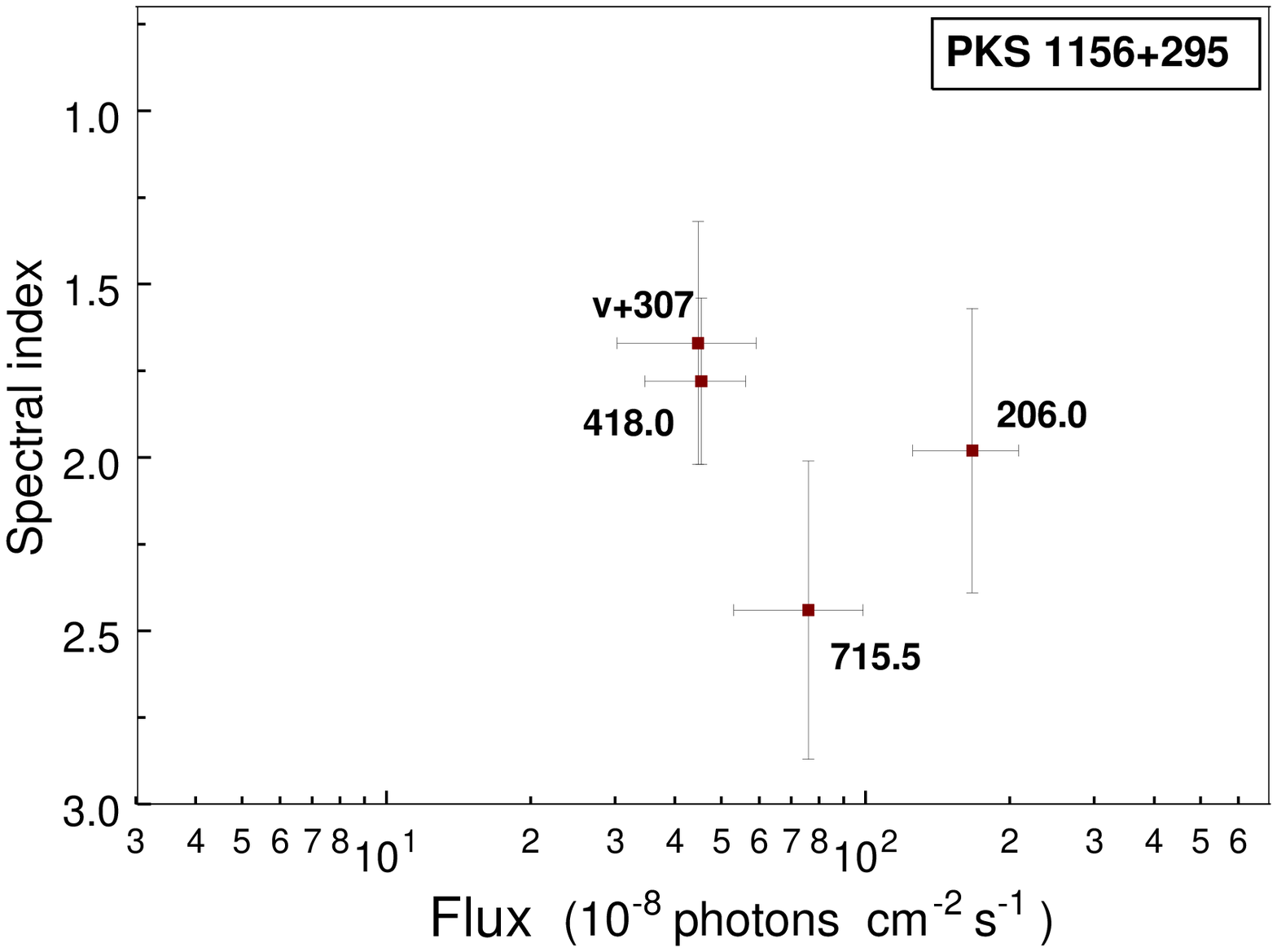}

\end{minipage}
\hspace{0.8in}
\begin{minipage}[b]{0.4\textwidth}
\centering
\includegraphics[width=3.6in]{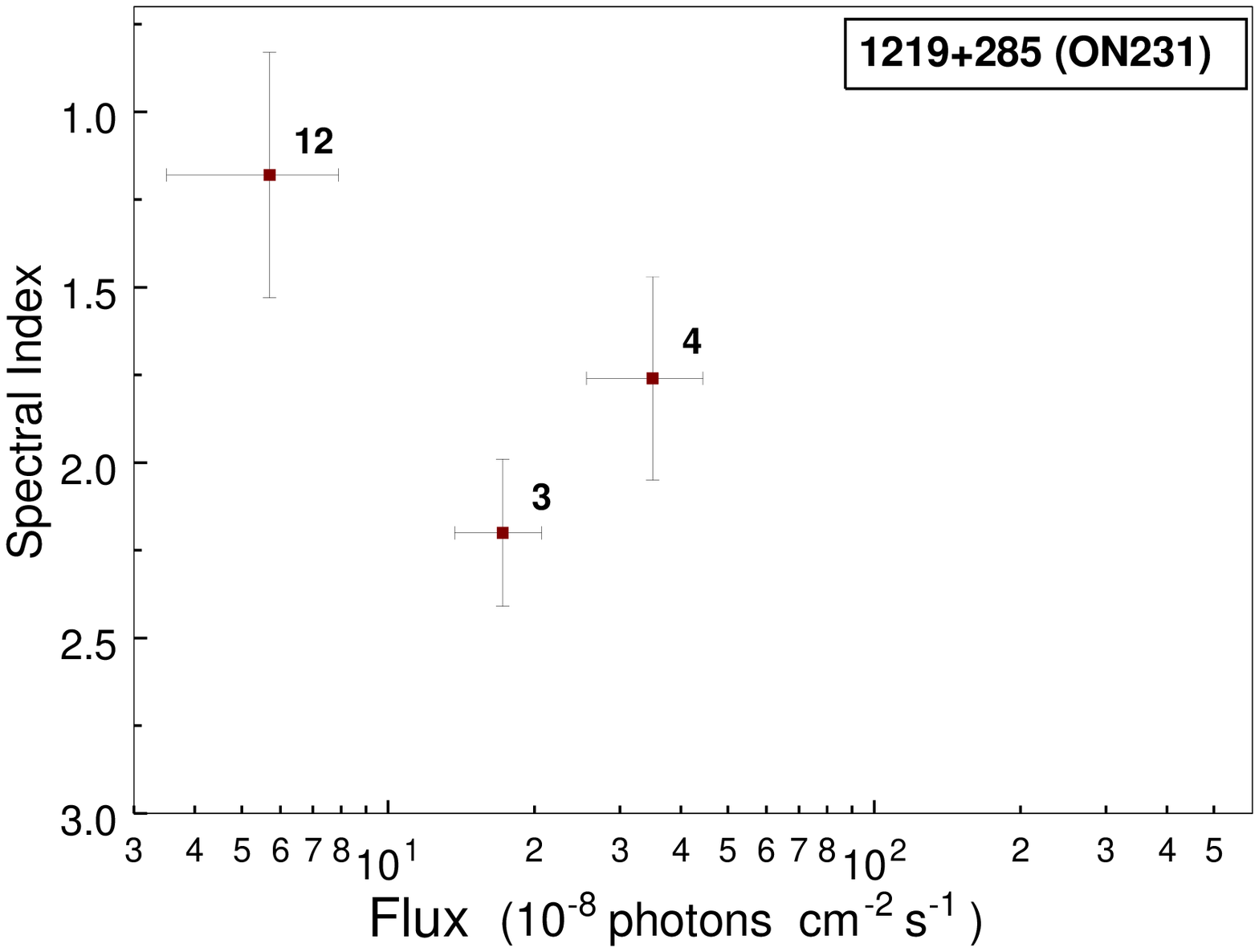}

\end{minipage}
\figurenum{3}
\caption{\it{Continued}}
\end{figure}

\begin{figure}
\begin{minipage}[b]{0.4\textwidth}
\centering
\includegraphics[width=3.6in]{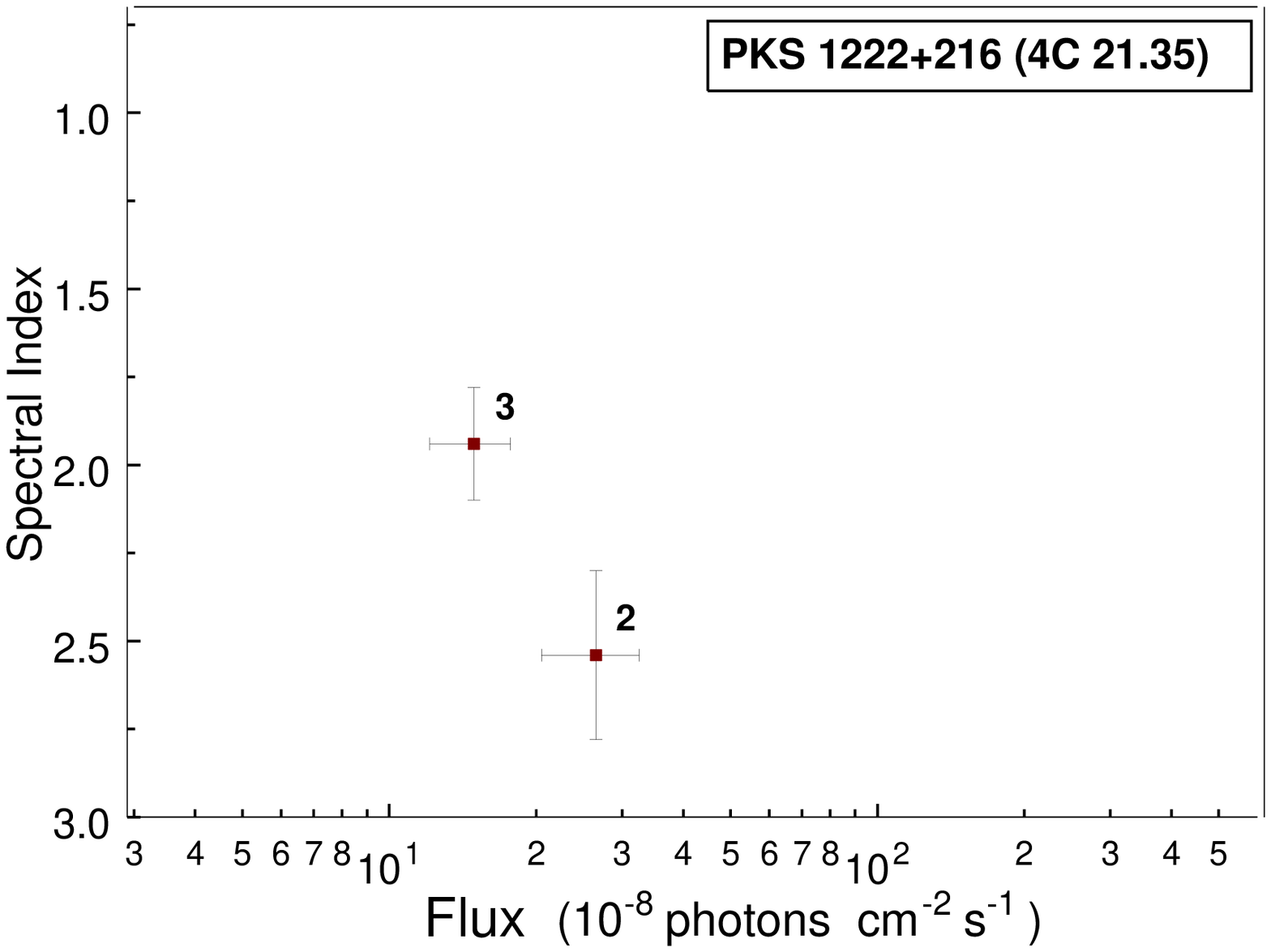}

\end{minipage}
\hspace{0.8in}
\begin{minipage}[b]{0.4\textwidth}
\centering
\includegraphics[width=3.6in]{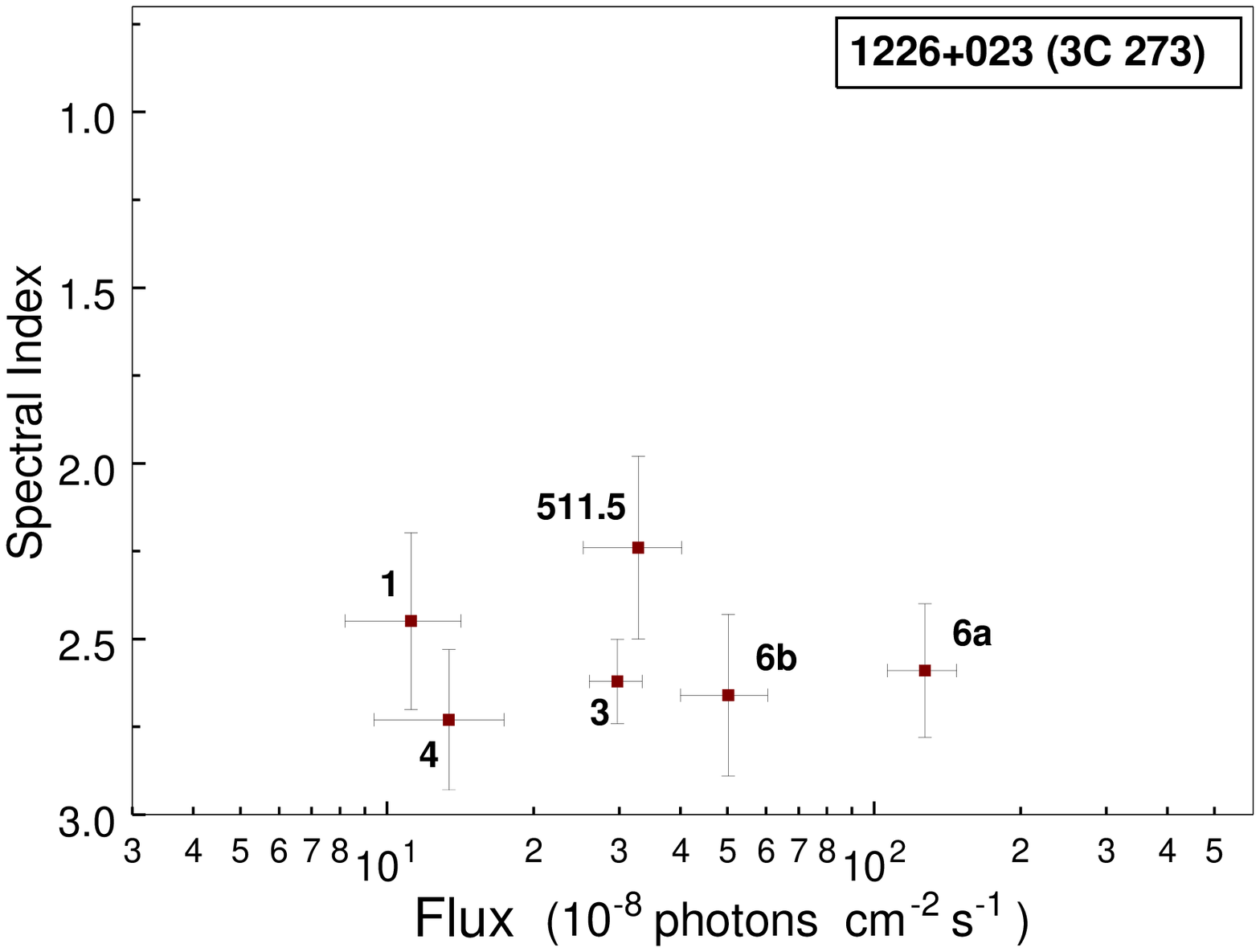}

\end{minipage}

\begin{minipage}[b]{0.4\textwidth}
\centering
\includegraphics[width=3.6in]{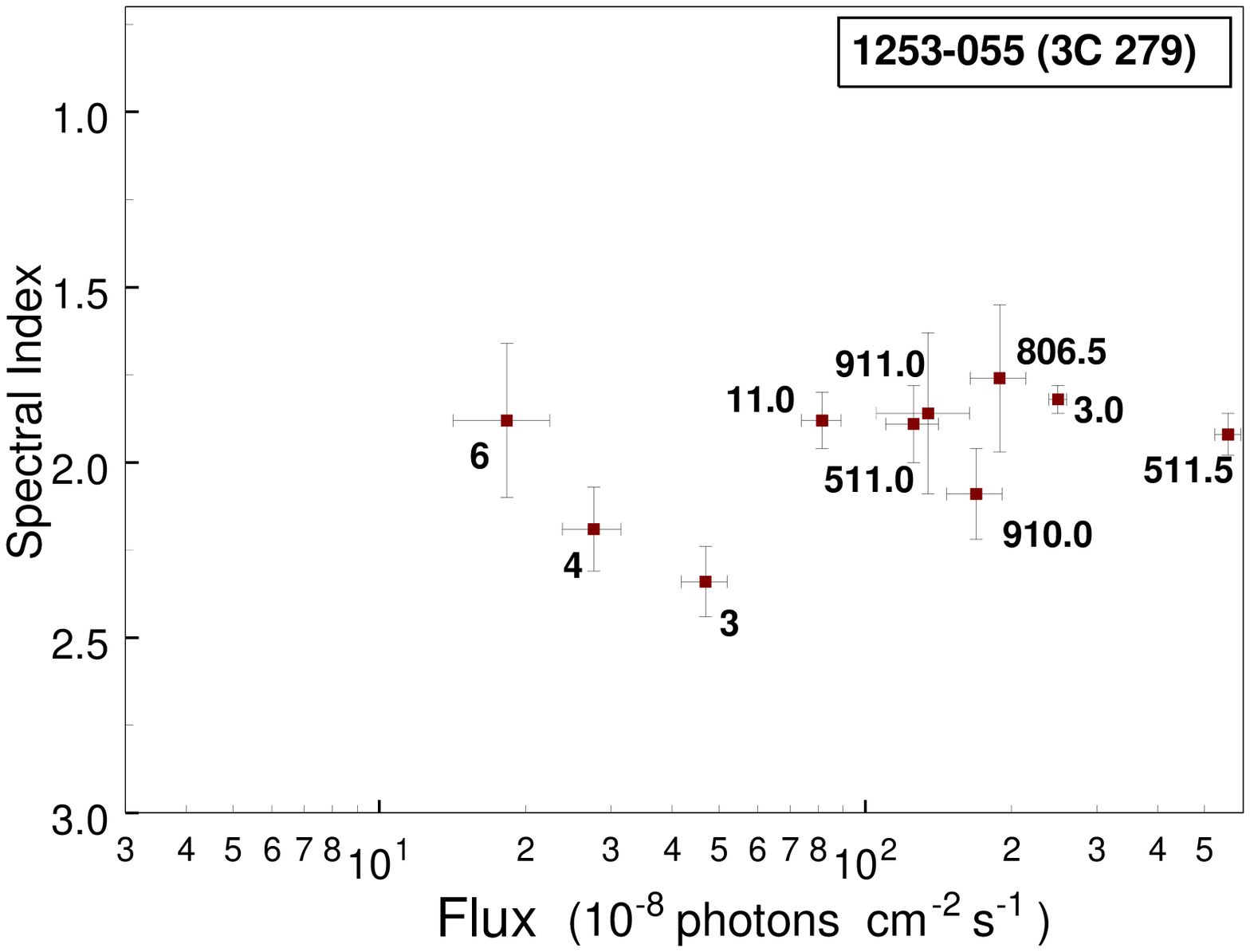}

\end{minipage}
\hspace{0.8in}
\begin{minipage}[b]{0.4\textwidth}
\centering
\includegraphics[width=3.6in]{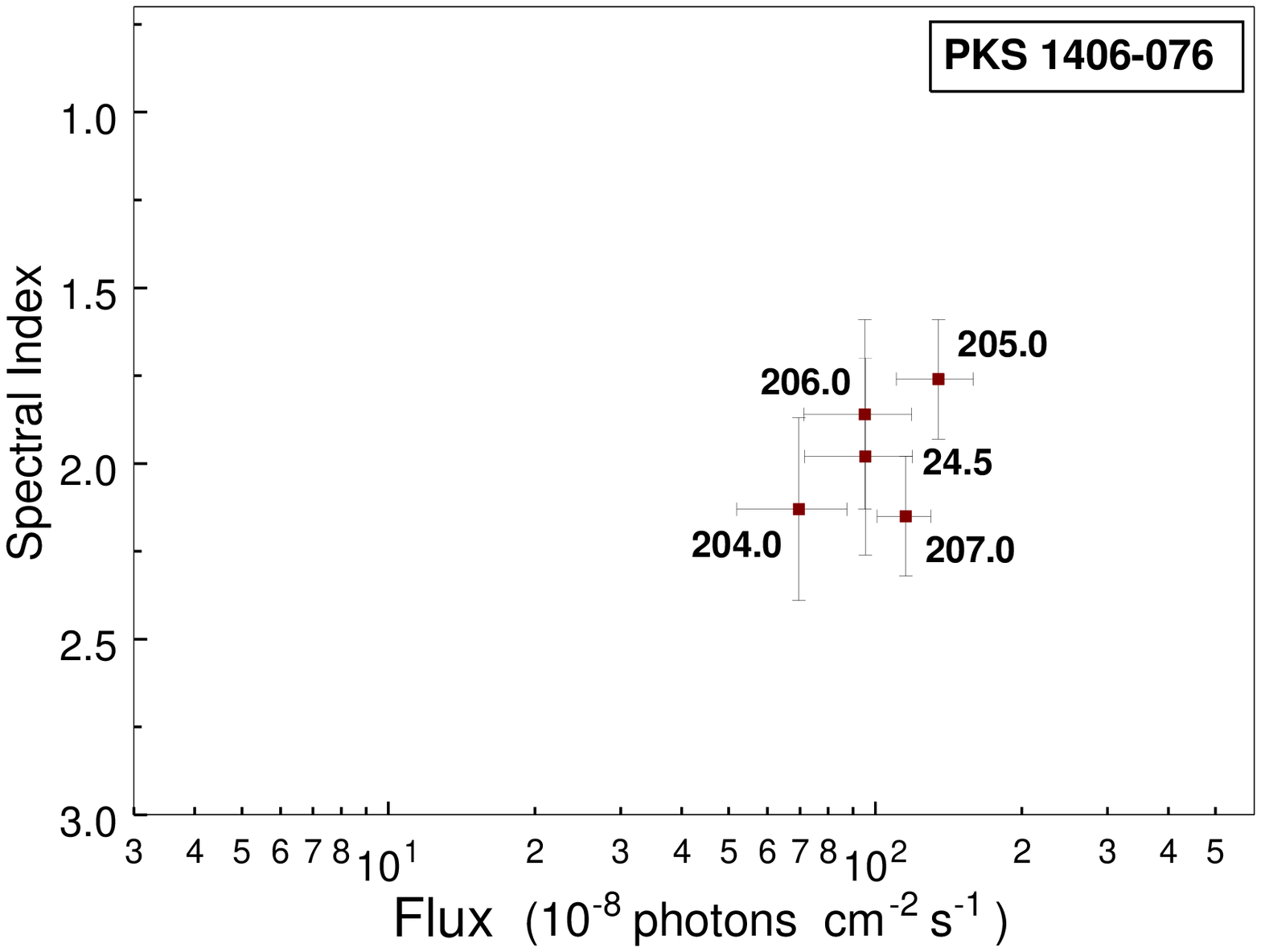}

\end{minipage}

\begin{minipage}[b]{0.4\textwidth}
\centering
\includegraphics[width=3.6in]{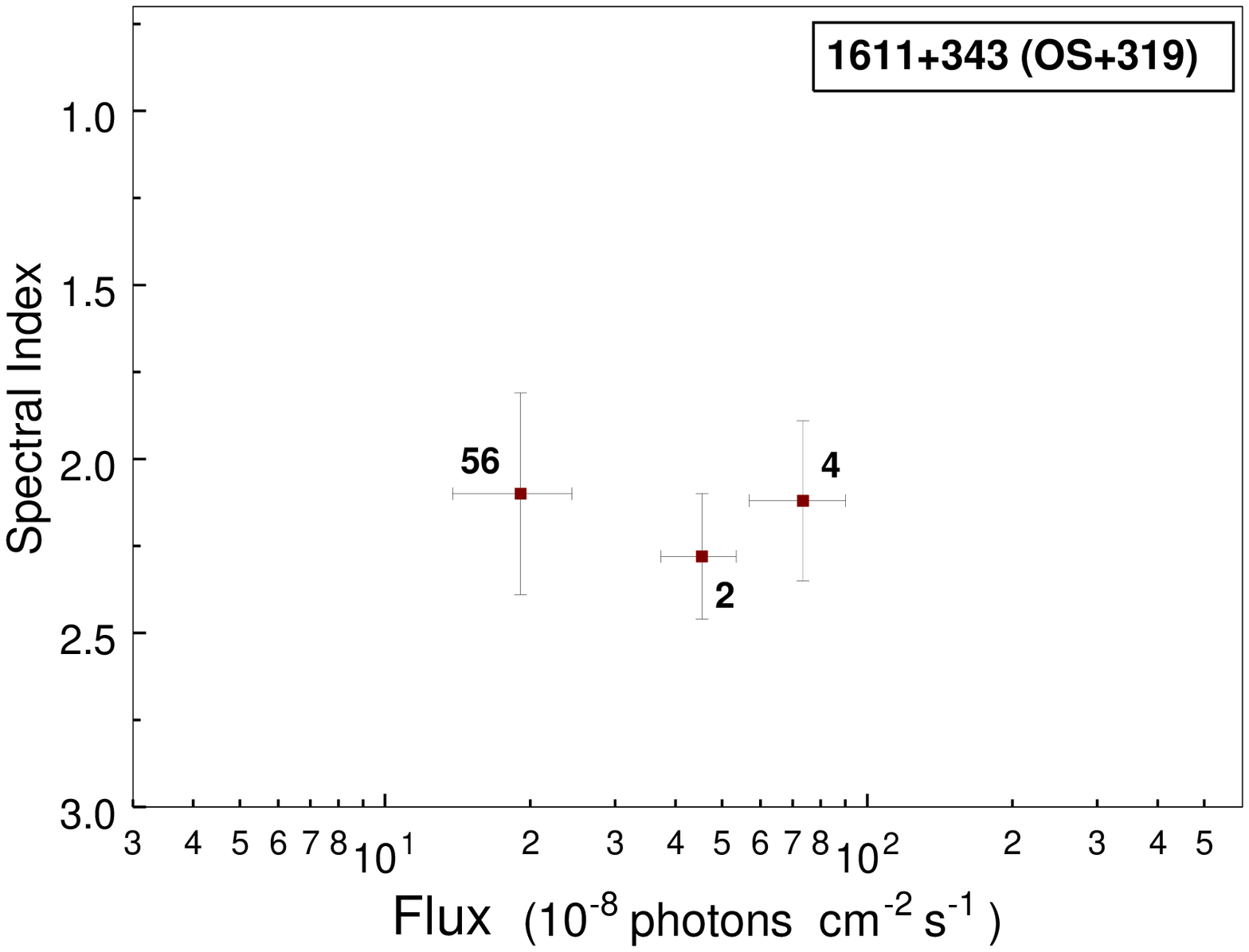}

\end{minipage}
\hspace{0.8in}
\begin{minipage}[b]{0.4\textwidth}
\centering
\includegraphics[width=3.6in]{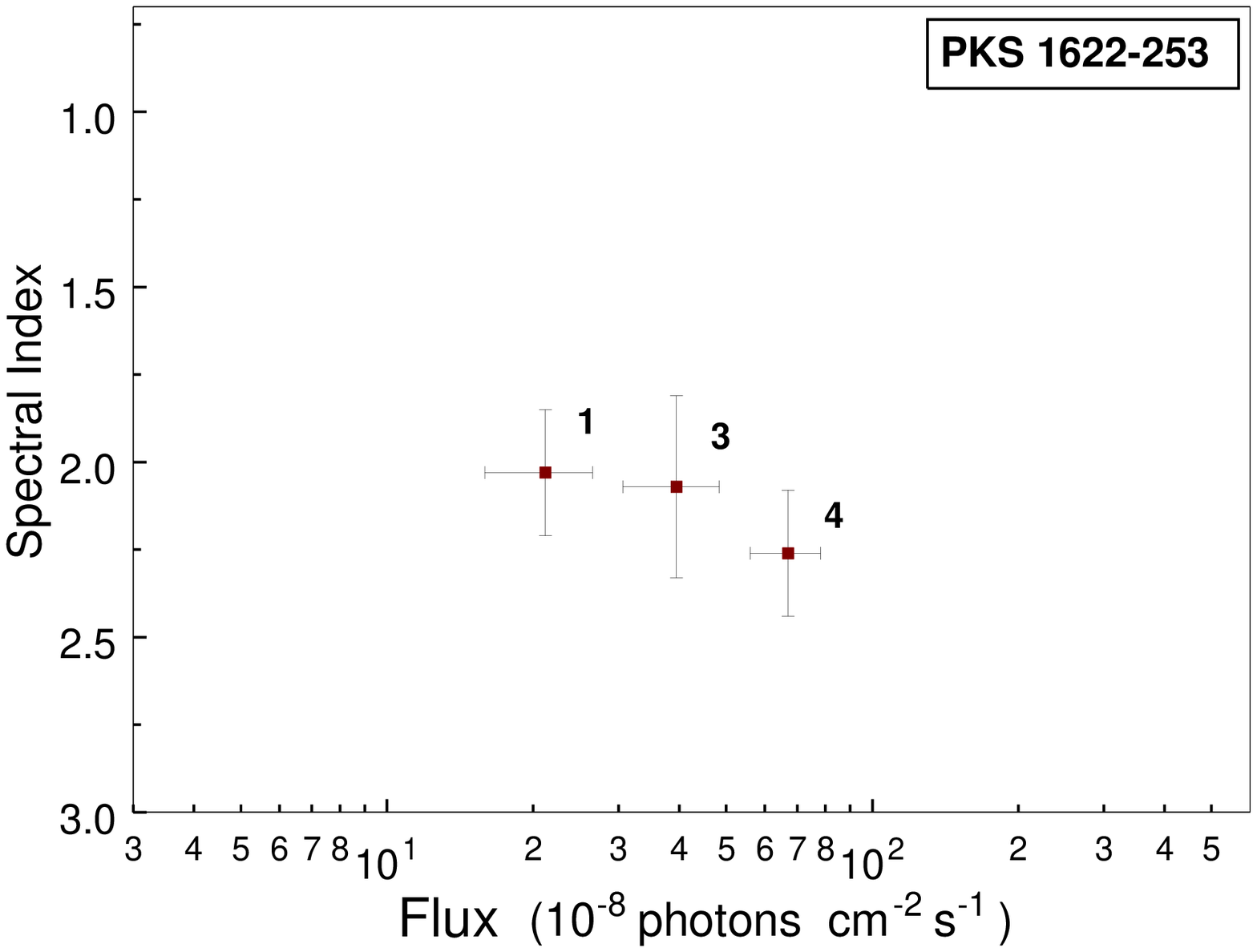}

\end{minipage}
\figurenum{3}
\begin{center}
\caption{\it{Continued}}
\end{center}
\end{figure}

\begin{figure}
\begin{minipage}[b]{0.4\textwidth}
\centering
\includegraphics[width=3.6in]{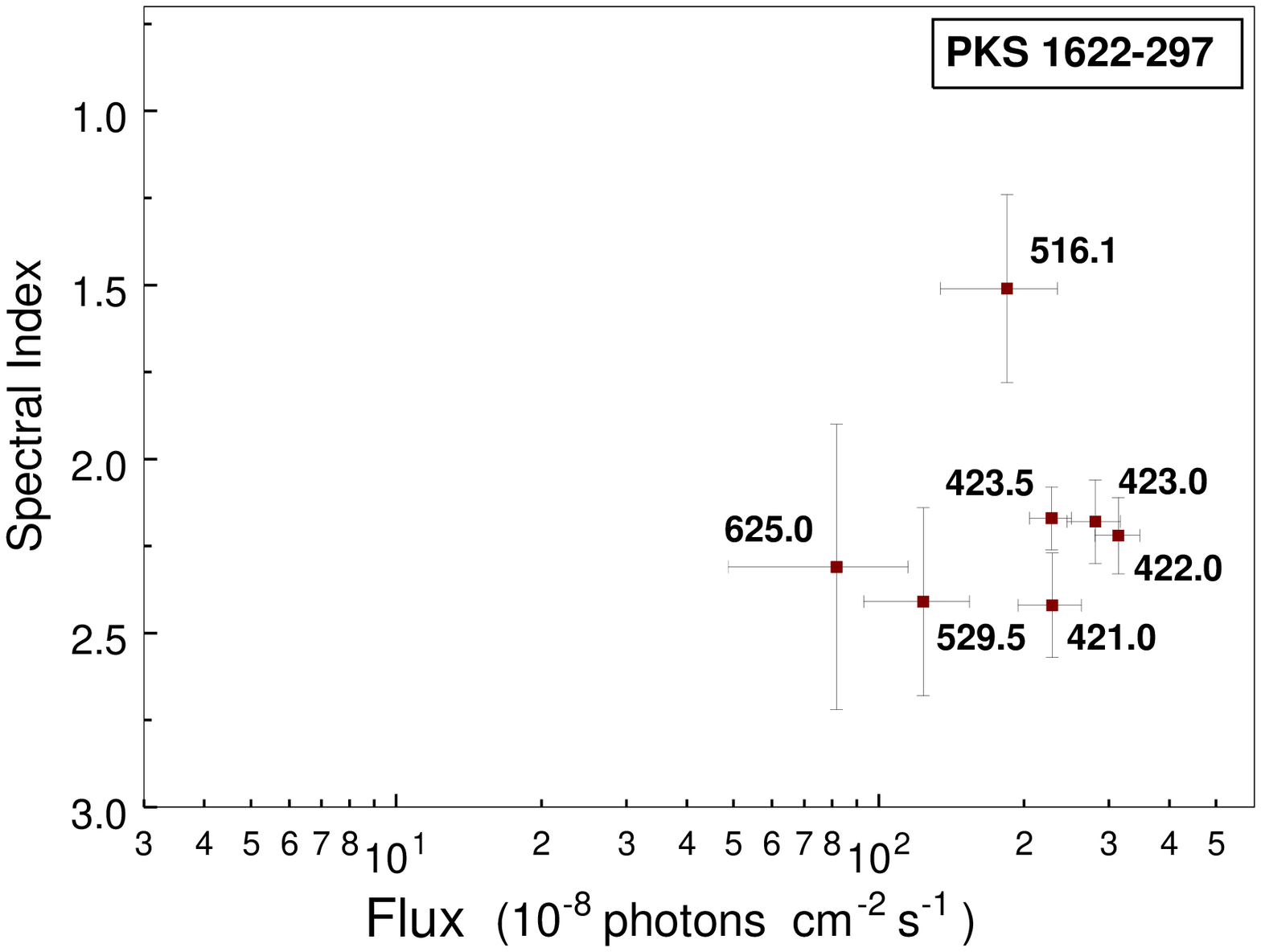}
\end{minipage}
\hspace{0.8in}
\begin{minipage}[b]{0.4\textwidth}
\centering
\includegraphics[width=3.6in]{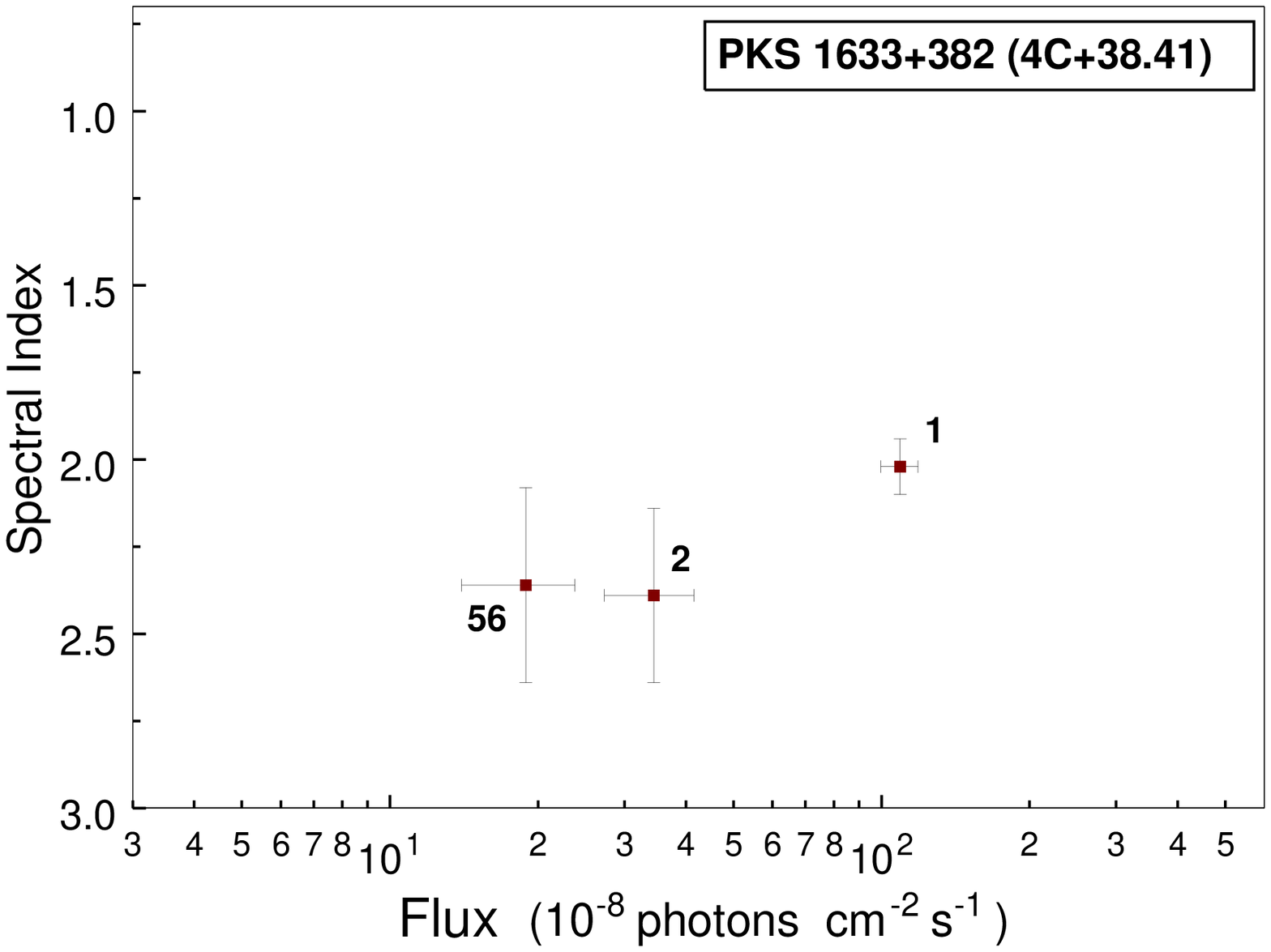}
\end{minipage}

\begin{minipage}[b]{0.4\textwidth}
\centering
\includegraphics[width=3.6in]{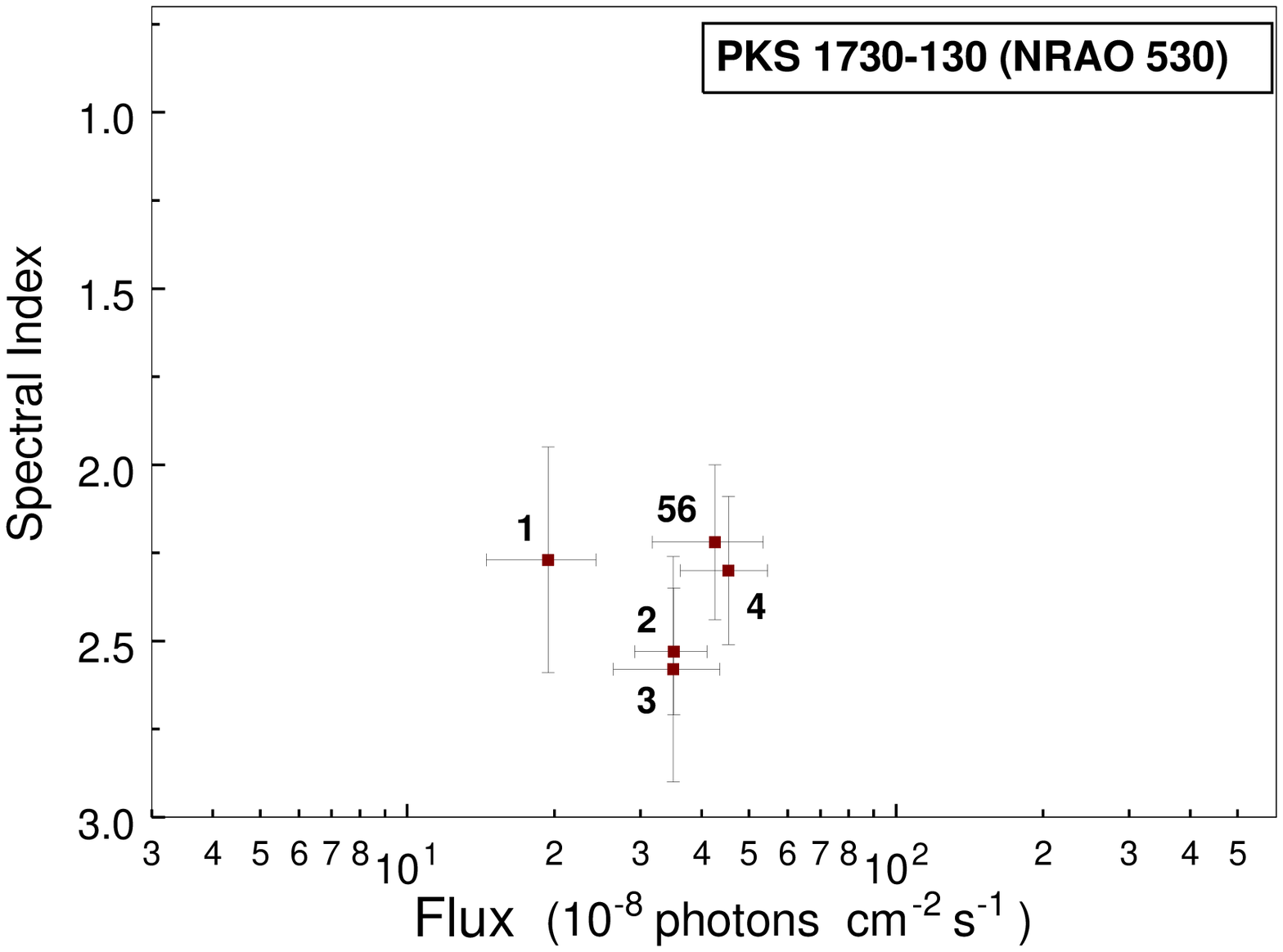}
\end{minipage}
\hspace{0.8in}
\begin{minipage}[b]{0.4\textwidth}
\centering
\includegraphics[width=3.6in]{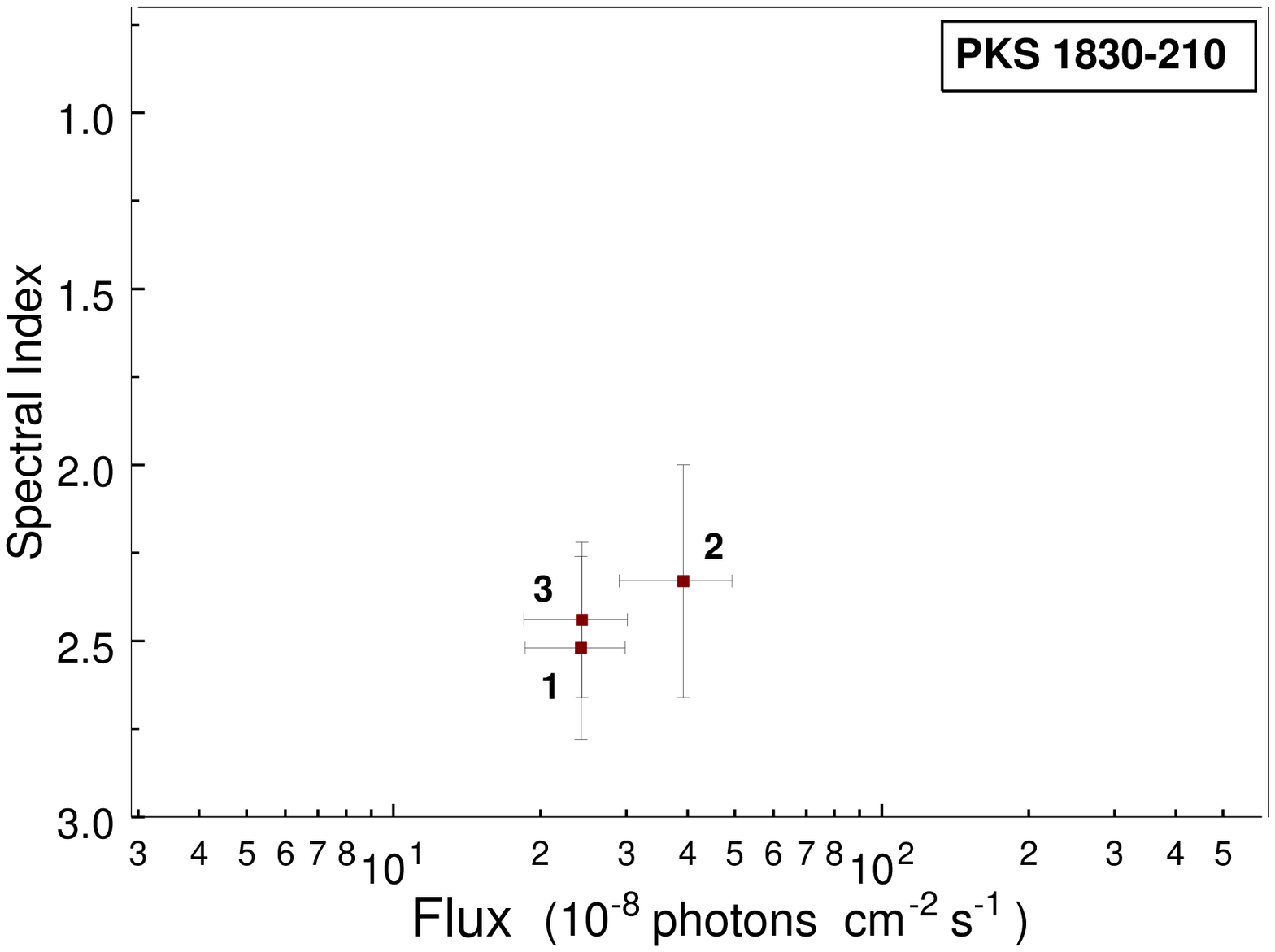}
\end{minipage}

\begin{minipage}[b]{0.4\textwidth}
\centering
\includegraphics[width=3.6in]{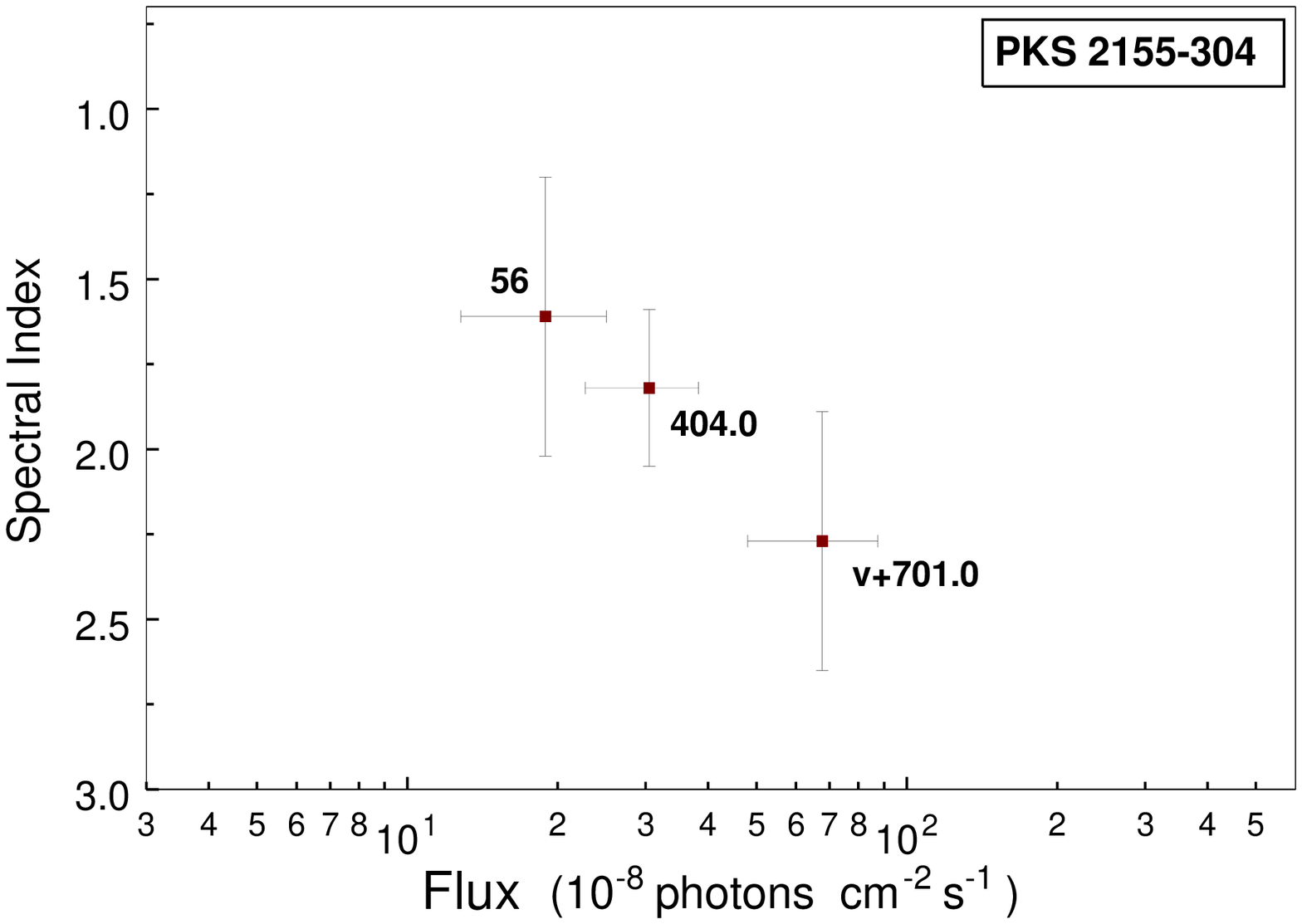}
\end{minipage}
\hspace{0.8in}
\begin{minipage}[b]{0.4\textwidth}
\centering
\includegraphics[width=3.6in]{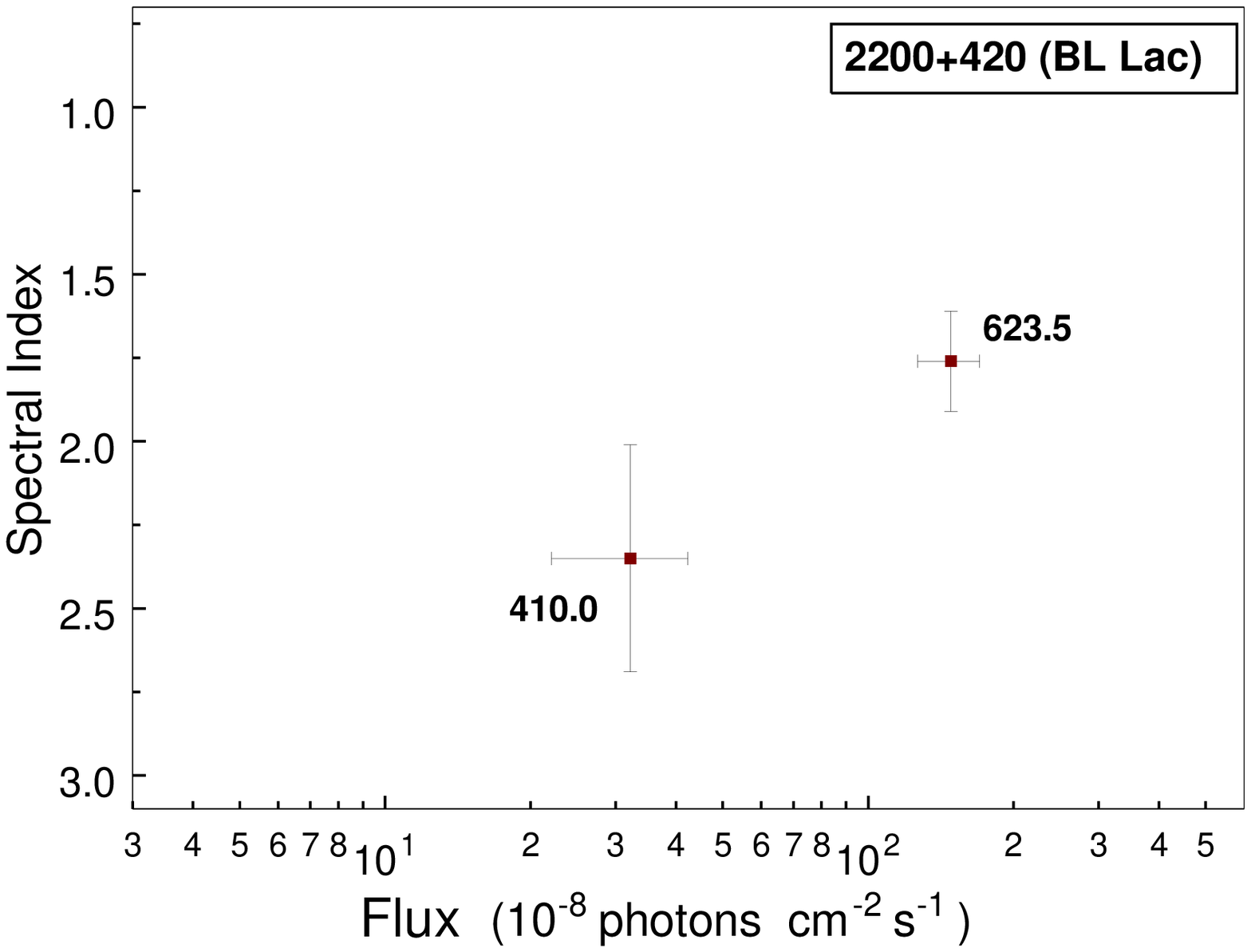}
\end{minipage}
\figurenum{3}
\caption{\it{Continued}}
\end{figure}

\begin{figure}
\begin{minipage}[b]{0.4\textwidth}
\centering
\includegraphics[width=3.6in]{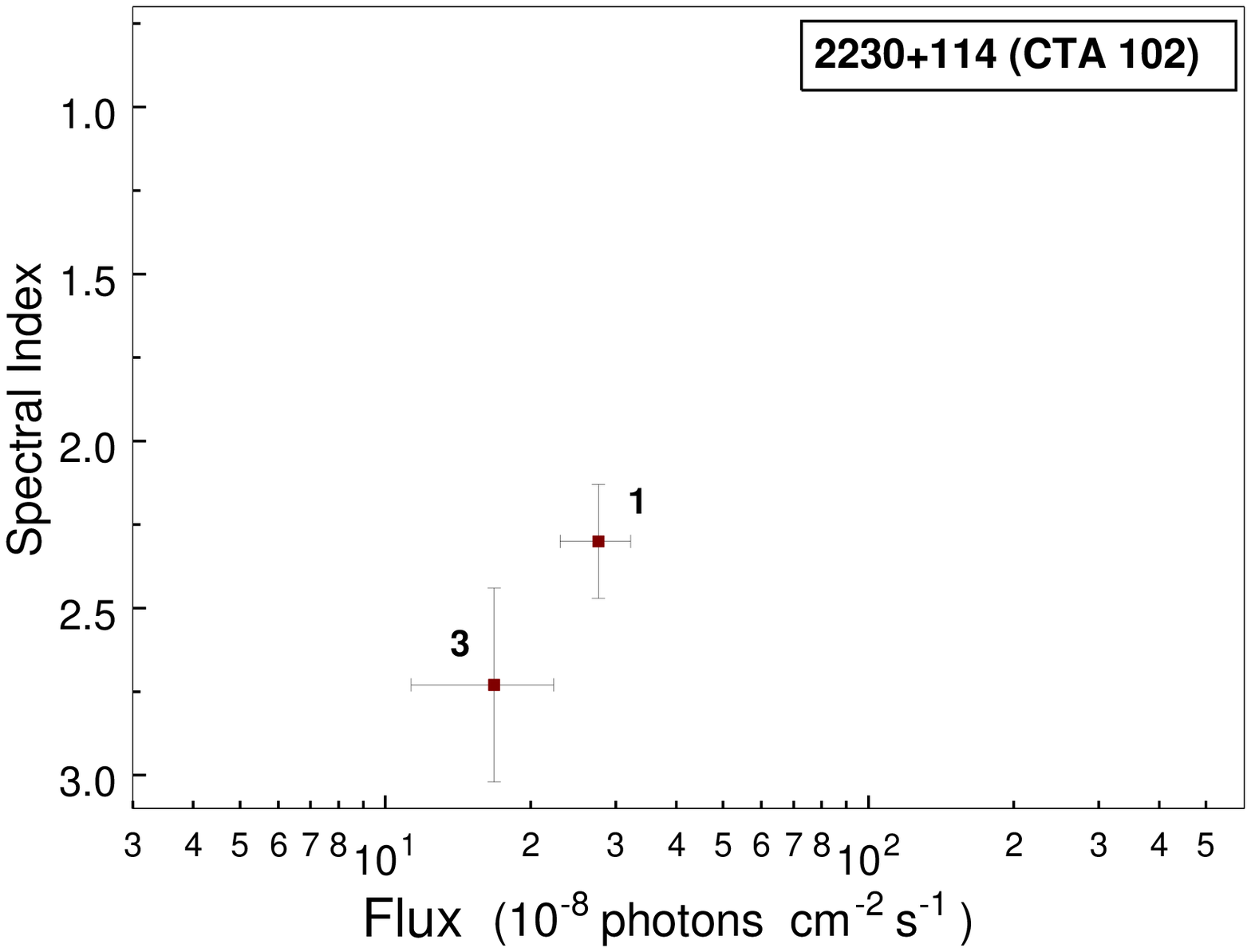}
\end{minipage}
\hspace{0.8in}
\begin{minipage}[b]{0.4\textwidth}
\centering
\includegraphics[width=3.6in]{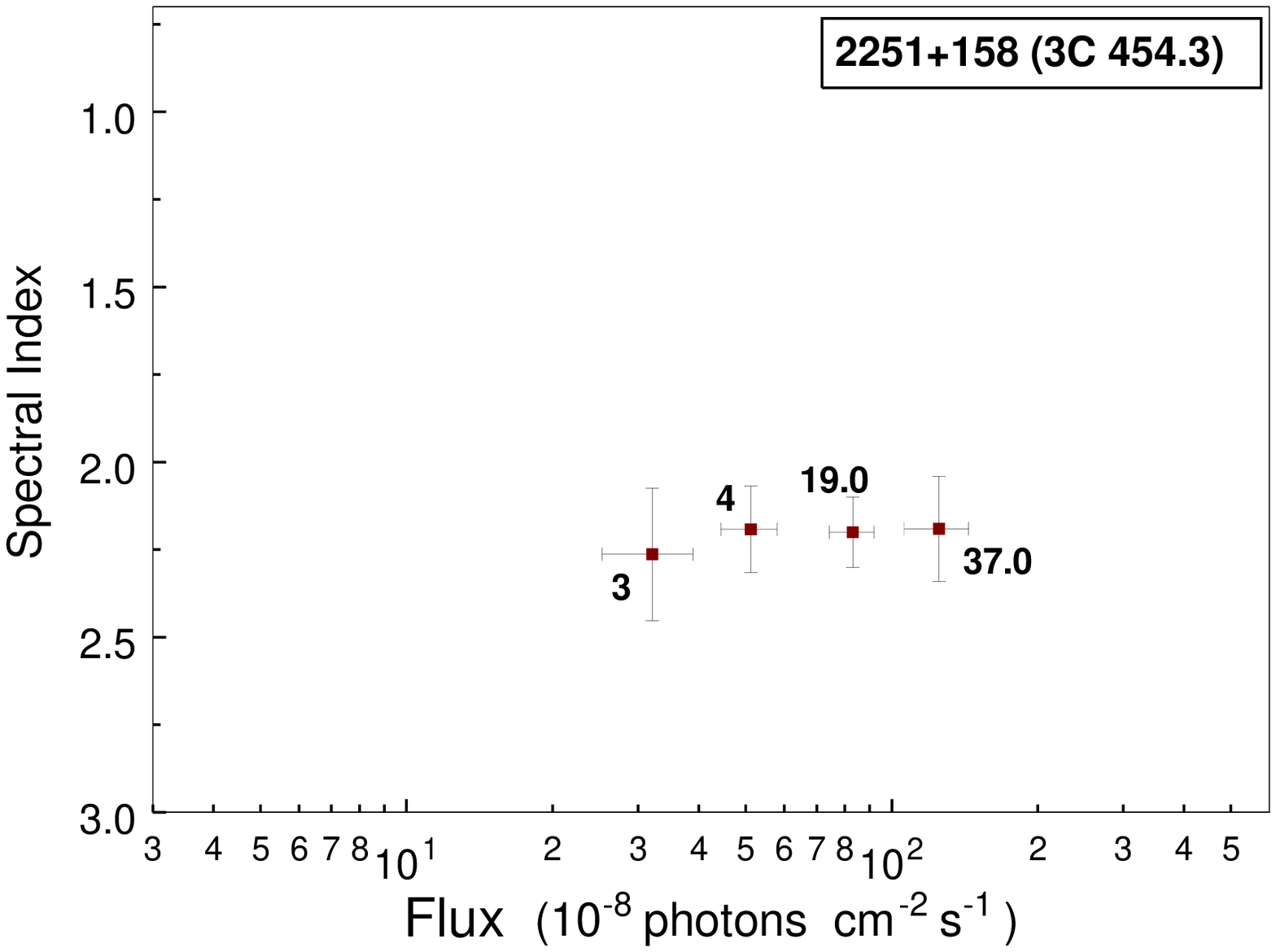}
\end{minipage}
\figurenum{3}
\caption{\it{Continued}}
\end{figure}

\begin{figure}
\begin{minipage}[b]{0.4\textwidth}
\centering
\includegraphics[width=3.6in]{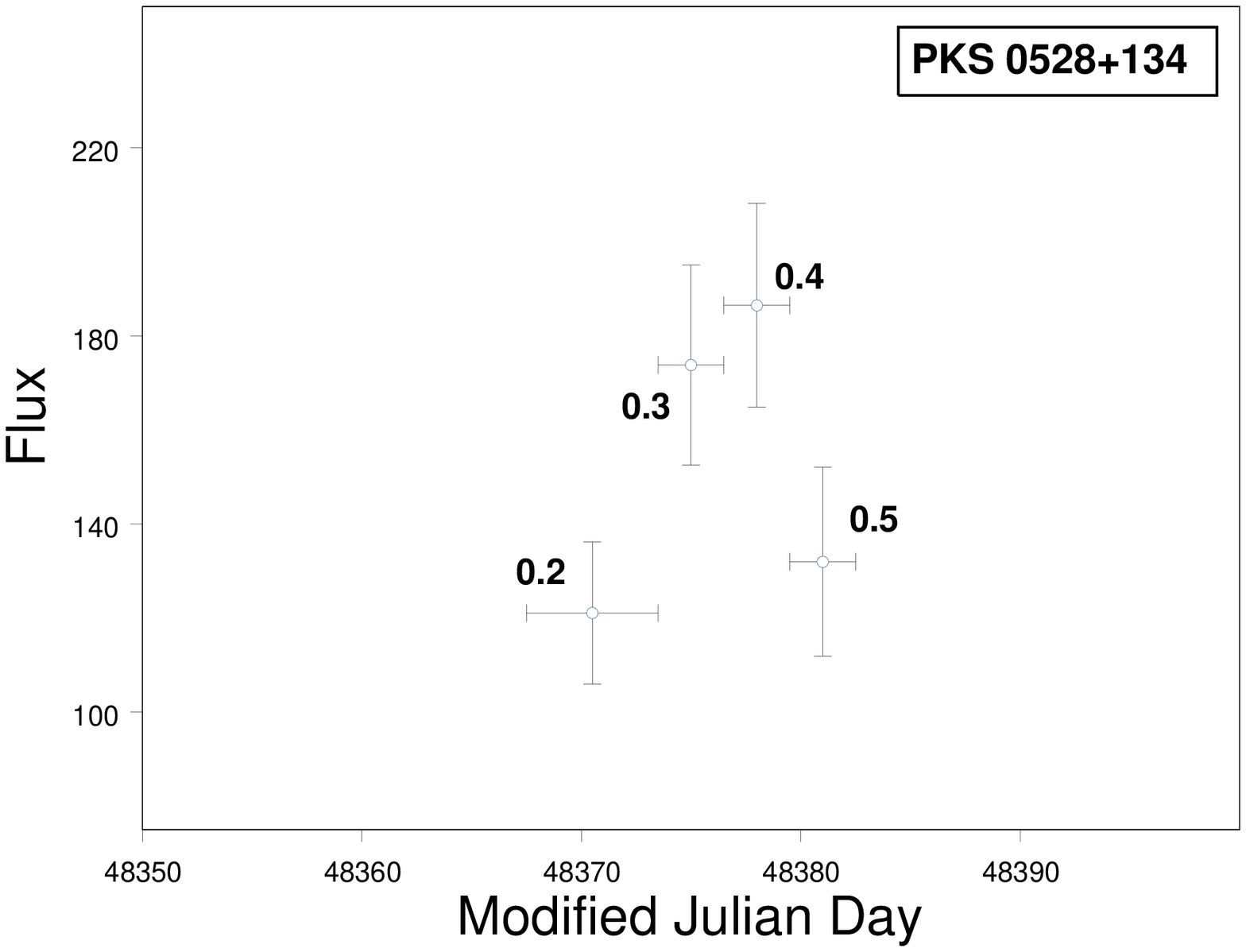}
\end{minipage}
\hspace{0.8in}
\begin{minipage}[b]{0.4\textwidth}
\centering
\includegraphics[width=3.6in]{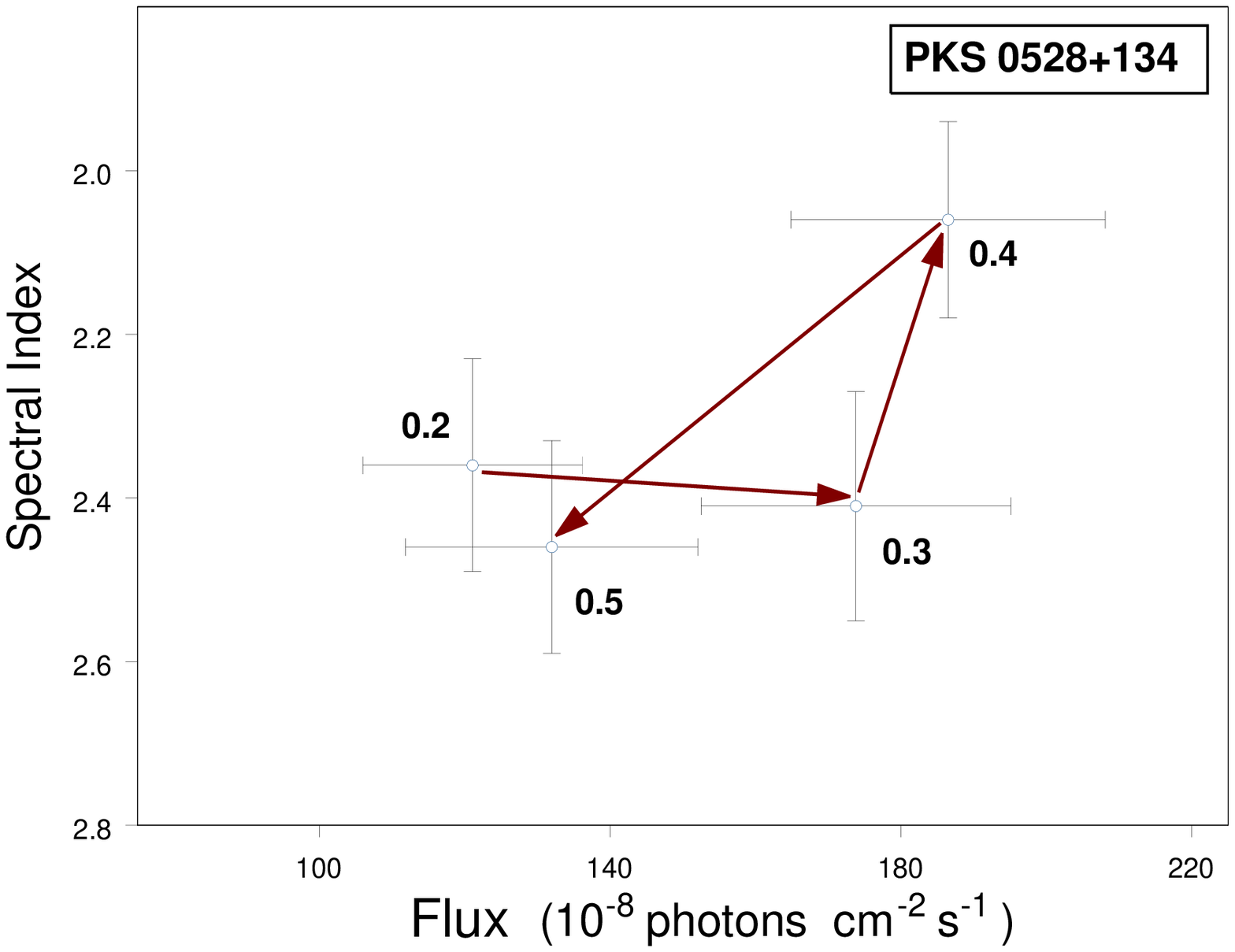}
\end{minipage}

\begin{minipage}[b]{0.4\textwidth}
\centering
\includegraphics[width=3.6in]{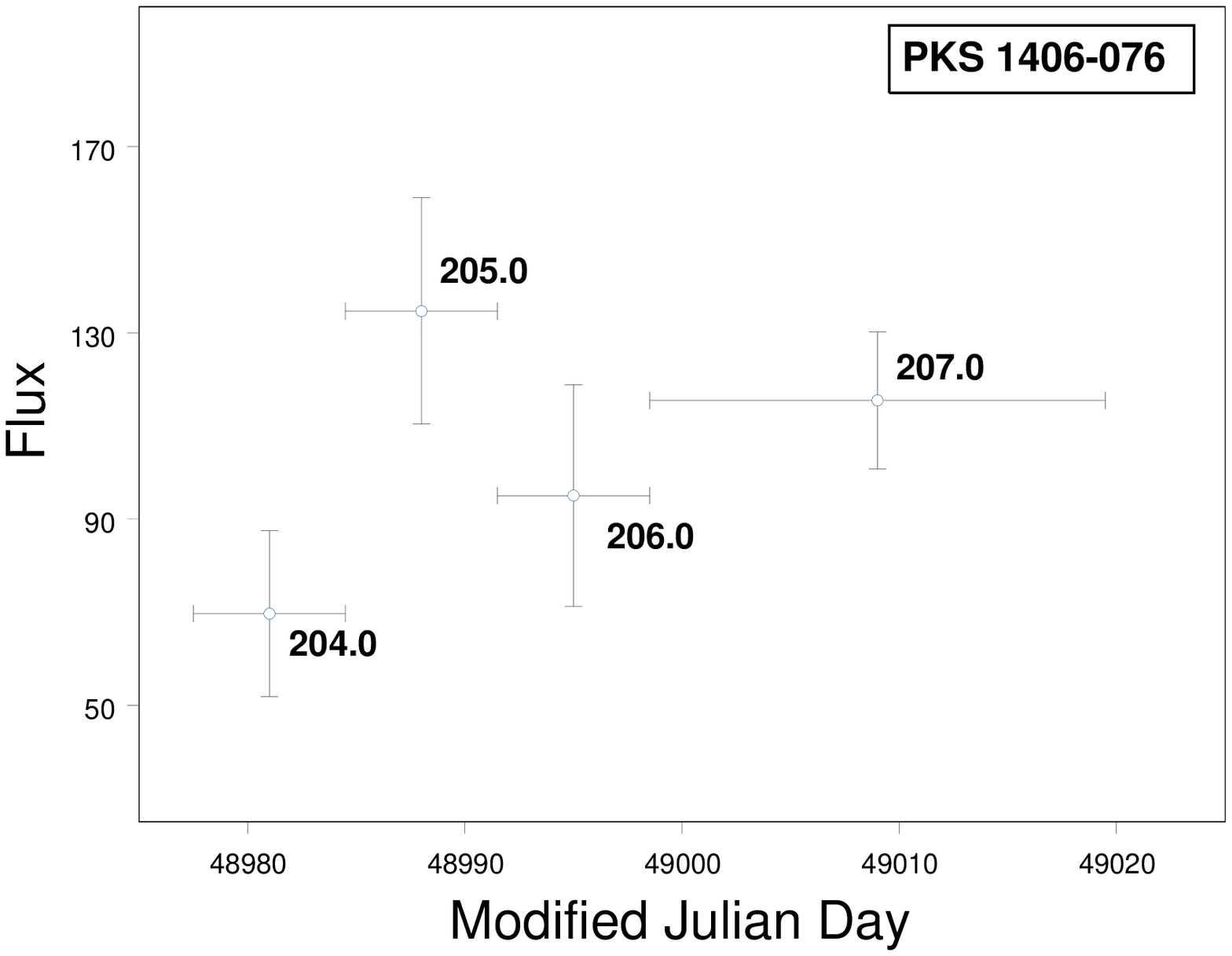}
\end{minipage}
\hspace{0.8in}
\begin{minipage}[b]{0.55\textwidth}
\includegraphics[width=3.6in]{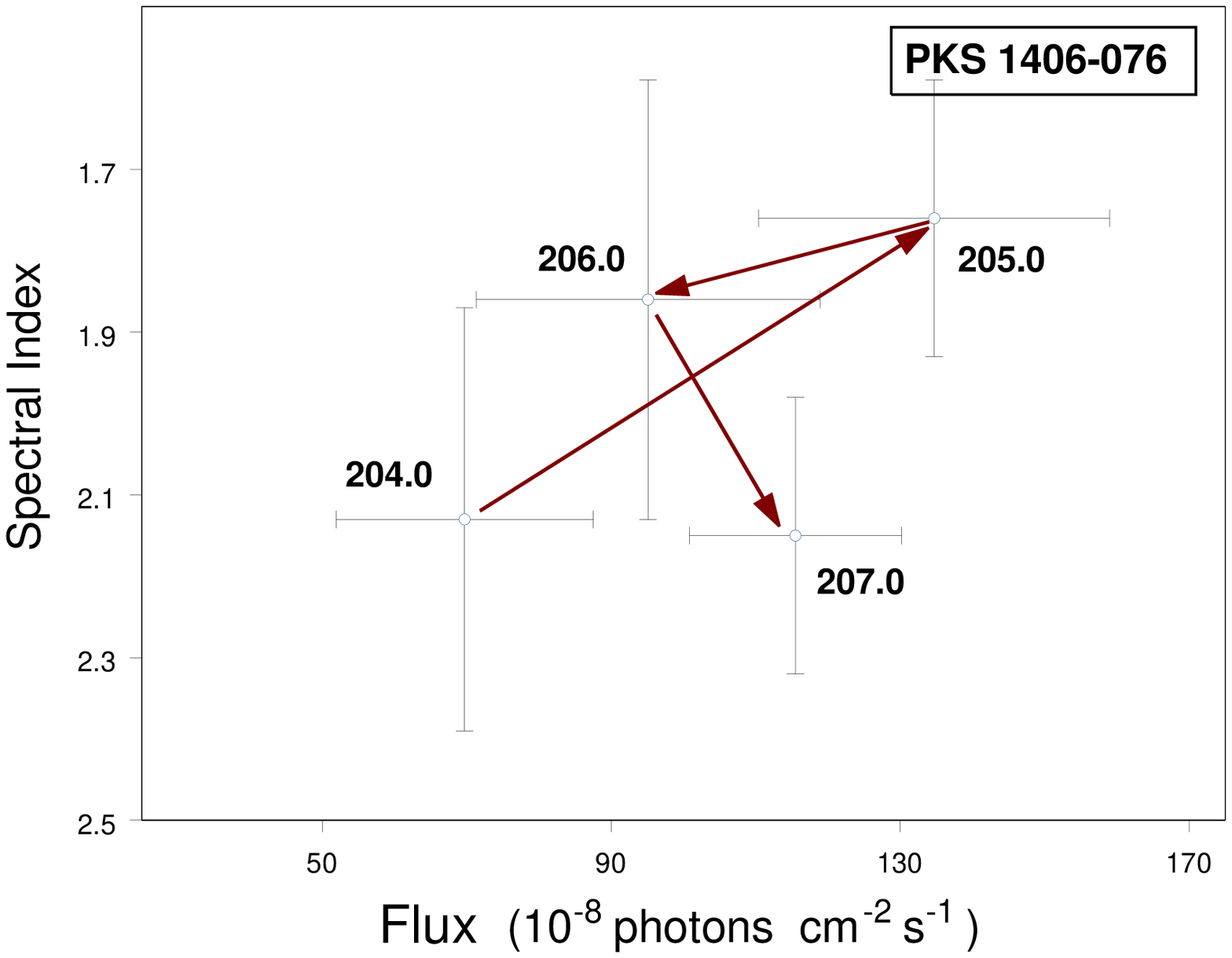}
\end{minipage}
\begin{minipage}[b]{0.4\textwidth}
\centering
\includegraphics[width=3.6in]{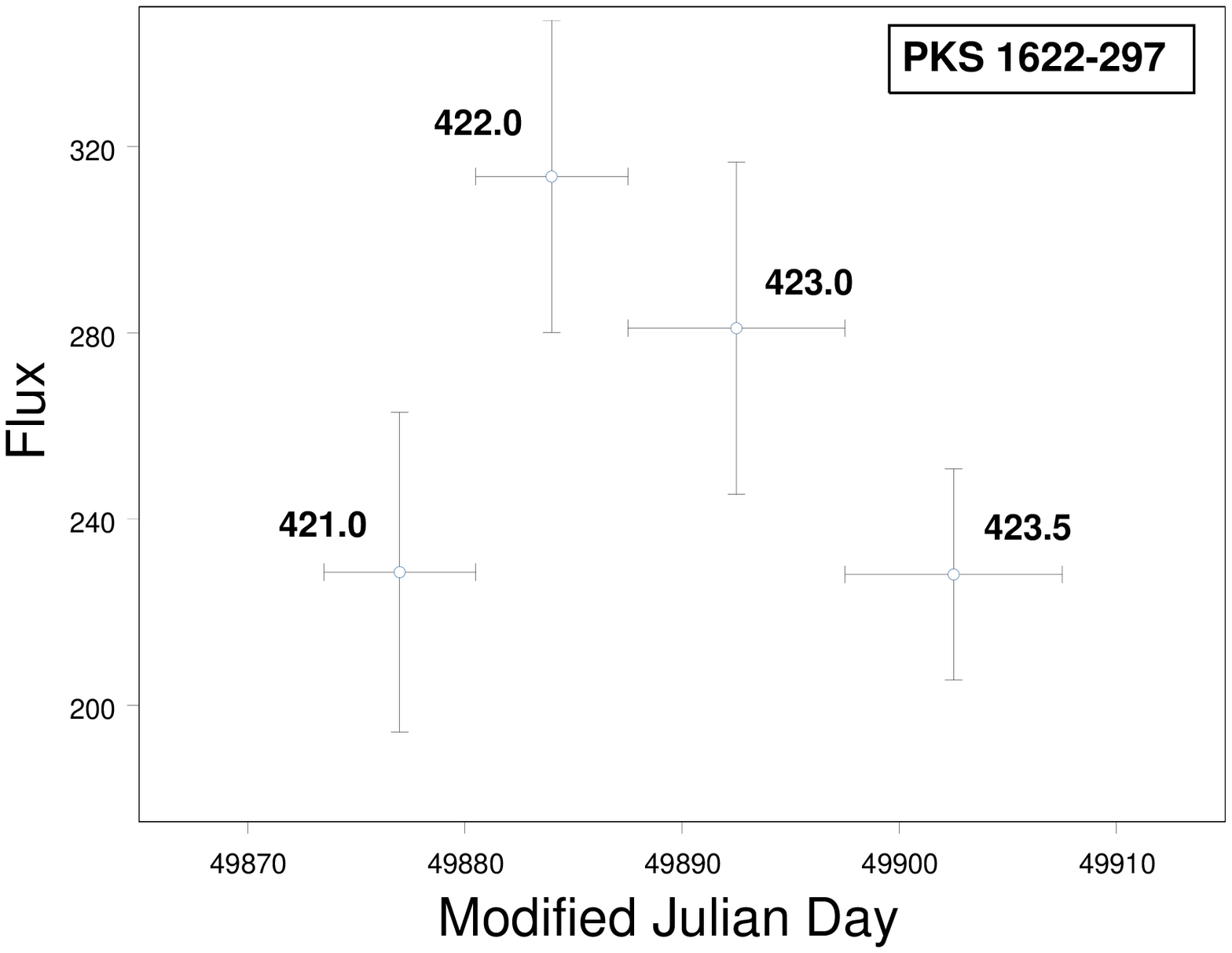}
\end{minipage}
\hspace{0.8in}
\begin{minipage}[b]{0.4\textwidth}
\centering
\includegraphics[width=3.6in]{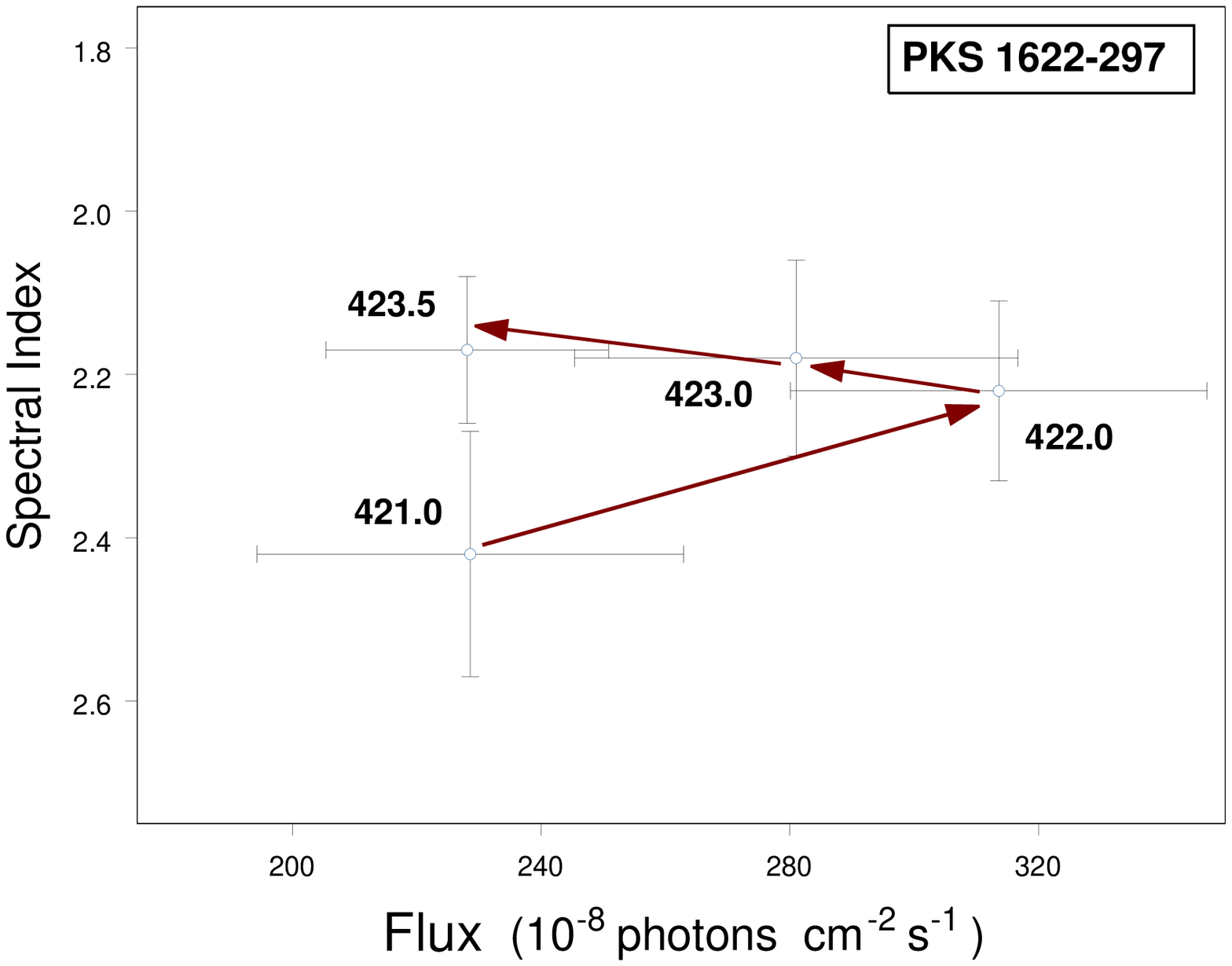}
\end{minipage}

\caption{Spectral hysteresis during blazar flares that lasted 3~4 weeks. Graphs on the left show the variability
of flux $>100 MeV$ in units of $10^{-8}$ photons cm$^{-2}$ sec$^{-1}$ with time (in MJD). The ranges for the time-axis 
and the flux axis are set to 50 days and 150 units respectively.  The graphs on the right show hysteresis 
in the photon spectral index vs. flux space. The arrows show a chronological progression.}

\label{egspec2}
\end{figure}

\begin{figure}
\plotone{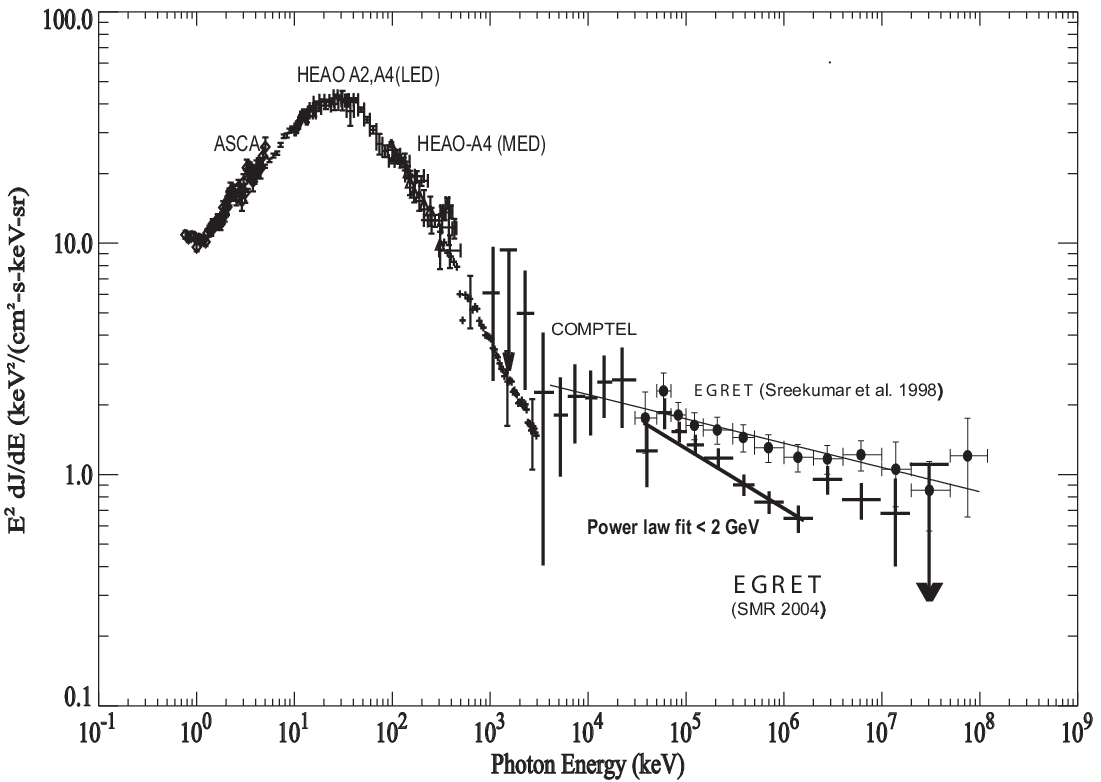}
\caption{The X-ray and Gamma-ray background broadband spectrum. The data are taken from \citet{stro04}. The extragalactic gamma-ray background (EGRB) calculated from the recalibrated EGRET data by \citet{stro04} (labeled SMR 2004) shows a break at 2~GeV and is steeper than the spectrum calculated by \citet{sre98} in the energy range 30~Mev - 2~GeV. We were
able to fit the points below 2~GeV with a power law of index 2.24$\pm$0.01, which is very close to the average blazar spectral index of 2.25$\pm$0.03 obtained with the recalibrated EGRET data. }
\label{egrb}
\end{figure}

\begin{deluxetable}{llllr}
\tabletypesize{\scriptsize}
\tablecolumns{5}
\tablecaption{Details of the Viewing Periods(after Cycle 4) included in the analysis and the sources that were analyzed \label{Tab-2}}
\tablewidth{0pc}
\tablehead{
\colhead{Viewing }& \colhead{Start Date}   & \colhead{End Date} &\colhead{Sources}&\colhead{Viewing}\\
Period\tablenotemark{a}&&&&Angle\tablenotemark{b}}
\startdata
502.0 &  10/17/95&    10/31/95&PKS~0528+134&0.86\\
511.0 &	01/16/96&	01/30/96&1253-055(3C~279)&6.72\\
511.5 &	01/30/96&	2/06/96&1253-055(3C~279)&5.01\\
 &&& 1226+023(3C~273)&15.36\\
513.0&02/06/96&02/13/96&PKS~2155-304&0.16\\
515.0 & 02/20/96&03/05/96&1101+384(Mrk~421)&18.45\\
516.1&   03/18/96&    	03/21/96&PKS~1622-297&10.91\\ 
516.5&03/21/96 &04/03/96&PKS~1633+382(4C+38.41)&1.13\\
 &&& PKS~1611+343(OS+319)&6.68\\
&&& Mrk~501&0.07\\
517.0&03/05/96   & 03/18/96&PKS~0208-512&2.25\\
518.5&   04/03/96&    	04/23/96&S5~0716+714&0.00\\
519.0&04/23/96  &  05/07/96 &PKS~1633+382(4C+38.41)&2.75\\
&&&  PKS~1611+343(OS+319)&8.56\\
&&&Mrk~501&1.23\\
520.4& 05/21/96  &  05/28/96&PKS~2155-304&0.00\\
526.0&    07/30/96&    	08/13/96&PKS~0528+134&8.29\\
527.0&    08/13/96&    08/20/96&PKS~0528+134&9.27\\
528.0&    08/20/96&    08/27/96&PKS~0528+134&12.13\\
606.0&   12/10/96&	12/17/96&1226+023(3C~273)&11.34\\
&&&  1253-055(3C~279)&1.00\\
607.0&    12/17/96&	12/23/96&1226+023(3C~273)&11.33\\
&&&  1253-055(3C~279)&1.00\\
608.0&    12/23/96&    12/30/96&1226+023(3C~273)&11.24\\
&&&  1253-055(3C~279)&1.01\\
609.0&   12/30/96&    01/07/97&1226+023(3C~273)&11.19\\
&&& 1253-055(3C~279)&1.01\\
610.0&   01/07/97&    01/14/97&1226+023(3C~273)&11.19\\
 &&& 1253-055(3C~279)&1.01\\
610.5&   01/14/97&    01/21/97&1226+023(3C~273)&9.60\\
&&&  1253-055(3C~279)&1.99\\
611.1&   01/21/97&    01/28/97&1226+023(3C~273)&11.16\\
 &&& 1253-055(3C~279)&1.01\\
616.1& 02/18/97  &  03/18/97&PKS~0528+134&0.00\\
617.8& 04/09/97   & 04/15/97&PKS~1633+382(4C+38.41) &6.71\\
&&&  PKS~1611+343(0S+319)&12.57\\
&&& Mrk~501&2.99\\
621.5 &  06/17/97&    06/24/97&1226+023(3C~273)&8.64\\
&&&  1253-055(3C~279)&1.99\\
623.5& 07/15/97	&     07/22/97	&BL~Lac&0.00\\
625.0&   08/05/97&    08/19/97&PKS~1622-297&17.68\\
615.1&   08/19/97&    08/26/97&PKS~1622-297&0.00\\
701.0 &  11/11/97&    11/18/97&PKS~2155-304&4.92\\
708.0 & 12/30/97 &    01/06/98	&PKS~2255-282&12.95\\
709.1 &	01/06/98&     01/13/98	& PKS~2255-282&12.95\\
715.5 &  03/20/98&   03/27/98&PKS~1156+295&2.00\\
716.5 &  03/27/98&   04/02/98&PKS~1156+295&17.98\\
806.5 &  01/19/99&    01/26/99&1253-055(3C~279)&6.30\\
806.7 &  01/26/99&    02/02/99&1253-055(3C~279)&2.87\\
910.0 &  02/08/00&    02/23/00&1253-055(3C~279)&4.00\\
911.1 &  02/23/00&    03/01/00&1253-055(3C~279)&8.40\\
 \enddata
\tablenotetext{a}{EGRET was operated in the reduced field of view mode during all the viewing periods to conserve gas.}
\tablenotetext{b}{Viewing angle is in degrees}
\end{deluxetable}

\clearpage
\LongTables
\begin{deluxetable}{lcllllllll}
\tabletypesize{\tiny}
\setlength{\tabcolsep}{0.01in}

\tablecolumns{10}
\tablecaption{Photon Spectral index (30 MeV -10 GeV), average flux ($>$ 100 MeV) and Log(synchrotron peak frequency) of blazars detected by EGRET\label{Tab-5}}
\tablehead{
\colhead{Source}& \colhead{Possible(?)}&\colhead{Other names}&
\colhead{RA \tablenotemark{a}}&\colhead{DEC \tablenotemark{a}}&
\colhead{Spectral } &\colhead{Flux} & 
\colhead{Log($\nu_{sync}$) \tablenotemark{d}}   &
\colhead{Multiple}&\colhead{Classification}\\
&&&&&index \tablenotemark{b}&$>$100 Mev \tablenotemark{c}&&Obs&} 

\startdata
0119+041&?&PKS, OC+033	&19.60&2.81&2.24$\pm$0.34	&12.6$\pm$4.3&&N&FSRQ(HP)\\
0130-171&?&PKS		&22.70	&-17.97&2.37$\pm$0.29	&12.4$\pm$3.8&		&N&FSRQ\\
0202+149&&PKS		&31.11&14.97&1.98$\pm$0.21	&23.0$\pm$5.5&13$\pm$1	&N&FSRQ(HP)\\
0208-512&&PKS		&32.58  & -50.93&1.95$\pm$0.11	&88.6$\pm$4.2	&13.8$\pm$0.1	&Y&FSRQ(HP)\\
0219+428&&3C~66A	&35.70&42.9&1.95$\pm$0.14	&17.7$\pm$2.8&15.0$\pm$1.5&Y&LBL\\
0234+285&?&4C+28.07	&39.99&28.26&2.56$\pm$0.23	&12.7$\pm$2.9&13.5$\pm$1&N&FSRQ(HP)\\
0235+164&&PKS, OD~160	&39.36&16.59&1.86$\pm$0.11	&25.4$\pm$3.6&13.3$\pm$0.3&N&LBL\\
0336-019&&CTA~026	&55.04&-2.02&1.87$\pm$0.22	&15.5$\pm$3.5&		&N&FSRQ(HP)\\
0414-189&&		&63.14& -18.88&1.96$\pm$0.45	&44.2$\pm$15.5&		&N&FSRQ\\
0415+379&?&3C111	&64.04&36.84&2.55$\pm$0.24	&12.2$\pm$2.6&		&N&FSRQ\\
0420-014&&PKS		&65.65&-1.04&2.59$\pm$0.17	&15.0$\pm$3.0&13.2$\pm$0.3&N&FSRQ(HP)\\
0430+2859&?&		&68.40&29.14&1.97$\pm$0.10	&21.30$\pm$2.8&	\nodata	&Y&LBL\\
0440-003&&PKS, NRAO~190	&70.55&-0.55&2.23$\pm$0.12	&11.6$\pm$2.7&13.5$\pm$0.5&N&FSRQ(HP)\\
0446+112&&PKS		&72.61&11.09&2.19$\pm$0.15	&14.0$\pm$2.0&13$\pm$1	&N&FSRQ\\
0454-234&&PKS		&74.24&-23.64&2.27$\pm$0.32	&7.5$\pm$2.5&		&N&LBL\\
0454-463&&PKS		&74.57&-46.60&2.56$\pm$0.37	&7.8$\pm$2.1&13$\pm$1	&N&FSRQ(LP)\\
0458-020&?&PKS		&75.10&-1.99&2.45$\pm$0.22	&10.2$\pm$2.2&		&N&FSRQ(HP)\\
0459+060&?&		&74.93&5.75&2.06$\pm$0.36	&9.7$\pm$3.1&		&N&FSRQ\\
0506-612&?&PKS		&78.15&-61.84&2.37$\pm$0.29	&5.7$\pm$1.8&		&N&FSRQ\\
0521-365&?&PKS		&82.54&-36.44&2.36$\pm$0.24	&19.3$\pm$3.5&14.0$\pm$0.5&N&FSRQ(HP)\\
0528+134&&PKS		&82.74&13.38&2.36$\pm$0.03	&95.8$\pm$3.7&13.0$\pm$0.2&Y&FSRQ(LP)\\
0537-286&? &PKS, OG-263 &82.91& -29.68&2.23$\pm$0.50	&32.6$\pm$11.5&		&N&FSRQ\\
0537-441&&PKS		&85.02&-44.05&2.36$\pm$0.12	&24.3$\pm$3.0&13.5$\pm$0.5&Y&LBL\\
0539-057&?&PKS		&85.57&-6.93&1.88$\pm$0.34	&65.0$\pm$19.8&&N&FSRQ\\
0616-116&&		&95.58&-11.66&2.74$\pm$0.31	&17.8$\pm$4.7&		&N&FSRS\\
0716+714&&S5		&110.43&71.35&2.19$\pm$0.11	&18.4$\pm$2.1&15.0$\pm$0.1&Y&LBL\\
0735+178&&		&114.47&17.35&2.44$\pm$0.23	&14.8$\pm$3.1&14.5$\pm$0.5&N&LBL\\
0738+5451&&		&115.83&54.80&2.00$\pm$0.19	&11.1$\pm$2.1&		&N&FSRS\\
0803+5126&?&		&122.15&51.24&2.78$\pm$0.27	&9.4$\pm$2.4 &		&N&FSRQ\\
0805-077&?&		&123.14&-6.78&2.39$\pm$0.31	&24.6$\pm$5.5&13.2$\pm$1&N&FSRQ\\
0804+499&?&		&122.18&48.75&2.13$\pm$0.42	&9.9$\pm$2.4&13.5$\pm$1&N&FSRQ(HP)\\
OR 0809+483&&&&&&&&&\\ 
0827+243&&OJ~248	&127.49&24.22&2.38$\pm$0.2	&25.4$\pm$3.9&13.3$\pm$1&Y&FSRQ(LP)\\
 0829+046&&OJ+049	&127.03&5.14&2.42$\pm$0.42	&16.1$\pm$4.9&		&N&LBL\\
 0836+710&&4C +71.07	&131.46&70.83&2.69$\pm$0.16	&9.7$\pm$1.7&13$\pm$0.5&N&FSRQ(LP)\\
 0847-120&&		&133.16&-12.27&1.48$\pm$0.26	&44.4$\pm$11.8&&N&FSRQ\\
 0851+202&&PKS, OJ~287	&133.42&19.68&1.91$\pm$0.28	&10.4$\pm$3.0	&	&N&LBL\\
 0917+449&?&		&139.33&44.45&2.06$\pm$0.14	&14.1$\pm$2.1&13.0$\pm$0.5&Y&FSRQ(LP)\\
 0954+556&&4C~55.17	&148.01&55.02&2.07$\pm$0.18	&9.5$\pm$1.6&14.0$\pm$1.0&Y&FSRQ(HP)\\
 0954+658&&S4		&149.62&65.56&2.03$\pm$0.18	&6$\pm$1.6&13$\pm$1&N&LBL\\
 1011+496&?&		&152.29&48.93&1.85$\pm$0.32	&4.7$\pm$1.4&&N&LBL\\
 1055+567&?&		&163.21&57.31&2.15$\pm$0.37	&9.0$\pm$2.4&&N&LBL\\
 1101+384&&Mrk 421	&166.10&38.15&1.57$\pm$0.14	&13.6$\pm$1.8&16.7$\pm$0.5&Y&HBL\\
 1127-145&?&PKS		&173.66&-15.50&2.30$\pm$0.25	&37.8$\pm$8.2&12.7$\pm$1&N&FSRQ(LP)\\
 1156+295&&PKS		&180.12&28.80&1.97$\pm$0.17	&8.7$\pm$1.8&14.5$\pm$0.3&Y&FSRQ(HP)\\
 1219+285&&PKS, W Comae, ON~231	&185.75&28.70&1.80$\pm$0.17	&11.5$\pm$1.5&14.5$\pm$0.5&Y&LBL\\
 1222+216&&4C~21.35	&186.11&21.31&2.33$\pm$0.1	&15.1$\pm$1.8&13$\pm$1&Y&FSRQ(LP)\\
 1226+023&&3C~273	&185.25&2.17&2.56$\pm$0.07	&19.5$\pm$1.7&13$\pm$1&Y&FSRQ(LP)\\
 1229-021&&PKS, 4C-02.55,ON-049	&187.65&-2.79&2.64$\pm$0.36&5.5$\pm$1.5	&13$\pm$1&N&FSRQ(LP)\\
1237+0459&?&		&188.91&4.97&2.78$\pm$0.32	&4.8$\pm$1.5&&N&FSRS\\
1243-072&&PKS, ON-073	&191.75&-6.86&2.75$\pm$0.17	&8.4$\pm$1.9&&N&FSRQ\\
1253-055&&3C~279	&193.98&-5.82&1.98$\pm$0.02	&81.8$\pm$2.5&13.2$\pm$0.2&Y&FSRQ(HP)\\
1313-333&?&PKS, OP-322	&198.51&-34.52&2.09$\pm$0.18	&35.5$\pm$3.3&12.7$\pm$0.1&N&FSRQ\\
1322-428&?&PKS, Cen.~A	&201.15&-43.25&2.54$\pm$0.23	&13.5$\pm$2.5&&N&Radio Gal.\\
1324+224&?&		&200.80&22.01&1.62$\pm$0.24	&17.9$\pm$4.1&&N&FSRQ\\
1331+170&&OP~151	&202.39&17.14&2.38$\pm$0.38	&7.6$\pm$2.6&&N&FSRQ\\
1334-127&&PKS		&204.84&-14.32&1.92$\pm$0.25	&11.9$\pm$3.3&&N&FSRQ(HP)\\
1406-076&&PKS		&212.42&-7.75&2.21$\pm$0.10	& 31.3$\pm$3.0&14.8$\pm$0.2&Y&FSRQ(LP)\\
1424-418&&PKS		&217.39&-42.30&2.10$\pm$0.16    &11.0$\pm$2.6&13.0$\pm$0.5&N&FSRQ(HP)\\
1504-166&?&PKS		&226.20&-15.63&1.79$\pm$0.34	&32.1$\pm$10.1&&N&FSRQ\\
1510-089&&PKS		&228.17&-8.83&2.45$\pm$0.21	&18.1$\pm$3.7&13.0$\pm$0.5&N&FSRQ(HP)\\
1514-241&?&PKS		&229.34&-25.65&2.67$\pm$0.41	&26.6$\pm$8.1&&N&LBL\\
 1604+159&&4C+15.54	&241.30&15.89&2.02$\pm$0.31	&11.0$\pm$3.9&14$\pm$1&N&LBL\\
 1606+106&&4C~10.45	&242.12&10.93&2.44$\pm$0.18	&25.4$\pm$4.5&13.5$\pm$1&N&FSRQ(LP)\\
 1611+343&&OS+319	&243.54&34.40&2.35$\pm$0.15	&27.6$\pm$4.0&&Y&FSRQ\\
 1622-253&&PKS		&246.50&-25.32&2.12$\pm$0.14	&24.2$\pm$3.5&12.5$\pm$1&Y&FSRQ(LP)\\
 1622-297&&PKS		&246.36&-29.92&2.17$\pm$0.11	&47.7$\pm$3.5&13.2$\pm$0.4&Y&FSRQ(LP)\\
 1633+382&&4C+38.41	&248.92&38.22&2.15$\pm$0.08	&59.$\pm$5.2&13$\pm$1&Y&FSRQ(LP)\\
 1652+398&&Mrk~501	& 253.47&  39.76&1.48$\pm$0.44	&10.1$\pm$4.1&18.5$\pm$1.0&N&HBL\\
1716-771&?&		&260.22&-78.34&2.08$\pm$0.47	&19.8$\pm$6.9&&N&FSRS\\
 1725+044&&PKS		&261.97&4.50&2.63$\pm$0.26	&16.2$\pm$3.9&&N&FSRQ\\
 1730-130&&NRAO~530	&263.46&-13.23&2.38$\pm$0.08	&35.0 $\pm$  3.3&13$\pm$1&Y&FSRQ\\
 1739+522&&4C+51.37	&264.64&52.05&2.49$\pm$0.21	&21.0 $\pm$3.9&13$\pm$1&N&FSRQ\\
1741-038&&PKS		&266.02&-3.18&2.59$\pm$0.33	&18.4 $\pm$5.2&14$\pm$1&N&FSRQ(HP)\\
 1759-396&&		&270.22&-39.93&2.96$\pm$0.26	&10.3$\pm$2.8&&N&FSRS\\
 1804-502&?&J1808-5011	&271.55& -50.10&2.86$\pm$0.34	&6.2$\pm$2.7&&N&FSRS\\
1830-210&&		&278.10&-21.18&2.62$\pm$0.13	&26.6$\pm$3.6&&Y&FSRQ\\
 1908-201&&		&287.93&-20.00&2.31$\pm$0.18	&16.0$\pm$2.6&&N&FSRS\\
 1920-211&?&		&290.50&-20.26&2.37$\pm$0.48	&28.3$\pm$8.0&&N&FSRS\\
 1933-400&&PKS		&293.98&-40.38&2.69$\pm$0.32	&8.3$\pm$2.6&13.3$\pm$1&N&FSRQ\\
1936-155&&PKS		&294.47&-15.49&2.32$\pm$0.42	&55.4$\pm$18.7&&N&FSRQ\\
2002-233&&TXS		&301.54&-23.35&2.35$\pm$0.27	&16.7$\pm$4.2&&N&FSRS\\
2005-489&?&PKS		&302.35&-48.83&\nodata          &11.0$\pm$4.4&&N&HBL\\
2022-077&&		&306.36&-7.75&2.32$\pm$0.17	&20.0$\pm$3.5&&N&FSRQ\\
 2032+107&&PKS		&309.18&11.54&2.79$\pm$0.24	&14.4$\pm$3.1&13.0$\pm$0.5&N&LBL\\
 2052-474&&PKS		&313.80&-47.28&1.85$\pm$0.26	&21.4$\pm$5.8 &13.5$\pm$1&N&FSRQ(LP)\\
 2105+598&?&		&315.18&60.21&2.07$\pm$0.24	&19.4$\pm$4.1&&N&FSRQ\\
 2155-304&&PKS		&329.68&-30.40&1.88$\pm$0.17	&18.8$\pm$2.9&16.2$\pm$0.2&Y&HBL\\
 2200+420&&BL~Lac	&330.60&42.29&1.72$\pm$0.18	&20.3$\pm$3.2&14.6$\pm$0.4&Y&LBL\\
 2206+650&?&		&331.60&66.05&2.37$\pm$0.25	&25.9$\pm$5.2&&N&FSRS\\
 2209+236&&PKS		&332.41&24.03&2.31$\pm$0.32	&13.3$\pm$4.2 &&N&FSRQ\\
 2230+114&&CTA~102	&338.11&11.80&2.46$\pm$0.13	&19.0$\pm$2.8&13$\pm$1&Y&FSRQ(HP)\\
 2250+1926&?&		&343.99&19.73&1.87$\pm$0.43	&62.2$\pm$22.2&&N&FSRQ\\
 2251+158&&3C~454.3	&343.51&16.02&2.22$\pm$0.06	&56.5$\pm$4.0&13.5$\pm$0.5&Y&FSRQ(HP)\\
 2255-282&&PKS		&344.52& -27.97&1.69$\pm$0.18\tablenotemark{e}	&12.8$\pm$2.8&12.7$\pm$0.3&N&FSRQ(HP)\\
 2320-035&&PKS		&350.41&-3.48&2.17$\pm$0.45	&30.5$\pm$9.6&&N&FSRQ\\
 2346+385&?&		&358.10&37.88&2.70$\pm$0.33	&35.5$\pm$10.2&&N&FSRQ\\
 2351+456&&		&359.57&46.07&2.57$\pm$0.35	&13.7$\pm$3.6&&N&FSRQ\\
 2356+196&&OZ+193	&359.99&20.70&2.24$\pm$0.33	&8.5$\pm$2.8&&N&FSRQ\\  

\enddata
\tablenotetext{a}{These are EGRET positions from 3EG}
\tablenotetext{b}{Photon index measured in the 30 MeV - 10 GeV energy range}
\tablenotetext{c}{Flux is in units of $10^{-8}$ photons cm$^{-2}$ s$^{-1}$}
\tablenotetext{d}{$Log(\nu_{sync})$-Logarithm of the frequency (in Hz) of the synchrotron peak. These have been obtained from the literature cited in \S \ref{specgdist}}
\tablenotetext{e}{Spectral index is during a flare. This was the only spectral index that could be calculated. The quoted flux is the average flux observed.}
\tablenotetext{?}{Possible EGRET identification}
\tablecomments{FSRQ: Flat spectrum Radio Quasar; LBL: Low frequency-peaked BL~Lac object; HBL: High frequency-peaked BL~Lac object; HP: High polarization; LP: Low polarization; FSRS: Flat spectrum Radio Source. Classifications are based on \citet{hart97} and \citet{ghi98}}
\end{deluxetable}

\clearpage

\clearpage
\LongTables
\begin{deluxetable}{lllllll}
\tabletypesize{\tiny}

\tablecolumns{7}
\tablecaption{Gamma-ray photon spectral index (30 MeV - 10 GeV) and flux ($>$100MeV) of blazars that were bright and were observed multiple times by EGRET\label{Tab-3}. The corresponding plots of spectral index vs. flux are shown in Fig. \ref{egspec1}. }
\tablehead{\colhead{Source}& \colhead{Start Dates}& \colhead{Viewing Periods}& \colhead{Graph label \tablenotemark{a}} & \colhead{Spectral Index}&Flux& \colhead{Det. $\sigma$}\\
			   &			   &	Pooled		    &in Fig. \ref{egspec1}& (30 MeV-10 GeV)&($>100~MeV$)\tablenotemark{c}&}          
\startdata
0208-512	&09/05/91		&9.1		&9.1		&1.49$\pm$0.30	&39.4$\pm$13.4	&4.2	\\
		&09/19/91			&10.0			&10.0		&1.91$\pm$0.06	&111.8$\pm$8.2	&21.5	\\
		&05/08/93, 06/03/93		&220.0, 224.0		&2		&2.13$\pm$0.21	&57.4$\pm$10.8	&7.8	\\
		&05/31/94, 07/12/94, 07/25/94	&329.0, 335.0, 335.5	&3		&2.04$\pm$0.12	&98.6$\pm$12.0	&6.7	\\
		&01/10/95, 09/07/95		&409.0, 428.0		&4		&2.32$\pm$0.13  &75.0$\pm$9.8  &4.6    \\
		&03/05/96			&517.0			&517.0		&1.82$\pm$0.08	&139.8$\pm$12.4 &18.3	\\

0219+428(3C 66A)	&11/28/91, 08/11/92, 08/12/92,	&15.0, 36.0, 36.5,	&12	&1.88$\pm$0.21	&14.7$\pm$3.5	&4.9\\
			&09/01/92, 02/25/93		&39.0, 211.0		&	&		&		&\\	
			&04/26/94			&325.0			&3	&1.66$\pm$0.23	&22.9$\pm$5.7	&5.1\\
0430+2859	&04/22/91, 04/28/91, 05/01/91,		&0.2, 0.3, 0.4,  	&1	&1.85$\pm$0.16	&15.4$\pm$3.3	&5.2\\
		&05/04/91, 05/16/91, 06/08/91,		&0.5, 1.0, 2.1,		&	&	&	&\\
		&11/28/91, 06/11/92, 08/11/92,		&15.0, 31.0, 36.0,	&	&	&	&\\	
		&08/12/92, 09/01/92			&36.5, 39.0		&	&	&	&\\
		&12/01/93, 02/08/94, 02/15/94,		&310.0, 321.1, 321.5,	&3	&1.65$\pm$0.26	&25.8  $\pm$ 8.0	&4.1\\
		&04/26/94				&325.0			&	&	&	&\\
		&02/28/95, 03/07/95, 05/23/95,		&412.0, 413.0, 420.0,	&4	&2.44$\pm$0.22	&37.6 $\pm$7.8	&5.8\\
		&08/08/95, 08/22/95 			&426.0, 427.0		&	&	&	&\\
		
0528+134	&04/22/91			&0.2			&0.2		&2.36$\pm$0.13	&121.0$\pm$15.1	&10.4	\\
	&04/28/91			&0.3			&0.3		&2.41$\pm$0.14	&173.8$\pm$21.3	&11.0	\\
	&05/01/91			&0.4			&0.4		&2.06$\pm$0.12	&186.5$\pm$21.6	&12.0	\\
	&05/04/91			&0.5			&0.5		&2.46$\pm$0.13	&132.0$\pm$20.1 &8.5	\\
	&05/16/91		 	&1.0			&1.0		&2.31$\pm$0.09	&102.1$\pm$9.1	&14.5	\\
	&06/08/91			&2.1			&2.1		&2.36$\pm$0.25	&68.7$\pm$13.1	&6.2\\
	&03/23/93			&213.0			&213.0		&2.30$\pm$0.10	&356.7$\pm$37.7	&14.0	\\
	&12/01/93, 02/08/94, 02/15/94, &310.0, 321.1, 321.5,	&3		&2.48$\pm$0.18	&44.0$\pm$7.4&6.3	\\
	& 08/09/94 		       &337.0			&		&		&		&\\	
	&02/28/95, 03/07/95, 04/04/95, 		&412.0, 413.0, 419.1, &4		&2.44$\pm$0.07	&100.4$\pm$7.4&17.3	\\
	&05/09/95,				&419.5, 420.0, 426.0	&	&	&	&\\	
	&10/17/95, 07/30/96, 08/13/96,		&502.0, 526.0, 527.0, 	&56		&2.28$\pm$0.12	&68.1$\pm$6.9&11.7\\
	&08/20/96, 02/18/97			&528.0, 616.1					&	&	&	&\\

0537-441	&07/26/91, 08/22/91, 12/27/91,	&6.0,8.0,17.0,			&1		&2.64$\pm$0.23	&18.0$\pm$4.3	&5.1\\
		&05/14/92			&29.0			&	&	&	&\\
		&05/31/94, 07/12/94, 07/25/94	&329.0, 335.0, 335.5	&3		&2.59$\pm$0.27	&16.4$\pm$4.5	&4.5\\	
		&01/10/95, 04/11/95		&409.0, 415.0		&4		&2.16$\pm$0.17	&57.3$\pm$9.1	&9.3\\

0716+714	&05/07/91, 01/10/92, 03/05/92,	&0.6, 18.0, 22.0, 			&1		&2.31$\pm$0.20	&22.4$\pm$3.9	&7.3\\
		&06/11/92 			&31.0	&		&		&		&	\\
		&04/06/93, 04/06/93, 07/13/93	&216.0, 227.0, 228.0		&2		&2.07$\pm$0.23	&13.8$\pm$3.4	&5.0\\		
		&03/01/94, 02/21/95		&411.1, 411.5		&4		&2.47$\pm$0.28	&26.7$\pm$6.4	&5.4\\	
		&04/03/96, 09/06/96		&518.5, 530.0		&5		&1.81$\pm$0.26	&25.4$\pm$5.2	&6.2\\

0827+343 (OJ~248)	&09/17/92			&40.0			&1		&2.13$\pm$0.28	&22.0$\pm$5.4	&5.2\\
		&11/09/94			&403.5			&4		&2.22$\pm$0.25	&70.6$\pm$14.7	&7.1\\
0917+499	&05/07/91, 06/28/91, 01/10/92,   &0.6, 4.0, 18.0,		&1		&1.97$\pm$0.25&14.0$\pm$3.2	&5.5\\
		&09/17/92			&40.0			&		&	&	&\\
		&04/06/93, 04/20/93, 05/24/93,	&216.0, 218.0, 222.0,	&2		&1.98$\pm$0.36	&11.2$\pm$3.40	&4.1\\		
		&06/29/93, 07/13/93		&227.0, 228.0		&		&		&		&\\
		&04/05/94, 05/10/94		&322.0, 326.0		&3		&2.19$\pm$0.24	&20.8$\pm$5.1	&5.5\\
0954+556 (4C~55.17)	&05/07/91, 06/28/91, 01/10/92, 	&0.6, 4.0, 18.0,	&1		&2.32$\pm$0.37	&6.50$\pm$2.5	&3.1\\	
		&09/17/92			&40.0			&		&		&		&\\
		&04/06/93, 04/20/93, 05/24/93,	&216.0, 218.0, 222.0,	&2		&1.87$\pm$0.29	&8.4$\pm$2.4	&4.1\\	
		&06/29/93, 07/13/93		&227.0, 228.0		&		&		&		&\\
		&03/01/94, 03/15/94, 04/05/94	&319.0, 319.5, 322.0,	&3		&1.75$\pm$0.28	&18.3$\pm$5.0	&4.8\\
		&05/10/94			&326.0			&		&		&		&\\
1101+384 (Mrk~421)	&05/07/91, 06/28/91, 09/17/92	&0.6, 4.0, 40.0		&1		&1.67$\pm$0.2	&17.7$\pm$3.2	&7.2\\
		&04/20/93, 05/24/93, 06/29/93,	&218.0, 222.0, 227.0,	&2		&1.83$\pm$0.29	&12.4$\pm$3.5	&4.6\\
		&07/13/93 			&228.0			&		&		&		&	\\
		&05/10/94			&326.0			&326.0		&1.51$\pm$0.26	&24.5$\pm$6.7	&5.3\\
1156+295	&01/05/93			&206.0			&206.0		&1.98$\pm$0.41	&166.9$\pm$41.4	&6.8\\
		&11/09/93, 11/16/93		&307.0,308.0		&v+307\tablenotemark{b}		&1.67$\pm$0.35	&44.7$\pm$14.4	&4.1\\	
		&04/25/95			&418.0			&418.0 		&1.78$\pm$0.24&45.4$\pm$10.8&6.3\\	
		&03/20/98			&715.5			&715.5		&2.44$\pm$0.43	&76.0$\pm$22.9&5.1\\
1219+285 (ON~231)	&06/15/91, 06/28/91, 10/03/91,	&3.0, 4.0, 11.0,		&12		&1.18$\pm$0.35	&5.7$\pm$2.2&3.0\\
	        &12/22/92, 12/29/92, 01/05/93, 	&204.0, 205.0, 206.0,	&		&		&	\\
		&04/20/93, 05/24/93		&218.0, 222.0		&		&		&		&	\\
		&10/19/93, 10/25/93, 11/02/93,			&304.0, 305.0, 306.0,	&3		&2.2$\pm$0.21	&17.2$\pm$3.5	&6.1\\
		&11/09/93, 11/16/93, 11/23/93,			&307.0, 308.0, 308.6,	&		&		&		&	\\
		&12/13/93, 12/17/93, 12/20/93,			&311.0, 311.6, 312.0,	&		&		&		&	\\
		&12/27/93, 04/05/94, 05/10/94			&313.0, 322.0, 326.0	&		&		&		&	\\

		&12/13/94, 04/25/95		&406.0, 418.0		&4 	&1.76$\pm$0.29	&35.0$\pm$9.4	&4.9\\
	
1222+216 (4C~21.35)	&12/22/92, 12/29/92, 01/05/93, 		&204.0, 205.0, 206.0,	&2		&2.54$\pm$0.24	&26.5$\pm$6.0 &5.7\\
		&04/20/93, 05/24/93			&218.0, 222.0		&		&		&		&	\\	
		&10/19/93, 10/25/93, 11/02/93,		&304.0, 305.0, 306.0,	&3		&1.94$\pm$0.16	&14.9$\pm$2.8 &6.8\\			
		&11/09/93, 11/16/93, 11/23/93,			&307.0, 308.0, 308.6,	&		&		&		&	\\
		&12/13/93, 12/17/93, 12/20/93,			&311.0, 311.6, 312.0,	&		&		&		&	\\
		&12/27/93, 04/05/94, 05/10/94			&313.0, 322.0, 326.0	&		&		&		&	\\

1226+023 (3C~273)&06/15/91, 10/03/91			&3.0, 11.0		&1		&2.45$\pm$0.25	&11.2$\pm$3.0	&4.3\\
		&10/19/93, 10/25/93, 11/02/93,			&304.0, 305.0, 306.0,	&3		&2.62$\pm$0.12	&29.7$\pm$3.7	&10.6\\
		&11/09/93, 11/16/93, 11/23/93,			&307.0, 308.0, 308.6,	&		&		&		&	\\
		&12/13/93, 12/17/93, 12/20/93,			&311.0, 311.6, 312.0,	&		&		&		&	\\
		&12/27/93, 04/05/94, 05/10/94			&313.0, 322.0, 326.0,	&		&		&		&	\\
		&11/29/94, 12/13/94, 12/20/94			&405.0, 406.0, 407.0,			&4		&2.73$\pm$0.2	&13.4$\pm$4.0	&4.0\\
		&01/03/95			&408.0			&		&		&		&	\\
		&01/30/96			&511.5			&511.5		&2.24$\pm$0.26	&32.8$\pm$7.5	&5.8\\
		&12/30/96, 01/07/97			&609.0, 610.0			&6a		&2.60$\pm$0.19	&127.1$\pm$20.7 &9.2\\
		&12/10/96, 12/17/96, 12/23/96,			&606.0, 607.0, 608.0,		&6b		&2.66$\pm$0.23	&50.2$\pm$10.2	&6.60\\
		&01/14/97,  01/21/97			&610.5, 611.1	&		&		&		&	\\
1253-055 (3C~279)&06/15/91			&3.0			&3.0		&1.78$\pm$0.04	&249.5$\pm$10.7	&37.1\\
		&10/03/91			&11.0			&11.0		&1.88$\pm$0.08	&81.5$\pm$7.6	&15.2\\
		&10/19/93, 10/25/93, 11/02/93,			&304.0, 305.0, 306.0,	&3		&2.34$\pm$0.10	&46.9$\pm$5.10	&12.5\\
		&11/09/93, 11/16/93, 11/23/93,			&307.0, 308.0, 308.6,	&		&		&		&	\\
		&12/13/93, 12/17/93, 12/20/93,			&311.0, 311.6, 312.0,	&		&		&		&	\\
		&12/27/93, 04/05/94, 05/10/94 			&313.0, 322.0, 326.0	&		&		&		&	\\
		&11/29/94, 12/13/94, 12/20/94			&405.0, 406.0, 407.0	&4		&2.19$\pm$0.12	&27.6$\pm$3.8	&9.2\\
		&01/16/96					&511.0			&511.0		&1.89$\pm$0.11	&125.7$\pm$15.6	&10.4\\
		&02/20/96					&511.5			&511.5		&1.92$\pm$0.06	&558.6$\pm$34.5 &27.6\\	
		&12/10/96, 12/17/96, 12/23/96,			&606.0, 607.0, 608.0,	&6		&1.88$\pm$0.22	&18.3$\pm$4.1	&5.1\\
		&12/30/96, 01/07/97, 01/14/97			&609.0, 610.0, 610.5	&		&		&		&	\\
		&01/21/97, 06/17/97				&611.0, 621.5		&		&		&		&	\\
		&01/19/99			&806.5			&806.5		&1.76$\pm$0.21	&189.2$\pm$24.8	&9.6\\
		&02/08/00			&910.0			&910.0		&2.09$\pm$0.13	&169.2$\pm$22.10&11.8\\
		&02/23/00			&911.1			&911.1		&1.87$\pm$0.24	&134.7$\pm$29.2 &7.0\\

1406-076	&04/09/92			&24.5			&24.5		&1.98$\pm$0.28	&95.4$\pm$23.8 &5.5\\
		&12/22/92			&204.0			&204.0		&2.13$\pm$0.29	&69.7$\pm$17.8&5.5\\
		&12/29/92			&205.0			&205.0		&1.76$\pm$0.17	&134.7$\pm$24.3&8.5\\
		&01/05/93			&206.0			&206.0		&1.86$\pm$0.27	&95.1$\pm$23.1&6.1\\
		&01/12/93			&207.0			&207.0		&2.15$\pm$0.17	&115.1$\pm$14.7&11.5\\

1611+343 (OS~+319)	&11/17/92, 11/24/92			&201.0, 202.0		&2		&2.28$\pm$0.18	&45.4$\pm$8.1&7.7\\
		&11/01/94			&403.0			&4		&2.12$\pm$0.23	&73.5$\pm$16.6&6.3\\
		&03/21/96, 04/23/96, 04/09/97			&516.5, 519.0, 617.8	&56		&2.10$\pm$0.29	&19.1$\pm$5.3&4.6\\		
1622-253	&07/12/91, 12/12/91, 04/28/92,	&5.0, 16.0, 27.0	&1	&2.03$\pm$0.18	&21.2$\pm$5.3	&4.4\\
		&09/09/93, 03/22/94, 04/19/94,	&302.3, 323.0, 324.0,	&3	&2.07$\pm$0.26	&39.5$\pm$8.9	&4.5\\	
		&07/18/94, 08/04/94, 09/20/94	&334.0, 336.5, 339.0	&	&		&		&\\
		&06/06/95, 06/13/95, 06/20/95,	&421.0, 422.0, 423.0,	&4	&2.26$\pm$0.18	&67.10$\pm$11.10&7.9\\
		&06/30/95			&423.5			&	&		&		&\\
1622-297	&06/06/95			&421.0			&421.0		&2.42$\pm$0.15	&228.6$\pm$34.30&9.3\\
		&06/13/95			&422.0			&422.0		&2.22$\pm$0.11	&313.6$\pm$33.50&14.3\\
		&06/20/95			&423.0			&423.0		&2.18$\pm$0.12	&281.0$\pm$35.7&11.2\\
		&06/30/95			&423.5			&423.5		&2.17$\pm$0.09	&228.1$\pm$22.7&14.9\\	
		&03/18/96			&516.1			&516.1		&1.51$\pm$0.27	&184.2$\pm$50.2&5.3\\
		&08/27/96			&529.5			&529.5		&2.41$\pm$0.27	&123.6$\pm$30.6&5.3\\
		&08/05/97			&625.0			&625.0		&2.31$\pm$0.42	&81.8$\pm$33.1&3.3\\

1633+382 (4C~+38.41)	&09/12/91			&9.2			&1		&2.02$\pm$0.08	&109.1$\pm$9.5&17.8\\
		&11/17/92, 11/24/92 			&201.0, 202.0		&2		&2.39$\pm$0.25	&34.4$\pm$7.1&6.4\\
		&03/21/96, 04/23/96, 04/09/97			&516.5, 519.0, 617.8	&56		&2.36$\pm$0.8	&18.9$\pm$4.9&4.7\\

1730-130 (NRAO~530)	&07/12/91, 08/15/91, 10/31/91 			&5.0, 7.2, 13.1			&1		&2.27$\pm$0.32	&19.40$\pm$4.9	&4.4\\
		&12/12/91, 02/06/92, 			&16.0, 20.0	&		&		&		&	\\
		&02/22/93, 03/29/93, 05/05/93			&210.0, 214.0, 219.0		&2		&2.58$\pm$0.32	&35.0$\pm$8.6	&4.6\\
		&05/31/93, 06/19/93, 08/03/93			&223.0, 226.0, 231.0 &		&		&		&	\\
		&08/10/93, 08/12/93, 08/24/93			&229.0, 229.5, 232.0	&		&		&		&	\\
		&09/09/93, 03/22/94, 04/19/94			&302.3, 323.0, 324.0 	&3		&2.53$\pm$0.18	&35.1$\pm$5.9	&6.8\\
		&06/10/94, 06/18/94, 07/18/94			&330.0, 332.0, 334.0	&		&		&		&	\\
		&06/06/95, 06/13/95, 06/20/95			&421.0, 422.0, 423.0	&4	&2.47$\pm$0.09	&34.1$\pm$3.3	&11.8\\
		&06/30/95, 09/20/95			&423.5, 429.0		&		&		&		&	\\
1830-210	&07/12/91, 08/15/91, 10/31/91,		&5.0, 7.2, 13.1,	&1		&2.52$\pm$0.26	&24.2$\pm$5.6	&4.6\\
		&12/12/91, 02/06/92			&16.0, 20.0		&		&		&		&\\
		&02/22/93, 03/29/93, 			&210.0, 214.0, 		&2		&2.33$\pm$0.33	&39.2$\pm$10.2	&4.3\\
		&05/31/93, 06/19/93, 08/03/93		&223.0, 226.0, 231.0 &		&		&		&	\\
		&08/10/93, 08/12/93, 08/24/93		&229.0, 229.5, 232.0	&		&		&		&	\\		
		&09/09/93, 03/22/94, 04/19/94		&302.3, 323.0, 324.0 	&3		&2.44$\pm$0.22	&24.3$\pm$5.8	&4.6\\
		&06/10/94, 06/18/94, 07/18/94		&330.0, 332.0, 334.0	&		&		&		&	\\
2155-304	&11/15/94			&404.0			&404.0		&1.82$\pm$0.23	&30.5$\pm$7.8	&5.9\\
		&02/06/96, 05/21/96			&513.0, 520.4		&56		&1.61$\pm$0.41	&18.9$\pm$6.0	&4.1\\
		&11/11/97, 11/18/97			&701.0,702.0			&v+701.0\tablenotemark{b}		&2.27$\pm$0.38	&67.7.0$\pm$19.7	&4.9\\
2200+420 (BL~Lac)	&01/24/95			&410.0			&410.0		&2.35$\pm$0.34	&32.2$\pm$10.10	&4.0\\
		&07/15/97			&623.5			&623.5		&1.76$\pm$0.15	&148.4$\pm$21.7&10.2\\
2230+114 (CTA~102)	&01/23/92, 04/23/92, 05/07/92,	&19.0, 26.0, 28.0,	&1		&2.30$\pm$0.17	&27.60$\pm$4.60&7.9\\
		&08/20/92			&37.0			&		&    	&		&\\	
		&03/08/94, 05/17/94, 08/01/94	&320.0, 327.0, 336.0	&3		&2.73$\pm$0.29	&16.8$\pm$5.5&3.3\\
2251+158 (3C 454.3)	&01/23/92			&19.0			&19.0		&2.20$\pm$0.10	&83.1$\pm$8.9	&13.5\\	
		&08/20/92			&37.0			&37.0		&2.19$\pm$0.15	&125.1$\pm$19.0	&10.0\\
		&03/08/94, 05/17/94, 08/01/94			&320.0, 327.0, 336.0	&3		&2.26$\pm$0.19	&32.1$\pm$6.8	&5.9\\
		&01/24/95			&410.0			&4		&2.19$\pm$0.12	&51.2$\pm$6.8	&10.9\\		

\enddata
\tablenotetext{a} {The label used to identify each point in the spectral index vs. flux plots shown in Fig. \ref{egspec1}. The labels are represented by a decimal number if the data is from a single viewing period, and an integer if the data from one ore more EGRET Cycle(s) is combined.}
\tablenotetext{b} {Any label starting with v+ indicates a combination of only a fraction of the viewing periods during a cycle. The source was comparatively faint during the other viewing periods of the cycle.}
\tablenotetext{c} {Flux is in units of $10^{-8}$ phot. cm$^{-2}$ s$^{-1}$}
\end{deluxetable}

\clearpage

\begin{deluxetable}{llcccl}
\tablecolumns{6}
\tablecaption{Spectral Variability results. Columns 3, 4 \& 5 are results from the $\chi^2$ test and column 5 lists the correlation coefficient between gamma-ray spectral index and Flux~($>100~MeV$)\label{Tab-4}. The coefficient is
negative when the spectral index (positive) hardens with increasing flux. }
\tablewidth{0pc}
\tablehead{
\colhead{Source} & \colhead{Mean $\pm$ Stdev}     &
\colhead{$\chi^2_{red}$} & \colhead{DOF}\tablenotemark{b}&\colhead{Confidence Level~($\%$)}& \colhead{Pearson's} \\ 
 &($\Gamma_{\mu}\pm\sigma$)  &   &  &   of variability& Corr. Coeff}
\startdata
0208-512&1.95$\pm$0.28&2.93&5&99&+0.10\\
0219+428 (3C~66A)&1.77$\pm$0.16&0.50&1&52&\tablenotemark{a}\\
0430+2859&1.98$\pm$0.41&3.32&2&97&+0.62\\
0528+134&2.36$\pm$0.12&1.10&9&62&-0.34\\
0537-441&2.46$\pm$0.26&2.04&2&87&-0.99\\
0716+714&2.17$\pm$0.29&1.25&3&71&+0.21\\
0827+243 (OJ~248)&2.17$\pm$0.06&0.06&1&19&\tablenotemark{a}\\
0917+499&2.05$\pm$0.12&0.24&2&21&+0.95\\
0954+556 (4C~55.17)&1.98$\pm$0.3&0.83&2&56&-0.77\\
1101+384 (Mrk~421)&1.67$\pm$0.16&0.34&2&28&-0.99\\
1156+295&1.97$\pm$0.34&0.85&3&53&+0.28\\
1219+285 (ON~231)&1.71$\pm$0.51&3.86&2&98&+0.46\\
1222+216 (4C~21.35)&2.24$\pm$0.42&5.07&1&98&\tablenotemark{a}\\
1226+023 (3C~273)&2.55$\pm$0.17&0.60&5&31&+0.11\\
1253-055 (3C~279)&1.98$\pm$0.18&3.9&9&100&-0.32\\
1406-076&1.98$\pm$0.17&0.79&4&46&-0.54\\
1611+343 (OS+319)&2.16$\pm$0.10&0.24&2&22&+0.08\\
1622-253&2.12$\pm$0.12&0.44&2&36&+0.97\\
1622-297&2.17$\pm$0.31&1.63&6&87&-0.07\\
1633+382 (4C+38.41)&2.25$\pm$0.20&4.59&2&99&-0.97\\
1730-130 (NRAO~530)&2.42$\pm$0.16&0.47&4&24&-0.05\\
1830-210&2.43$\pm$0.10&0.1&2&10&-0.91\\
2155-304&1.88$\pm$0.30&0.78&2&54&+0.99\\
2200+420 (BL~Lac)&1.94$\pm$0.25&1.75&1&81&\tablenotemark{a}\\
2230+114 (CTA~102)&2.51$\pm$0.30&2.15&1&86&\tablenotemark{a}\\
2251+158 (3C~454.3)&2.22$\pm$0.03&0.04&3&1&-0.69\\

 \enddata
\tablenotetext{a}{Correlation coefficient was not determined since the sample size was less than 3}
\tablenotetext{b}{The DOF (degrees of freedom) is for the $\chi^2$ test of variabilty and is one less than the sample size.}
\end{deluxetable}

\end{document}